\documentclass[a4paper,fleqn,10pt]{cas-sc}
\usepackage[utf8]{inputenc}
\usepackage[english]{babel}
\usepackage{textcomp}

\usepackage[authoryear]{natbib}

\usepackage{mathtools, amsmath, amssymb}
\usepackage{subcaption}
\usepackage{siunitx}
\usepackage{tabularx}
\usepackage{url}
\usepackage{hyperref}
\usepackage[capitalise]{cleveref}
\crefrangelabelformat{equation}{(#3#1#4-#5#2#6)}
\crefformat{appendix}{#2#1#3}
\usepackage[dvipsnames]{xcolor}
\usepackage{graphicx}
\usepackage{placeins, afterpage}
\usepackage{setspace}
\usepackage{multirow}
\newcommand{\specialcell}[2][c]{\begin{tabular}[#1]{@{}c@{}}#2\end{tabular}}

\newcommand{\abar}{\bar{\alpha}}
\newcommand{\bbar}{\bar{\beta}}
\newcommand{\gbar}{\bar{\gamma}}
\newcommand{\umin}{u_{\rm min}}
\newcommand{\umax}{u_{\rm max}}
\newcommand{\unmin}{u_{n, \rm min}}

\newcommand{\smin}{s_{\rm min}}
\newcommand{\smax}{s_{\rm max}}
\newcommand{\rmin}{r_{\rm min}}
\newcommand{\rmax}{r_{\rm max}}

\def\tsc#1{\csdef{#1}{\textsc{\lowercase{#1}}\xspace}}
\tsc{WGM}
\tsc{QE}

\begin{document}
\let\WriteBookmarks\relax
\def\floatpagepagefraction{1}
\def\textpagefraction{.001}

\shorttitle{ }    

\shortauthors{Trisolini et al.}  

\title [mode = title]{Ejecta cloud distributions for the statistical analysis of impact cratering events onto asteroids' surfaces: a sensitivity analysis}  



%

\author[1]{Mirko Trisolini}[
	orcid=0000-0001-9552-3565,
	linkedin=linkedin.com/in/mirkotrisolini/
]
\cormark[1]
\fnmark[1]
\ead{mirko.trisolini@polimi.it}
\ead[url]{www.cradle.polimi.it}
\credit{Conceptualisation, Methodology, Software, Formal analysis, Writing - Original Draft, Visualisation}

\affiliation[1]{organization={Politecnico di Milano},
            addressline={ Via La Masa 34}, 
            city={Milano},
            postcode={20156},
            country={Italy}}

\author[1]{Camilla Colombo}[
	orcid=0000-0001-9636-9360
]
\fnmark[2]
\ead{camilla.colombo@polimi.it}
\ead[url]{www.compass.polimi.it}
\credit{Writing - Review and Editing, Supervision}

\author[2]{Yuichi Tsuda}  
\fnmark[3]
\ead{tsuda.yuichi@jaxa.jp}
\credit{Writing - Review and Editing, Supervision}
\affiliation[2]{organization={Institute of Space and Astronautical Science (ISAS)/Japan Aerospace Exploration Agency (JAXA)},
	addressline={3-1-1 Yoshinodai, Chuo-ku}, 
	city={Sagamihara, Kanagawa},
	postcode={252-5210}, 
	country={Japan}}

\cortext[1]{Corresponding author}



\begin{abstract}
This work presents the model of an ejecta cloud distribution to characterise the plume generated by the impact of a projectile onto asteroids surfaces. A continuum distribution based on the combination of probability density functions is developed to describe the size, ejection speed, and ejection angles of the fragments. The ejecta distribution is used to statistically analyse the fate of the ejecta. By combining the ejecta distribution with a space-filling sampling technique, we draw samples from the distribution and assigned them a number of \emph{representative fragments} so that the evolution in time of a single sample is representative of an ensemble of fragments. Using this methodology, we analyse the fate of the ejecta as a function of different modelling techniques and assumptions. We evaluate the effect of different types of distributions, ejection speed models, coefficients, etc. The results show that some modelling assumptions are more influential than others and, in some cases, they influence different aspects of the ejecta evolution such as the share of impacting and escaping fragments or the distribution of impacting fragments on the asteroid surface.
\end{abstract}


\begin{highlights}
	\item Development of an ejecta cloud distribution-based model
	\item Characterisation of asteroid ejecta fate via a representative fragments sampling
	\item Sensitivity analysis of ejecta fate to modelling techniques and assumptions
\end{highlights}


\maketitle

\section{Introduction}
\label{sec:intro}

The study of impact cratering and ejecta generation has been at the forefront of recent exploration missions toward small bodies of the Solar System. In 2005, NASA's mission Deep Impact collided with comet Tempel1 at a speed of more than 6 km/s \citep{blume2003deep}; in 2019, JAXA's mission Hayabusa2 carried out an impact experiment on asteroid Ryugu \citep{tsuda2013system,tsuda2019hayabusa2,tsuda2020hayabusa2}, obtaining images and videos of the impact event and the subsequent crater formation. In 2021, the NASA Double Asteroid Redirection Test (DART) mission \citep{CHENG_DART_2018} has been launched toward the Didymos binary system as the first part of the Asteroid Impact and Deflection Assessment (AIDA) joint mission between ESA and NASA \citep{CHENG_AIDA_2015,Rivkin_2021}. DART has impacted Dimorphos on the 26th of September 2022, while the CubeSat LICIACube \citep{2019TortoraLICIA} has recorded the event, sending back to Earth the first images of the impact. The ESA HERA mission \citep{2020EPSC_Michel,2018cospar_Michel} will follow DART as the second part of the AIDA mission to investigate in depth the effects of the impact event, analysing the crater formation, the deflection, and the fate of the ejecta.

Ejecta models are of utmost importance in studying impact phenomena such as the ones associated with these missions. They are used to predict the size of the crater, the number of generated fragments, and the initial conditions of the ejected particles. This information is exploited to predict the fate of ejecta in time and estimate the share of particles re-impacting and escaping the target body. Most of the ejecta models currently available are based on scaling relationships that have been observed in experimental impacts. These relationships are based on point-source solutions, which link impacts of different sizes, velocities, and gravitational accelerations. These scaling relationships are based upon the Buckingham $\pi$ theorem \citep{buckingham1914physically} of dimensional analysis and have had an extensive development over the years \citep{holsapple2007crater,holsapple2012momentum,housen2011ejecta}. Also based on scaling relationships and on principles derived by the Maxwell Z-model of crater excavation \citep{maxwell1975modeling,maxwell1977simple}, Richardson developed a further set of scaling relationships \citep{richardson2007ballistics,richardson2009cratering,richardson2011modeling} for the analysis of the Deep Impact mission to comet Tempel1. In his work, Richardson also derived relationships to describe the in-plane and out-of-plane components of the ejection angles and how they vary with the impact angle.
The work of Housen and Holsapple has been extensively used in recent years to model impact craters. Several works have been dedicated to predict the dynamical fate of the ejecta resulting from the impact of the DART spacecraft with Dimorphos. \cite{yu2017ejecta,yu2018ejecta} performed full-scale simulations of the DART impact, modelling the shape of Didymos as a combination of tetrahedral simplices and Dimorphos as an ellipsoid. In their work, the fate of the ejected particles is studied drawing a few hundred of thousand of samples, assuming a normal impact and studying two possible types of materials for the target asteroid. \cite{rossi2022dynamical} also analyse the evolution of the ejecta plume after the DART impact focusing on the fate of the particles at different time scales, in order to assess the possibility of the future ESA Hera mission \citep{2020EPSC_Michel} to find particles at its arrival. \cite{fahnestock2022pre}, simulate the ejecta plume to obtain synthetic images, representative of the camera view of LICIACube and an Earth-orbiting telescope. Ejecta models have also been used to analyse the effects on the safety and operations of a mission scenarios, as it was the case for Hayabusa2 \citep{soldini2017assessing}. Additionally, when the impact event can be observed, such as for Hayabusa2, ejecta models can help characterise the properties of the target body like the type and strength of the material by comparing the predicted and observed effects \citep{arakawa2020artificial}. The ejecta models are also fundamental to understanding and analysing the momentum transfer associated with an impact event, which can then characterise the deflection capabilities of the impact \citep{holsapple2012momentum,rossi2022dynamical}.

In this work, we present the development of an ejecta cloud distribution-based model used to describe the ejecta parameters (i.e., size, launch position, ejection speed, and ejection direction) via Probability Density Functions (PDFs). The formulation we present starts with a review of existing modelling techniques based on experimental correlations \citep{holsapple2007crater,holsapple2012momentum,housen2011ejecta,richardson2007ballistics,richardson2009cratering,richardson2011modeling,sachse2015correlation}. After the review process, we identified common aspects and differences between the various modelling techniques. The work of Housen and Holsapple \citep{holsapple2007crater,holsapple2012momentum,housen2011ejecta} is mainly dedicated to scaling relationships describing normal impacts. As we are interested in a generic model, also valid for oblique impacts, we integrated the work of Richardson \citep{richardson2007ballistics,richardson2009cratering,richardson2011modeling} to extend the distribution-based model to a generic oblique impact. In addition, following the work of \cite{sachse2015correlation}, we introduce a correlation between the particle size and the ejection speed, which consider larger particles more likely to be ejected at lower speeds.

The work of synthesis we have performed on previous experimental correlations has resulted in a modular formulation in which the ejecta cloud distribution is a combination of probability distribution functions and conditional distributions that describe the parameters characterising the ejecta. By introducing such a modularity, we allow different models to be plugged-in, as long as they can be described as PDFs \citep{trisolini2022scitech}. In this work, we present three different formulations of the ejecta model and assess how they affect the prediction of the ejecta fate. To do so, the ejecta correlations have been reformulated into probability distribution functions and analytical solutions for the corresponding Cumulative Distribution Functions (CDFs) have been obtained. Leveraging the knowledge of the CDFs of the ejecta cloud distribution, we can directly generate sample that follow the cloud distribution. In addition, we can exploit the CDF to analytically integrate the ejecta distribution to estimate the number and mass of particles within specific initial conditions. This feature is exploited in this work to introduce a sampling methodology based on \emph{representative fragments}. With this methodology, each sample represents an ensemble of fragments, to better characterise the overall behaviour of the ejecta cloud. The associated fragments are estimated integrating the distribution in a neighbourhood of the selected sample.

The ability to predict the fate of the ejecta generated by an impact relies on the knowledge of the impactor and target properties, and the impact scenario; however, it also depends on the selection of the ejecta model and the definition of its parameters. Previous works have considered impacts on specific target asteroids and comets (e.g., comet Tempel1, asteroid Ryugu, and the Didymos system) and focused on the effect of the target material and equivalent strength \citep{richardson2007ballistics,holsapple2007crater}, the impact location \citep{yu2017ejecta,yu2018ejecta}, or the asteroid environment. Petit and Farinella \citep{petit1993modelling} combined different scaling laws to model the outcome of the impact between asteroids or other small bodies in the solar system.
In this work, instead, we focus on the ejecta modelling decisions and study how they can affect the overall evolution of the ejecta \citep{trisolini2022iac}. In fact, as several formulations are available in literature for the initialisation of the initial conditions of the ejecta, it is interesting to compare the different modelling techniques available, integrated within the distribution-based formulation we developed. Specifically, we investigate, among others, the effect of the particle size range, different models for the ejection speed, and different distributions for the in-plane and out-of-plane components of the ejection angles. As the study of the motion and fate of the ejecta relies on the models used for their initialisation, it is important to understand to what extent different modelling techniques and different assumptions can affect our predictions.


\bigbreak
\cref{sec:distributions} describes the developed distribution-based formulation for the ejecta model and its mathematical derivation. \cref{sec:sampling} introduces the sampling methodology based on the representative fragments, while \cref{sec:dynamics} describes the dynamical environment for the simulations. \cref{sec:sensitivity} shows the results of the sensitivity analysis on the modelling techniques and \cref{sec:conclusions} summarises the conclusions of the study.

\section{Ejecta field distribution}
\label{sec:distributions}

This section provides the description of the modelling technique used to represent the characteristics of an ejecta field after a hypervelocity impact onto celestial bodies. Specifically, we focus the attention on the impact of small-kinetic impactor onto small-bodies surfaces. The proposed modelling principles is based on the scaling relations experimentally derived by \cite{housen2011ejecta,holsapple2007crater,richardson2007ballistics} and leverages their work to build a continuous model, where the ejecta field is described using a combination of probability density functions so that the particle number density is recovered as a function of the ejection variables. In addition, the relevant cumulative distributions are derived, which can be used to directly sample the distribution and obtain the relevant initial conditions after the impact event. Finally, the distribution can be integrated to estimate the number of particles having specified ejection conditions, allowing a better understanding of the overall fate of the ejected particles.

\begin{figure}[!htb]
	\centering
	\includegraphics[width=2.6in]{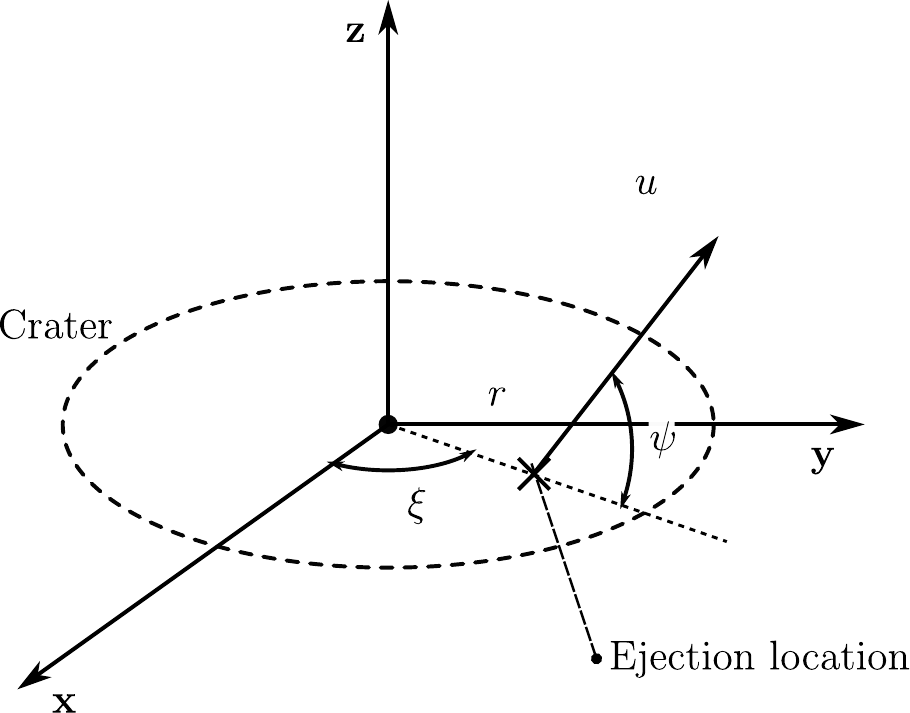}
	\caption{Schematics of the ejection parameters used for the ejecta distributions. The reference frame \emph{xyz} is a local horizontal frame tangent to the asteroid surface.}
	\label{fig:ejection_params}
\end{figure}

The most general expression of the ejecta distribution is a function $p(\mathbf{x})$ that represents the particles number density as a function of the ejection parameters, $\mathbf{x}$. The ejection parameters considered in this work are the particle radius, $s$, the particle launch position (i.e., the radial distance from the centre of the crater to the ejection location), $r$, the particle speed, $u$, and the in-plane and out-of-plane components of the ejection angle, $\xi$ and $\psi$, respectively. \cref{fig:ejection_params} shows the physical meaning of the ejection parameters in a local horizontal reference frame, tangent to the asteroid surface at the impact point. In general, the size and ejection angles are always present in the models; speed and position are instead mutually exclusive as they can be related according to experimental correlations \citep{housen2011ejecta,richardson2007ballistics}. Both cases have been analysed: \cref{subsec:position-dist} describes a model that considers the particle's launch position, $r$, while \cref{subsec:speed-dist,subsec:correlated-dist} both describe a model based on the ejection speed, $u$.

The following sections describes the different formulations developed for the ejecta distributions. Three main types of formulations have been considered. A first formulation, identified as \emph{position-based} where the distribution variables are $\mathbf{x} = \left\{ s, r, \xi, \psi \right\}$ and the speed is then obtained from the particle's launch position, $r$. A second formulation, identified as \emph{speed-based}, which instead drops the dependency on $r$ and for which the distribution variables are $\mathbf{x} = \left\{ s, u, \xi, \psi \right\}$. A last formulation that is a \emph{correlated} version of the \emph{speed-based} formulation in which the size and speed of the ejected particles are correlated. In fact, it is reasonable to expect that, on average, smaller particles have higher velocities and vice versa \citep{sachse2015correlation}. In the other two formulations, size and speed are not correlated and all particle sizes can assume any ejection velocity in the distribution range. \cref{sec:sensitivity} will present a comparison between the different formulations.

\subsection{Position-based distribution}
\label{subsec:position-dist}
The \emph{position-based} distribution formulation, derives the particle number density as a function of $s$, $r$, $\xi$, and $\psi$. As previously mentioned the model is based on experimental correlations, which have been mainly developed for normal impacts. In this work, we wish to derive a general distribution for oblique impacts by combining the results from \cite{housen2011ejecta} and \cite{richardson2007ballistics}. For a normal impact, we consider the distribution in $\mathbf{x}$ to have the following expression:

\begin{equation} \label{eq:normal-dist}
	p(\mathbf{x}) = p(s, r, \xi_n, \psi_n) = p_s (s) \cdot p_{\xi_n} (\xi_n) \cdot p_{\psi_n|r} (\psi_n|r) \cdot p_r (r) \rm,
\end{equation}

\noindent where $\xi_n$ and $\psi_n$ are the in-plane and out-of-plane ejection angles relative to a normal impact (i.e., an impact perpendicular to the surface of the target). We observe that, even for a normal impact, the out-of-plane ejection angle, $\psi_n$, depends on the launch position, $r$ \citep{richardson2007ballistics,richardson2011modeling}; therefore, we represent it with a conditional distribution.

However, to increase the generality of our formulation, we derive the distribution for a generic oblique impact; therefore we introduce the following transformation \citep{richardson2007ballistics,richardson2011modeling}:

\begin{equation}  \label{eq:transformation-position-based}
	\mathcal{T} : \begin{cases}
		\xi = \xi_n \\
		\psi = \psi_n - \frac{\pi}{6}  \cos \phi  \left( \frac{1 -\cos \xi}{2} \right) \left(1 - \frac{r}{\rmax} \right)^2
	\end{cases} \rm ,
\end{equation}

\noindent where $\phi$ is the impact angle, measured from a plane tangent to the target surface, and $\rmax$ is the maximum launch position, which does not necessarily coincide with the transient crater radius \citep{housen2011ejecta}. The variable $\xi$ is measured from the x-axis (as shown in \cref{fig:ejection_params}), which is defined as the projection of the impactor's incoming surface-relative velocity vector onto the plane tangent to the target surface. The expression of $\psi$ of \cref{eq:transformation-position-based} is derived from the work of \cite{richardson2007ballistics,richardson2011modeling} and expresses the variation of the out-of-plane ejection angle as function of the distance from the crater's centre and the in-plane ejection angle, $\xi$. By using this transformation, we can obtain the ejecta distribution for an oblique impact, starting from a normal impact as follows:

\begin{equation}
	p(s, r, \xi, \psi) = \frac{p(s, r, \xi_n, \psi_n)}{|J(\mathcal{T})|} \rm ,	
\end{equation}

\noindent where $|J(\mathcal{T})|$ is the determinant of the Jacobian of the transformation of \cref{eq:transformation-position-based}, which, in this case, equals to one. Alongside the transformation of variables, we need to take into account the dependency of the distributions from other variables. \cref{subsubsec:inplane-dist-position-based,subsubsec:outplane-dist-position-based} will show that the in-plane and out-of-plane ejection angles depend on a subset of the ejection variables so that the expression for the ejecta distribution can be re-written as:

\begin{equation} \label{eq:conditional-position-based}
	p(s, r, \xi, \psi) = p_s(s) \cdot p_{\psi | \xi, r} (\psi | \xi, r; \phi) 	\cdot p_{\xi | r} (\xi | r; \phi) \cdot p_r (r) \rm ,
\end{equation}

\noindent where $p_{\psi | \xi, r} (\psi | \xi, r; \phi)$ is the conditional distribution of $\psi$ given $\xi$ and $r$, and having fixed the impact angle $\phi$; $p_{\xi | r} (\xi | r; \phi)$ is the conditional distribution of $\xi$ given $r$ and fixing $\phi$. Therefore, we have a combination of probability distributions in $s$ and $r$, and conditional distributions in $\psi$ and $\xi$. We will now describe in detail all these contributions to the overall ejecta distribution.

\subsubsection{Particle size}
\label{subsubsec:size-dist}
In this formulation, the size distribution is considered to be independent from all the other variables. The distribution derives from the power law expression of the reverse cumulative distribution of the number of particles as a function of their size \citep{krivov2003impact}.

\begin{equation}  \label{eq:cum_size_unc}
	G(> s) = N_r \cdot s^{-\abar} \rm .
\end{equation}

Here, $N_r$ is a multiplicative factor that can be determined from mass conservation. Following the notation of Sachse \citep{sachse2015correlation}, $\abar$ is the exponent defining the slope of the power law. 

%

Differentiating \cref{eq:cum_size_unc}, we obtain the differential density distribution function, which has the following expression:

\begin{equation} \label{eq:size-distribution}
	p_s(s) = \frac{d G (< s)}{ds} = \frac{d \left( N_{\rm tot} - G (> s) \right)}{ds} = \abar N_r s^{-1 - \abar} \rm ,
\end{equation}

\noindent with $\smin \leq s \leq \smax$. We can then obtain $N_r$ from mass conservation as follows:

\begin{equation} \label{eq:nr}
	M_{\rm tot} = \frac{4}{3} \pi \rho \int_{\smin}^{\smax} s^3 p_s(s) \, d\!s \quad \rightarrow \quad N_r = \frac{3 (3 - \abar) M_{\rm tot}}{4 \abar \left( \smax^{3-\abar} - \smin^{3-\abar}\right) \pi \rho}	\rm ,
\end{equation}

\noindent where $M_{\rm tot}$ is the total mass ejected from the crater, $\rho$ is the density of the asteroid, and $\smin$ and $\smax$ are the minimum and maximum particle radii, respectively. The minimum and maximum radii are free parameters of the model that can be selected by the user. Commonly used values are 10-100 \si{\micro\meter} for the minimum diameter and 1-10 \si{\centi\meter} for the maximum one \citep{yu2018ejecta}. The total mass ejected can instead be derived from experimental correlations as follows \citep{housen2011ejecta}:

\begin{equation}  \label{eq:mtot}
	M_{\rm tot} = k \rho \left[ (n_2 R_c)^3 - (n_1 a)^3 \right] \rm ,
\end{equation}

\noindent where $a$ is the impactor diameter, and $k$, $n_1$, and $n_2$ are coefficients depending on the type of material and impact derived from experimental correlations. The works of Housen and Holsapple contain extensive coverage for the derivation and usage of these parameters. The interested reader is referred to their work \citep{housen2011ejecta,holsapple2007crater,holsapple2012momentum}. 

\subsubsection{Launch position}
\label{subsubsec:launch-dist}
The probability distribution of the launch position can be derived from the expression of the mass ejected within a distance $r$ from the crater origin. This expression has been derived by Housen \citep{housen2011ejecta} and has the following form:

\begin{equation}  \label{eq:mx}
	M(<r) = k \rho \left[r^3 - (n_1 a)^3 \right] \quad\quad \text{with} \quad n_1 a \leq r \leq n_2 R_c \rm .
\end{equation}

We can thus obtain the CDF in $r$ by simply dividing by the total mass (\cref{eq:mtot}).

\begin{equation}  \label{eq:r-cdf}
	P_r (<r) = \frac{k \rho}{M_{\rm tot}} \left( r^3 - \rmin^3 \right) \rm ,
\end{equation}

\noindent where $\rmin = n_1 a$. Analogously, we identify $\rmax = n_2 R_c$ as the maximum launch position (i.e., the maximum radial distance from the centre of the crater). By differentiating the CDF, we get the probability distribution of \cref{eq:r-distribution} \citep{pishro2016introduction}.

\begin{equation}  \label{eq:r-distribution}
	p_r (r) = \frac{3 k \rho}{M_{\rm tot}} r^2 \quad\quad \text{with} \quad \rmin \leq r \leq \rmax \rm .
\end{equation}

\subsubsection{In-plane ejection angle}
\label{subsubsec:inplane-dist-position-based}
\begin{figure}[htb!]
	\centering
	\includegraphics[width=2.6in]{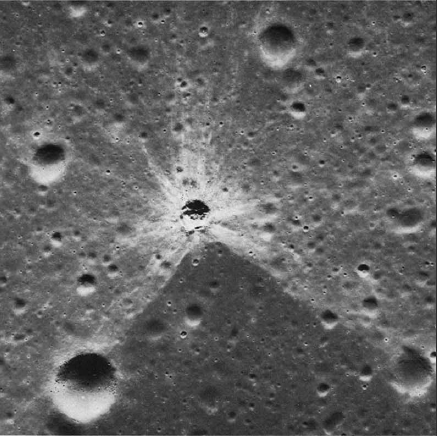}
	\caption{Image of the ejecta field resulting from an oblique impact on the Moon for a crater of about 0.5 \si{\kilo\meter} in diameter \citep{richardson2011modeling}.}
	\label{fig:crater-oblique}
\end{figure}

The distribution of the in-plane ejection angle expresses the azimuthal variation of the ejected samples. For normal impacts, this distribution is uniform and the fragments are ejected with the same probability within the range 0$^\circ$-360$^\circ$. For oblique impacts, instead, the uniformity is lost and the distribution starts to assume more complex patterns such as the one of \cref{fig:crater-oblique}.  

Richardson proposed an expression for the distribution of such patterns \citep{richardson2011modeling}. However, it was considered too complex for the implementation in the proposed framework as it is not integrable. Therefore, starting from the work of Richardson, we propose the following expression:

\begin{equation} \label{eq:xi-distribution}
	p_{\xi|r}(\xi | r; \phi) = \frac{1}{2 \pi} \left[ 1 - \cos \phi \left( \cos 2\xi \cdot \cos^3 \xi - \frac{1}{5} \cos \xi \cdot \cos^4 2\xi \right) \left( 1 - \frac{r^8}{\rmax^8} \right) \right] \rm .
\end{equation}

\begin{figure}[htb!]
	\centering
	\includegraphics[width=2.6in]{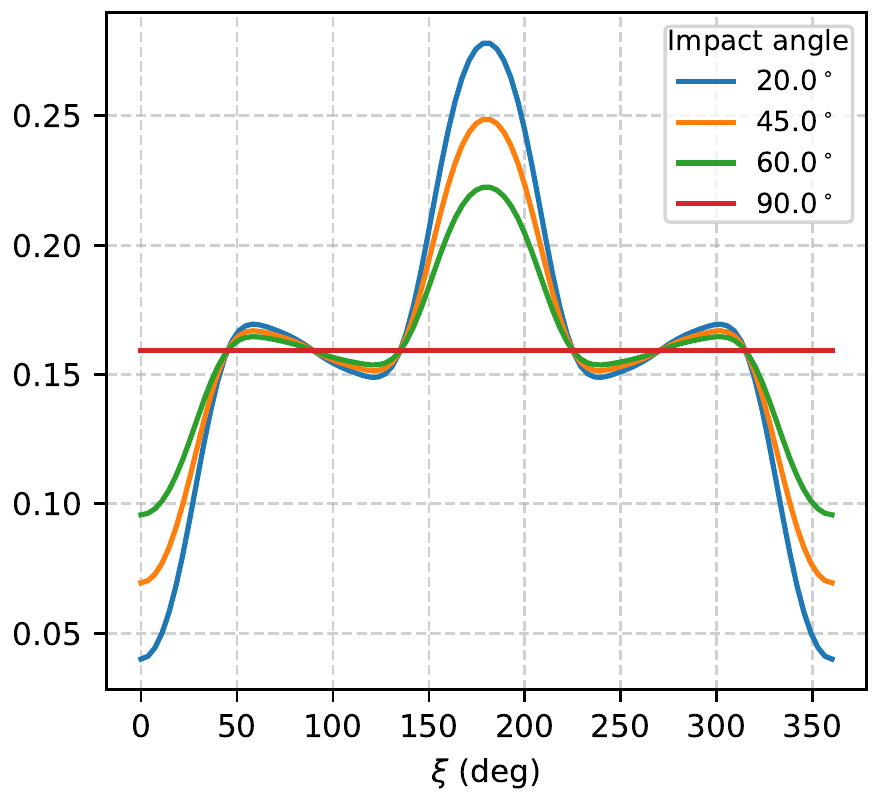}
	\caption{Variation of the in-plane ejection angle \emph{lobed} probability distribution function with respect to the impact angle.}
	\label{fig:xi-lobed-dist-variation}
\end{figure}

In addition, differently from Richardson, we have introduced $\rmax$, instead of the crater radius, to maintain the generality of the model among both Richardson \citep{richardson2011modeling}, and Housen and Holsapple \citep{housen2011ejecta} correlations. As it is possible to observe, \cref{eq:xi-distribution} is a conditional distribution of $\xi$ given $r$. The angle $\xi$ is measured from the direction of the incoming projectile; therefore, $\xi = 180^\circ$ is downstream to the incoming projectile. An example of the distribution as function of the impact angle is shown in \cref{fig:xi-lobed-dist-variation}, where we see the peak of the distribution along the projectile direction and two other smaller peaks on the sizes representing symmetrical \emph{lobes}. For a 90\si{\degree} impact angle, the distribution degenerates to uniform.

As discussed in \cite{richardson2011modeling}, a distribution of the type of \cref{eq:xi-distribution} cannot fully describe the complexity of the cratering and ejection phenomenon; however, it can be used for a first analysis of the behaviour of the ejected particles after the impact.

\begin{figure}[htb!]
	\centering
	\includegraphics[width=2.6in]{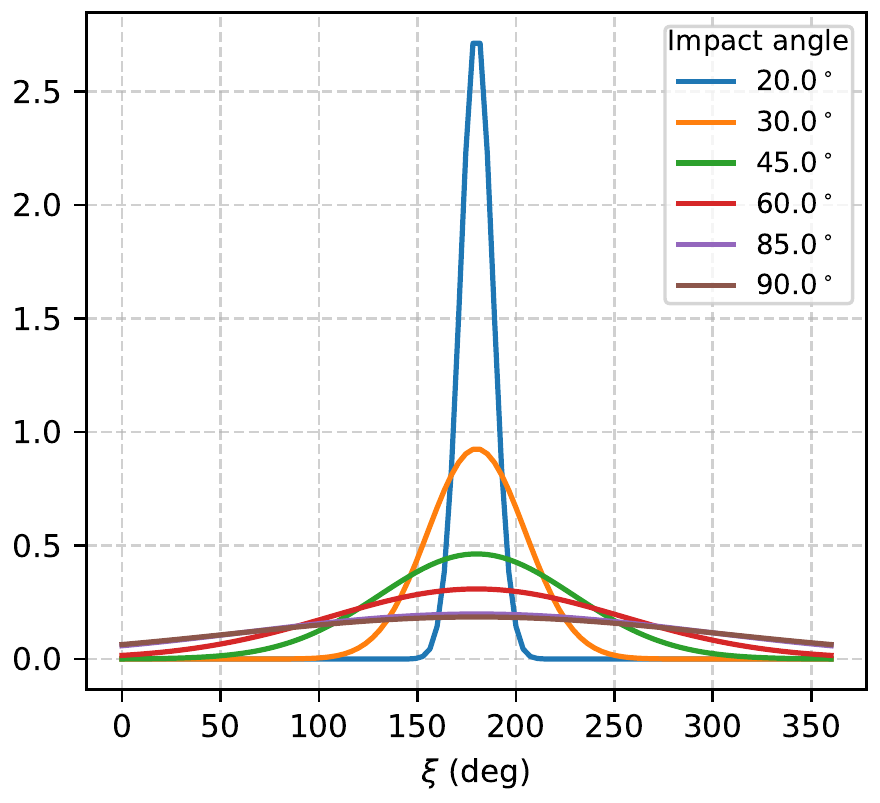}
	\caption{Variation of the in-plane ejection angle Gaussian probability distribution function with respect to the impact angle.}
	\label{fig:xi-dist-variation}
\end{figure}

Alongside the distribution of \cref{eq:xi-distribution}, we can also introduce a simpler and more manageable description, where only the main lobe is taken into account and the dependency on the launch position, $r$, is dropped. In this case, we model the in-plane ejection angle with a Gaussian distribution:

\begin{equation}  \label{eq:xi-distribution-gaussian}
	p_{\xi}(\xi) = \mathcal{N} (\mu_\xi, \sigma_\xi) \rm ,
\end{equation}

\noindent where $\mu_\xi$ and $\sigma_\xi$ are the mean and standard deviation of the distribution respectively. As the main lobe is in the downstream direction with respect to the incoming projectile, we have $\mu_\xi = 180^\circ$. For the standard deviation, instead, we assume it varies linearly with the impact angle as follows:

\begin{equation}  \label{eq:sigma-xi}
	\sigma_\xi = \frac{2}{5} \pi \cdot \frac{\phi - \phi_{\rm min}}{\phi_{\rm max} - \phi_{\rm min}} \quad \text{with} \quad  \phi_{\rm min} \leq \phi < \phi_{\rm max} \rm, 
\end{equation}

\noindent where $\phi_{\rm min}$ and $\phi_{\rm min}$ are the minimum and maximum impact angles, respectively. For the presented model, $\phi_{\rm max} = 90^\circ$, while $\phi_{\rm min} = 20^\circ$ as the experimental models on which the distribution formulation is based are valid only down to an impact angle of 20$^\circ$ (\cref{fig:xi-dist-variation}). The variation fo the standard deviation in \cref{eq:sigma-xi} was extrapolated from the work of \cite{yamamoto2002measurement}. In addition, care must be taken for normal impacts; in this case, the distribution is not Gaussian anymore and we fall back to a uniform distribution.

\cref{eq:xi-distribution,eq:xi-distribution-gaussian} describe two options for the characterisation of the in-plane component of the ejection angle after an oblique impact. These distributions can be seen as a starting point for the characterisation of the ejecta field. For example, by using images of the impact event these distributions may be tuned to better describe the event in exam. A different formulation may also be used; for example, a Gaussian mixture model may better be fitted to image data. Given the modularity of the presented formulation, such options can be explored in future works.

\subsubsection{Out-of-plane ejection angle}
\label{subsubsec:outplane-dist-position-based}
The out-of-plane component of the ejection angle, $\psi$, defines how steep the launch angle of the particles is with respect to the body surface. In several studies, this angle is assumed constant and equal to 45$^\circ$ for all the ejected particles. However, as also shown by experimental results, different particles will possess different ejection angles. According to \cite{richardson2007ballistics}, the ejection angle is typically within 27\si{\degree} and 63\si{\degree}. 

\bigbreak
We consider two different options for the distribution in $\psi$. For the distribution in exam, we first derive the distribution for normal impacts and then apply the transformation of \cref{eq:transformation-position-based} to obtain the oblique formulation. The first option is a spherically uniform distribution as follows:

\begin{equation}  \label{eq:psin-dist}
	p_{\psi_n} (\psi_n) = \frac{\cos \psi_n}{\sin \psi_{n,\rm max} - \sin \psi_{n,\rm min}} \quad \text{with} \:\: \psi_{n,\rm min} \leq \psi_n \leq \psi_{n,\rm max} \rm,
\end{equation}

\noindent which is defined within the aforementioned limits of $\psi_{n,\rm min} =$ 27\si{\degree} and $\psi_{n,\rm max} =$ 63\si{\degree} \citep{richardson2007ballistics}. Inverting the transformation of \cref{eq:transformation-position-based} and substituting into \cref{eq:psin-dist}, we obtain the distribution in $\psi$:

\begin{equation} \label{eq:psi-distribution-uniform}
	p_{\psi|\xi, r} (\psi| \xi, r; \phi) = \frac{\cos \left( \psi + K_\psi (\xi, r; \phi) \right) }{\sin \psi_{n,\rm max} - \sin \psi_{n,\rm min}} \quad \text{with} \:\: \psi_{\rm min} \leq \psi \leq \psi_{\rm max} \rm,
\end{equation}

\noindent where $K_\psi (\xi, r; \phi) = \frac{\pi}{6}  \cos \phi  \left( \frac{1 -\cos \xi}{2} \right) \left(1 - \frac{r}{\rmax} \right)^2$. Note that that now we have a conditional distribution in $\psi$, given $r$ and $\xi$. In addition, $\psi_{\rm min} =  \psi_{n,\rm min} - K_\psi(\xi, r; \phi)$ and $\psi_{\rm max} = \psi_{n,\rm max} - K_\psi(\xi, r; \phi)$; therefore, also the limits depend on $\xi$ and $r$.

\bigbreak
For the second option, we start from the work of \cite{richardson2007ballistics}. Here, it is observed that the ejection angle tends to decrease with the distance from the impact point, $r$. A simple linear scaling is considered as follows:

\begin{equation} \label{eq:psin}
	\psi_n (r) = \psi_0 - \psi_d \cdot \frac{r}{\rmax} \rm, 
\end{equation}

\noindent where $\psi_0 = 52.4^\circ \pm 6.1^\circ$ is the starting angle and $\psi_d = 18.4^\circ \pm 8.2^\circ$ is the total drop angle, using 2$\sigma$ errors \citep{richardson2007ballistics}. The values of $\psi_0$ and $\psi_d$ have been derived by laboratory shots performed by Cintala et al. \citep{cintala1999ejection}. These values can be used as starting points; however, they can be changed and tailored to the specific impact, using for example, direct imaging of the impact event. As the expression of \cref{eq:psin} is expressed as a combination of means and standard deviations, we can assume they can be treated as Gaussian distributions and combine them to have:

\begin{equation}
	p_{\psi_n|r} (\psi_n|r) = \mathcal{N} (\mu_n, \sigma_n) \rm,
\end{equation}

\noindent where $p_{\psi_n|r} (\psi_n|r)$ is a conditional probability distribution of $\psi_n$, given the position, $r$. The mean and standard deviation derive from the combination of the characteristics of the distribution of $\psi_0$ and $\psi_d$, as follows:

\begin{equation}
	\begin{cases}
		\mu_n = \mu_0 - \mu_d \cdot \frac{r}{\rmax} \\
		\sigma_n^2 = \sigma_0^2 + \sigma_d^2 \cdot \left( \frac{r}{\rmax} \right)^2
	\end{cases} \rm
\end{equation}

\noindent and $\mu_0 =$ 52.4$^\circ$, $\sigma_0 =$ 3.05$^\circ$, $\mu_d =$ 18.4$^\circ$, $\sigma_d =$ 4.1$^\circ$. Similarly to \cref{eq:psi-distribution-uniform}, we obtain the distribution for an oblique impact as follows:

\begin{equation}  \label{eq:psi-distribution-gaussian}
	p_{\psi|\xi, r} (\psi| \xi, r; \phi) = \mathcal{N} (\mu_n - K_\psi (\xi, r; \phi), \sigma_n)
\end{equation}

\subsubsection{Ejection speed}
\label{subsubsec:u-dist-position-based}
\begin{figure}[htb!]
	\centering
	\includegraphics[width=2.6in]{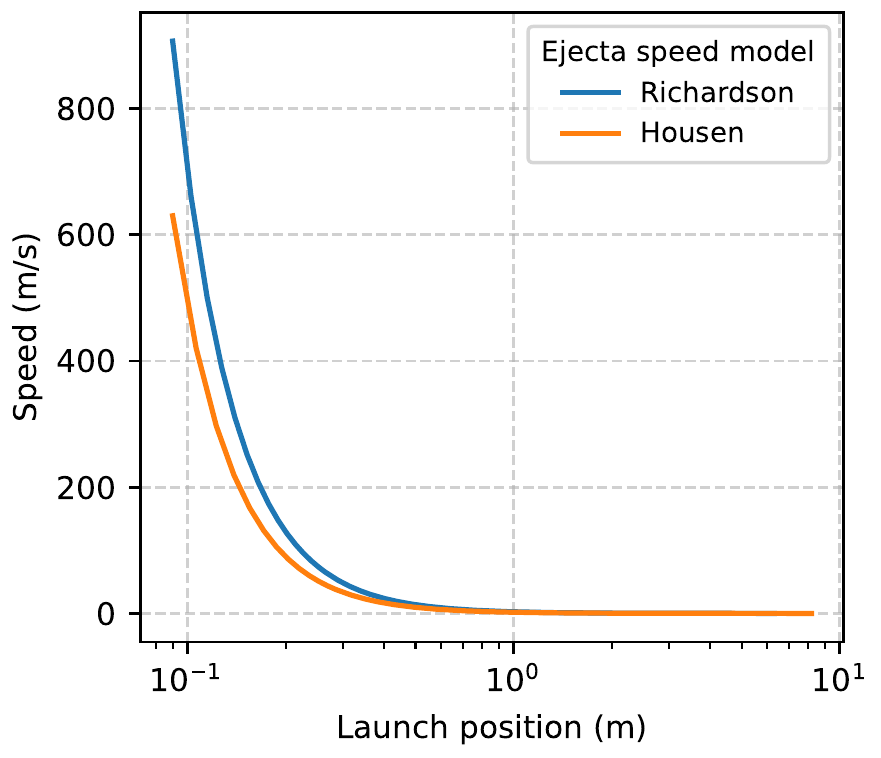}
	\caption{Example of the comparison between the ejection speed obtained with \cref{eq:speed-housen} (orange) and \cref{eq:speed-richardson} (blue).}
	\label{fig:speed-comparison}
\end{figure}

In this \emph{position-based} formulation, the ejection speed is a derived quantity that depends on the other variables, which are sampled from the distribution described in \cref{subsubsec:size-dist,subsubsec:launch-dist,subsubsec:inplane-dist-position-based,subsubsec:outplane-dist-position-based}. The expression for the ejection speed for oblique impacts has the following form \citep{richardson2007ballistics}:

\begin{equation}  \label{eq:ejection-speed-oblique}
	u_{\rm ej} = \sqrt{\left( u_n \sin \psi_n \right)^2 + \left( \frac{u_n \sin \psi_n}{\tan \psi} \right)^2} = u_n \cdot \frac{\sin \psi_n}{\sin \psi} \rm,
\end{equation}

\noindent where $u_n$ is the ejection speed associated to an equivalent normal impact. In principle, \cref{eq:ejection-speed-oblique} expresses the variation of the ejection speed as a function of the impact angle. This expression is a function of $r$, $\xi$, and $\psi$.
\bigbreak
The normal ejection speed, $u_n$ can have different expressions; in this work, we selected two commonly used experimentally derived expressions. The first one has been derived by Housen et al \citep{housen2011ejecta,holsapple2007crater} and has the following form:

\begin{equation}  \label{eq:speed-housen}
	u_n (r) = C_1 \cdot U \cdot \left[ \frac{r}{a} \cdot \left( \frac{\rho}{\delta} \right)^\nu \right]^{-\frac{1}{\mu}} \cdot \left( 1 - \frac{r}{\rmax} \right)^p \rm,
\end{equation}

\noindent where $C_1$, $\mu$, $\nu$, $p$, and $n_2$ are coefficients depending on the target material, $\delta$ is the density of the impactor and $U$ is the impactor speed. As \cref{eq:speed-housen} refers to normal impacts, $U$ is the impactor velocity normal to the impact surface. Therefore, for an oblique impact, $U = U_{\rm imp} \cdot \sin \phi$, where $U_{\rm imp}$ is the absolute magnitude of the impactor speed. The second expression, instead, has been derived by Richardson \citep{richardson2007ballistics} and is the following:

\begin{equation}
	\begin{cases}  \label{eq:speed-richardson}
		u_n (r) &= \sqrt{u_e^2 - C_{vpg}^2 \cdot g \cdot r - C_{vps}^2 \cdot \frac{Y}{\rho}} \\
		u_e &= C_{vpg} \cdot \sqrt{ g \cdot R_{c,g} } \cdot \left( \frac{r}{ n_2 R_{c,g} } \right)^{-\frac{1}{\mu}} 
	\end{cases} \rm,
\end{equation}

\noindent where $C_{vpg}$ and $C_{vps}$ are proportionality constants in the gravity and strength regime, respectively, $g$ is the net acceleration at the impact point on the surface of the small body accounting for both the small body's gravitation and its rotation, $ R_{c,g}$ is the crater radius in the gravity regime, and $Y$ is the strength of the small body material. \cref{fig:speed-comparison} compares the two velocity expression of \cref{eq:speed-housen} and \cref{eq:speed-richardson} for the same normal impact. The most noticeable feature is the different magnitude of the speed in the high-velocity region, where the model derived by Richardson returns higher speed up to about 200 \si{\meter\per\second}.

\subsection{Speed-based distribution}
\label{subsec:speed-dist}
The second formulation we propose is identified as \emph{speed-based}. Here, we remove the dependency on the launch position, $r$; therefore, the ejection location on the target surface is the same for all the particles, thus assuming that the crater size is small with respect to the target radius. The consequence of this assumption is the substitution of the position distribution, $p_r (r)$, with a speed distribution, $p_u (u)$. In addition, for the in-plane and out-of-plane ejection angles, we average out the contribution of the launch position (\cref{subsubsec:inplane-dist-speed-based,subsubsec:outplane-dist-speed-based}). In this case the transformation from normal to oblique impact is thus:

\begin{equation}  \label{eq:transformation-speed-based}
	\mathcal{T} : \begin{cases}
		u = \sqrt{\left( u_n \sin \psi_n \right)^2 + \left( \frac{u_n \sin \psi_n}{\tan \psi} \right)^2} = u_n \cdot \frac{\sin \psi_n}{\sin \psi} \\
		\xi = \xi_n \\
		\psi = \psi_n - \frac{\pi}{18}  \cos \phi  \left( \frac{1 -\cos \xi}{2} \right) \left(1 - \frac{\rmin}{\rmax} \right)^2 = \psi_n - \bar{K}_\psi (\xi; \phi)
	\end{cases} \rm ,
\end{equation}

\noindent where the first equation derives from \cref{eq:ejection-speed-oblique}, observing that $\sin \psi$ and $\sin \psi_n$ are always between zero and 1, because the out-of-plane ejection angle belongs to the interval $\left[ 0^\circ, 90^\circ \right]$. The third equation derives from the averaging over $r$ of the corresponding expression of \cref{eq:transformation-position-based}. We will see in the following sections that introducing the transformation of \cref{eq:transformation-speed-based}, the distribution can then be expressed as follows:

\begin{equation} \label{eq:conditional-speed-based}
	p(s, u, \xi, \psi) = p_s(s) \cdot p_{u | \xi, \psi} (u | \xi, \psi; \phi) \cdot p_{\psi | \xi} (\psi | \xi; \phi) \cdot p_\xi (\xi) \rm .
\end{equation}

In the following sections, the building blocks of \cref{eq:conditional-speed-based} will be described, except for the size distribution, which is equivalent to \cref{subsubsec:size-dist}.

\subsubsection{In-plane ejection angle}
\label{subsubsec:inplane-dist-speed-based}

In \cref{subsubsec:inplane-dist-position-based}, we introduced two different expressions for the in-plane ejection angle distribution. For the expression based on a single lobe and modelled by a Normal distribution, no further modification is needed as the expression is already independent from the launch position, $r$. The expression of \cref{eq:xi-distribution}, instead, depends on $r$; therefore, we remove this contribution by averaging out in $r$ as follows:

\begin{equation}  \label{eq:xi-dist-lobed-speed-based}
	\begin{split}
		p_\xi (\xi) &= \frac{1}{\rmax - \rmin} \int_{\rmin}^{\rmax} p_{\xi|r} (\xi | r) \, d\!r = \\ &=
		\frac{1}{2\pi} \left[1 - \bar{K}_{\xi r} \cdot \cos \phi \left( \cos 2\xi \cdot \cos^3 \xi - \frac{1}{5} \cos \xi \cdot \cos^4 2\xi \right) \right] \rm .
	\end{split}
\end{equation}

\noindent where $\bar{K}_{\xi r} = \left[ 8 - 9 \frac{\rmin}{\rmax} + \left( \frac{\rmin}{\rmax} \right)^9 \right] / \left[ 9 \left( \rmax - \rmin \right) \right]$.

\subsubsection{Out-of-plane ejection angle}
\label{subsubsec:outplane-dist-speed-based}
In a similar fashion, we can obtain the out-of-plane ejection angle distribution by averaging over $r$ the equations of \cref{subsubsec:outplane-dist-position-based}. For the spherically uniform formulation, the contribution of the launch position is limited to the transformation of \cref{eq:transformation-position-based}. Therefore, to remove the dependency, we simply use the averaged expression of \cref{eq:transformation-speed-based}. The resulting distribution is analogous to \cref{eq:psi-distribution-uniform}, with the only substitution of $K_\psi (\xi, r; \phi)$ with $\bar{K}_\psi (\xi; \phi)$. 
\bigbreak
For the Gaussian distribution formulation, we proceed in a similar fashion by averaging out \cref{eq:psin} as follows:

\begin{equation}
	\begin{split}
		\bar{\psi} &= \frac{1}{\rmax - \rmin} \int_{\rmin}^{\rmax} \left( \psi_0 - \psi_d \frac{r}{\rmax} \right) \, d\!r \\ &= \psi_0 -  \frac{\rmax + \rmin}{2 \cdot \rmax} \cdot \psi_d
	\end{split} \rm .
\end{equation}

Therefore, following the same procedure of \cref{subsubsec:outplane-dist-position-based}, we obtain the distribution of the out-of-plane ejection angle for a normal impact, averaged over the launch position:

\begin{equation}
	p_{\psi_n} (\psi_n) = \mathcal{N} (\bar{\mu}_n, \bar{\sigma}_n) \rm ,
\end{equation}

\noindent with modified values of the mean and standard deviations as follows:

\begin{equation}
	\begin{cases}
		\bar{\mu}_n &= \mu_0 -  \frac{\rmax + \rmin}{2 \cdot \rmax} \cdot \mu_d\\
		\bar{\sigma}_n^2 &= \sigma_0^2 + \left( \frac{\rmax + \rmin}{2 \cdot \rmax} \right)^2 \cdot \sigma_d^2
	\end{cases} \rm .
\end{equation}

Finally, the Gaussian model of the out-of-plane ejection angle, $\psi$, distribution of the \emph{speed-based} formulation for oblique impacts has the following expression:

\begin{equation}  \label{eq:psi-dist-gaussian-speed-based}
	p_{\psi|\xi} (\psi| \xi; \phi) = \mathcal{N} (\bar{\mu}_n - \bar{K}_\psi (\xi; \phi), \bar{\sigma}_n) \rm .
\end{equation}

\subsubsection{Ejection speed}
\label{subsubsec:u-dist-speed-based}
The ejection speed distribution is the main difference between the \emph{position-based} and the \emph{speed-based} formulations. To derive the speed distribution, we assume the distribution is of the form $p_{u_n} (u_n) = C_u \cdot u_n^{-1 -\gbar}$, where $u_n$ is the ejection speed after a normal impact \citep{sachse2015correlation} and $\gbar$ is an exponent that determines the slope of the speed distribution. The value of $\gbar$ depends on the characteristics of the target material. Comparing the speed distribution with experimental correlations, we derive that $\gbar = 3\mu$ \citep{trisolini2022scitech}. To compute the constant $C_u$, we simply impose that the integral of the probability density function is equal to unity, so that the final expression of the distribution is:

\begin{equation}  \label{eq:un-dist-speed-based}
	p_{u_n} (u_n) = \frac{\gbar}{u_{n, min}^{-\gbar} - u_{n, max}^{-\gbar}} u_n^{-1 - \gbar} \rm .
\end{equation}

The values of the minimum and maximum ejection speeds can be provided by the user. Possible values can be derived from \cref{eq:speed-housen}. Values for the minimum speed can also be derived from the simplified expressions for the \emph{knee velocity}, $v{*}$ that is the approximate value of the slowest ejecta velocity, which is found from laboratory experiment \citep{holsapple2012momentum}. The \emph{knee velocity} has two expressions depending whether the impact is in the strength ($v_g^{*}$) or in the gravity ($v_s^{*}$) regime:

\begin{equation}  \label{eq:knee-velocity}
	\begin{cases}
		v_g^{*} &= K_{\rm vs} \sqrt{\frac{Y}{\rho}}  \\
		v_s^{*} &= K_{\rm vg} U \left( \frac{g}{a U^2} \right)^{\frac{1}{2+\mu}}
	\end{cases} \rm .
\end{equation}

It is interesting to note here the main difference between the \emph{speed-based} and \emph{position-based} formulation. The speed distribution of the \emph{position-based} formulation (\cref{eq:speed-richardson}, \cref{eq:speed-housen}), has a minimum ejection speed, at the crater rim, equal to zero. The \emph{speed-based} formulation, instead, has a minimum speed, which cannot be equal to zero as \cref{eq:un-dist-speed-based} becomes indefinite. The speed distribution is thus characterised by a power-law with a presence of a low-speed cut-off, which eliminates the low-velocity transition after the \emph{knee velocity} (\cref{fig:dist_comparison}).
\bigbreak
The expression of \cref{eq:un-dist-speed-based} is for a normal impact. We need now to transform it to an oblique impact. The procedure is similar to the one of the ejection angles, so that we have: 

\begin{equation}  \label{eq:speed-based-tranformation}
	p_u (u) = \frac{p_{u_n} (u_n)}{|J(\mathcal{T})|} \rm .
\end{equation}

\noindent where $J(\mathcal{T}) = \frac{\sin \psi_n}{\sin \psi}$ is the Jacobian of the transformation of \cref{eq:transformation-speed-based}. Therefore, we have:

\begin{equation} \label{eq:u-dist-oblique-speed-based}
	\begin{split}
		p_{u|\psi \xi} (u | \psi, \xi; \phi) &= p_{u_n} (u_n) \cdot \frac{\sin \psi}{\sin \psi_n} = \\ &= p_{u_n} \left( \frac{u \sin \psi}{\sin \left( \psi - \bar{K}_\psi (\xi; \phi) \right)} \right) \cdot \frac{\sin \psi}{\sin \left( \psi - \bar{K}_\psi (\xi; \phi) \right)}
	\end{split} \rm ,
\end{equation}

\noindent where we have inverted the expressions of $u_n$ and $\psi_n$ as functions of $u$ and $\psi$ to complete the transformation. As in the previous cases, passing from a normal impact to an oblique impact, the distribution becomes conditioned, so that we now have a distribution in $u$, given $\psi$ and $\xi$.

\subsection{Correlated speed-based distribution}
\label{subsec:correlated-dist}
This third formulation directly derives form \cref{subsec:speed-dist}; therefore, they share several characteristics. Specifically, they share the same distributions of the in-plane and out-of-plane ejection angle components so that, also for this formulation, we can refer to \cref{subsubsec:inplane-dist-speed-based,subsubsec:outplane-dist-speed-based}. The main difference, instead, resides in the size and speed distribution, which in this case is correlated. Therefore, we can write the generic expression for our distribution as follows:

\begin{equation} \label{eq:conditional-speed-based-corr}
	p(s, u, \xi, \psi) = p_{su | \xi, \psi} (s, u | \xi, \psi; \phi) \cdot p_{\psi | \xi} (\psi | \xi; \phi) \cdot p_\xi (\xi) \rm .
\end{equation}

In \cref{subsubsec:size-speed-dist}, we focus on describing in detail the expression for the correlated size vs. speed distribution.

\subsubsection{Size vs. speed}
\label{subsubsec:size-speed-dist}
\begin{figure}[htb!]
	\centering
	\includegraphics[width=2.6in]{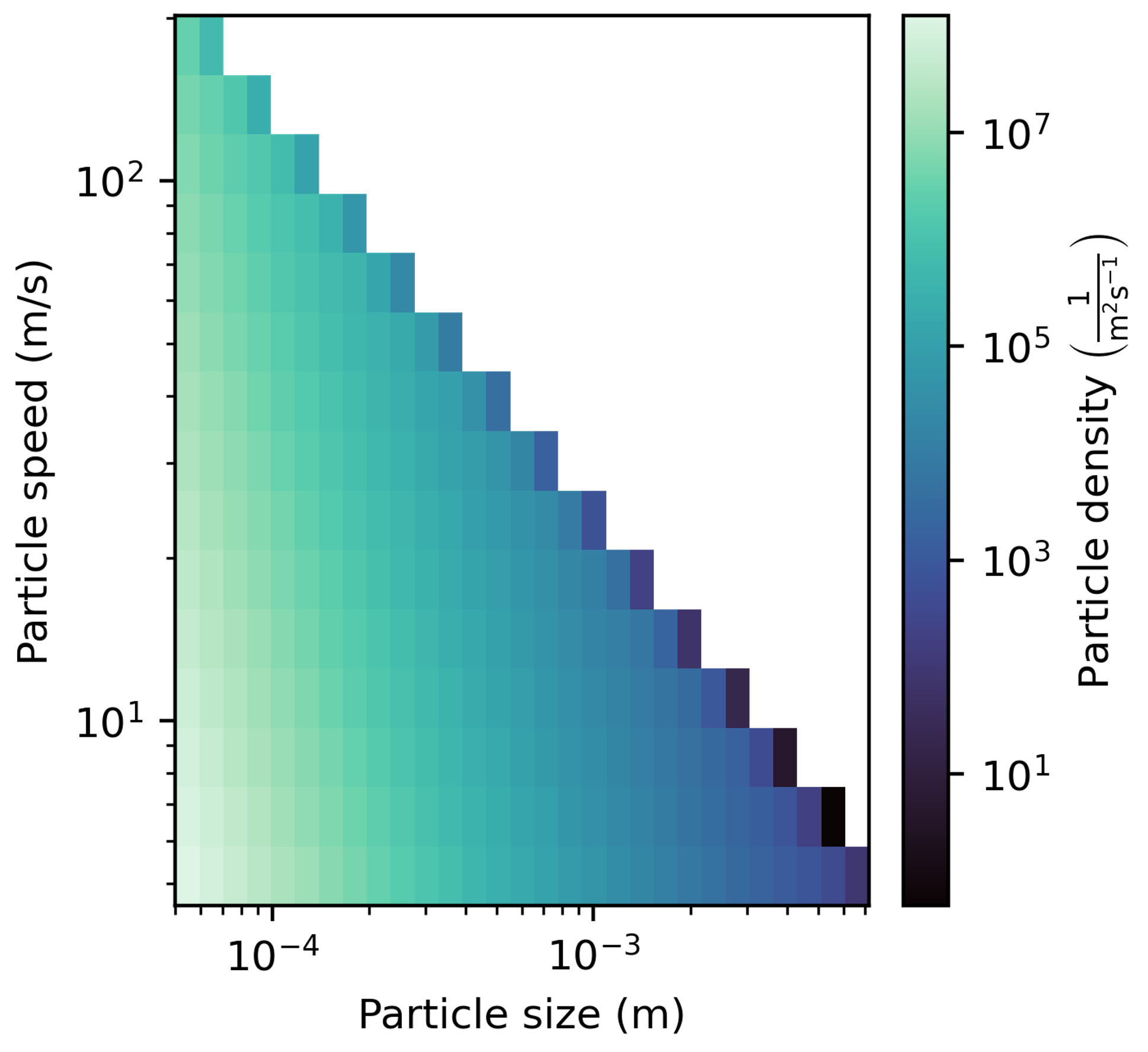}
	\caption{Example of the correlated distribution in size and speed for a normal impact.}
	\label{fig:dist_correlated_example}
\end{figure}

The size-speed correlated distribution is derived from the work of Sachse \citep{sachse2015correlation} and has the following expression:

\begin{equation}  \label{eq:su-corr-sachse}
	p_{s, u_n} (s, u_n) = A s^{-1-\abar} u_n^{-1-\gbar} \cdot \Theta \left[ b s^{-\bbar} - u_n \right]
\end{equation}

\noindent where $\Theta$ is the Heaviside step function, and $A$, $b$, $\abar$, $\bbar$, and $\gbar$ are parameters that characterise the shape of the distribution function. \cref{fig:dist_correlated_example} shows an example of the correlated distribution in size and speed for a normal impact. As we can notice, the maximum possible ejection speed decreases with increasing particle size. A detailed description of the procedure to select and derive these parameters for the correlated distribution is presented in Appendix \cref{app:corr-params}. 

Similarly to \cref{subsubsec:u-dist-speed-based}, the starting expression for the distribution only refers to normal impact. As we are interested in generic impacts, we must perform a transformation to obtain a distribution that is valid also for oblique impacts. The transformation is analogous to \cref{eq:speed-based-tranformation} of \cref{subsubsec:u-dist-speed-based} so that we have the following expression for the correlated distribution:

\begin{equation} \label{eq:su-corr-oblique}
	\begin{split}
		p_{s, u | \psi, \xi} (s, u | \psi, \xi) &= A \cdot s^{-1 - \abar} \cdot u^{-1 - \gbar} \cdot \left( \frac{\sin \psi}{\sin \left( \psi - \bar{K}_\psi (\xi; \phi) \right)} \right)^{-\gbar} \cdot \\ &\cdot \Theta \left[ b s^{-\bbar} - \frac{u \sin \psi}{\sin \left( \psi - \bar{K}_\psi (\xi; \phi) \right)} \right]
	\end{split}
\end{equation}

\subsection{Summary}
\label{subsec:summary-distributions}
At this point, it is useful to summarise the different formulations and their relevant models. As we have seen in \cref{subsec:position-dist,subsec:speed-dist,subsec:correlated-dist}, we have three distributions formulations (i.e., \emph{position-based}, \emph{speed-based}, and \emph{correlated}). For each of these formulations, we have derived the particle density distribution in the form of a combination of conditional distributions as function of size, $s$, position, $r$, speed, $u$, and ejection angles, $\xi$ and $\psi$. For some of these variables, namely the speed and the ejection angles, we have identified different modelling techniques based on the type of formulation and on past literature. \cref{tab:summary-models} presents a summary of the models used for the different formulations, along with the relevant equations, in order to increase the understanding of the structure of the different models proposed. In addition, this highlights the modular nature of the proposed distribution-based formulations, which can accommodate different types of models as long as they are expressed in terms of density distributions.

\begin{table}[htb!]
	\centering
	\caption{Summary of the ejecta distribution formulations and the of the available models. \label{tab:summary-models}}
	\begin{tabular}{lc|ccc}
		\hline
		Var.			& Model & Position-based 				& Speed-based 							& Correlated \\
		& & (\cref{subsec:position-dist}) & (\cref{subsec:speed-dist}) 			& (\cref{subsec:correlated-dist}) \\
		\hline
		\hline
		$s$ 		& & \cref{eq:size-distribution} 	& \cref{eq:size-distribution} 			& \cref{eq:su-corr-oblique} \\
		\hline
		$r$ 		& & \cref{eq:r-distribution}		& -	 									& - \\
		\hline
		$u$ & \specialcell{Richardson \\ Housen} & \specialcell{\cref{eq:ejection-speed-oblique,eq:speed-richardson} \\ \cref{eq:ejection-speed-oblique,eq:speed-housen}}
		& \cref{eq:u-dist-oblique-speed-based} 	& \cref{eq:su-corr-oblique} \\
		\hline
		$\xi$ & \specialcell{Lobed \\ Gaussian}		& \specialcell{\cref{eq:xi-distribution} \\ \cref{eq:xi-distribution-gaussian}} & \specialcell{\cref{eq:xi-dist-lobed-speed-based} \\ \cref{eq:xi-distribution-gaussian} } & \specialcell{\cref{eq:xi-dist-lobed-speed-based} \\ \cref{eq:xi-distribution-gaussian}} \\
		\hline
		$\psi$ & \specialcell{Uniform \\ Gaussian}		& \specialcell{\cref{eq:psi-distribution-uniform} \\ \cref{eq:psi-distribution-gaussian}} 	& \specialcell{\cref{eq:psi-distribution-uniform} \\ \cref{eq:psi-dist-gaussian-speed-based}} & \specialcell{\cref{eq:psi-distribution-uniform} \\ \cref{eq:psi-dist-gaussian-speed-based}} \\
		\hline
	\end{tabular}
\end{table}

\section{Sampling methodology}
\label{sec:sampling}
\begin{figure}[htb!]
	\centering
	\begin{subfigure}[b]{0.42\textwidth}
		\centering
		\includegraphics[width=\textwidth]{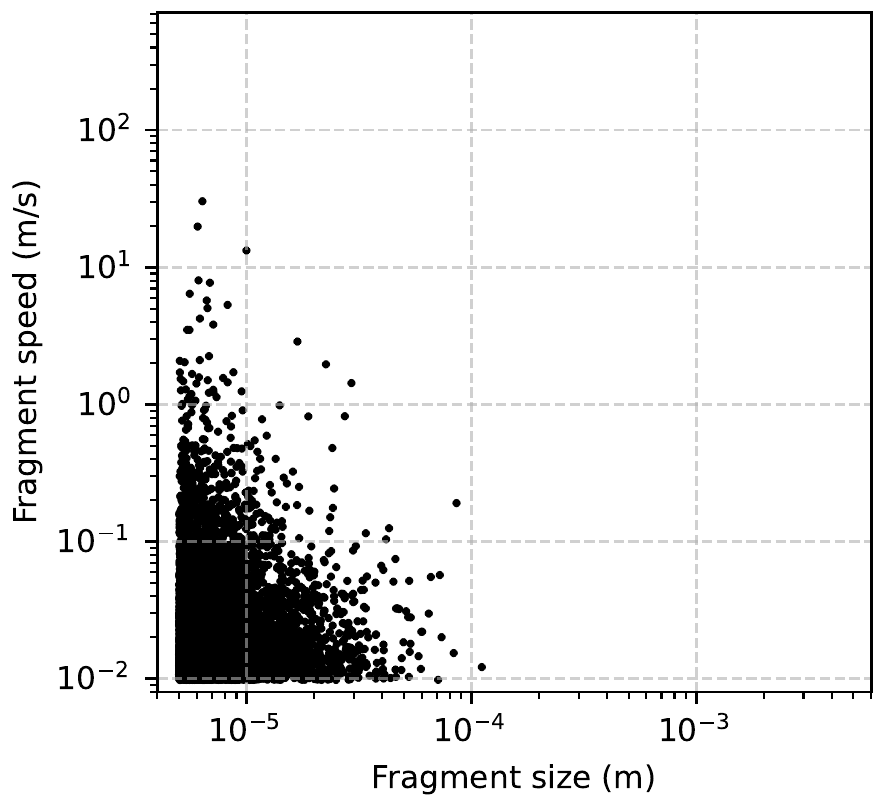}
		\caption{ }
		\label{fig:scatter-sampling}
	\end{subfigure}
	\hfill
	\begin{subfigure}[b]{0.42\textwidth}
		\centering
		\includegraphics[width=\textwidth]{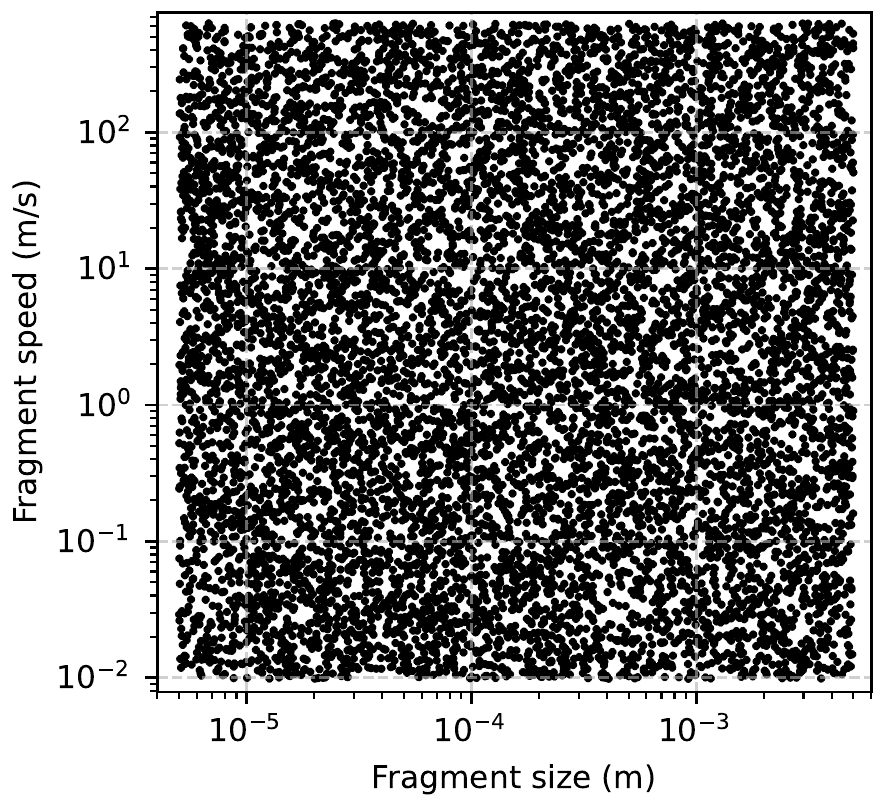}
		\caption{ }
		\label{fig:latin-sampling}
	\end{subfigure}
	\caption{Examples of sampling strategies of the size-speed distribution. (a) random variates (b) space-filling.}
\end{figure}

The sampling methodology is a key ingredient to the understanding of the fate of the ejecta after an impact with a small body. Having defined our ejecta models via continuous distributions, we can exploit this formulation to sample the ejecta distribution. Because we have expressed the distributions as the product of either independent or conditional probability distributions (\cref{eq:conditional-position-based,eq:conditional-speed-based,eq:conditional-speed-based-corr}), we can follow the order of these distributions to perform a random sampling of each of the presented formulations. As an example, let us consider the \emph{position-based} distribution of \cref{eq:conditional-position-based}. In this case, we can directly sample the distribution in size and position using the cumulative distributions derived from \cref{eq:size-distribution,eq:r-distribution}, because they are independent from the other variables. Then we can obtain the samples of the ejection angles. First, we perform the sampling of the in-plane distribution that is a conditional distribution in $r$. To do so, we use the already obtained launch positions and we draw samples from the distribution in $\xi$ inverting its cumulative distribution. Now that we have both the samples in $r$ and $\xi$, we can use the CDF of the conditional distribution in $\psi$ to sample the out-of-plane component of the ejection angle. Finally, we have the full set of random samples from the ejecta distribution. \cref{fig:scatter-sampling} shows an example of ten thousand samples drawn from a size-speed distribution. We can observe that most of the samples concentrate in the region of small diameters and small speeds. This is an expected behaviour as both the size and speed distributions are power laws. It is clear that, to fully characterise an impact, it is necessary to draw a large amount of samples, particularly if we are also interested in the dynamical fate of larger particles.

To address this drawback, different sampling strategies can be assessed. Specifically, in this work, a \emph{space-filling} strategy is adopted so that the entire domain defined by the ejecta distribution is covered. \cref{fig:latin-sampling} shows an example of such strategy, using a latin-hypercube algorithm \citep{SMT2019,Jin2003LatinHypercube} for the sample generation and a log-log filling technique. That is, the samples are created to best fill a space in log-log coordinates. A space-filling strategy in both logarithmic and linear coordinates can be performed. In this work. the logarithmic strategy is used for the ejecta size and speed, while a linear strategy for the ejection angles and ejection position. The idea behind this type of sampling is that we can have samples that better characterise the ejecta field and, therefore, a more representative prediction of the ejecta fate around asteroids and other small bodies. In addition, we can exploit the ejecta distribution to assign to each sample a number of \emph{representative fragments}. That is, the single sample we have drawn carries the information of an ensemble of fragments that we associate to it. In this way, also during the propagation we can understand the behaviour of the whole field of fragments generated by the impact. The computation of the number of \emph{representative fragments} comprise the following steps:

\begin{itemize}
	\item Subdivide the ejecta domain into a grid. The grid can be as fine as needed but such that at least few samples are in each bin. The grid can be specified either in a linear or in a logarithmic space.
	\item For each bin, we can integrate the ejecta distribution to compute the number of fragments associated to the bin (this procedure can be performed analytically using the cumulative distribution functions (Appendix \cref{app:cumulative-dist})).
	\item Assign to each sample inside the bin a number of \emph{representative fragments} by equally subdividing the total fragment associated to the bin by the number of samples in the bin:
	\begin{equation}
		\tilde{n}_r^{ij} = \frac{N_f^{j}}{N_s^{j}} \rm ,
	\end{equation}
	where $\tilde{n}_r^{ij}$ are the \emph{representative fragments} associated to the i-th sample belonging to the j-th bin, and $N_f^{j}$ and $N_s^{j}$ are the total fragments and samples of the j-th bin.
\end{itemize}

\begin{figure}[htb!]
	\centering
	\begin{subfigure}[b]{0.42\textwidth}
		\centering
		\includegraphics[width=\textwidth]{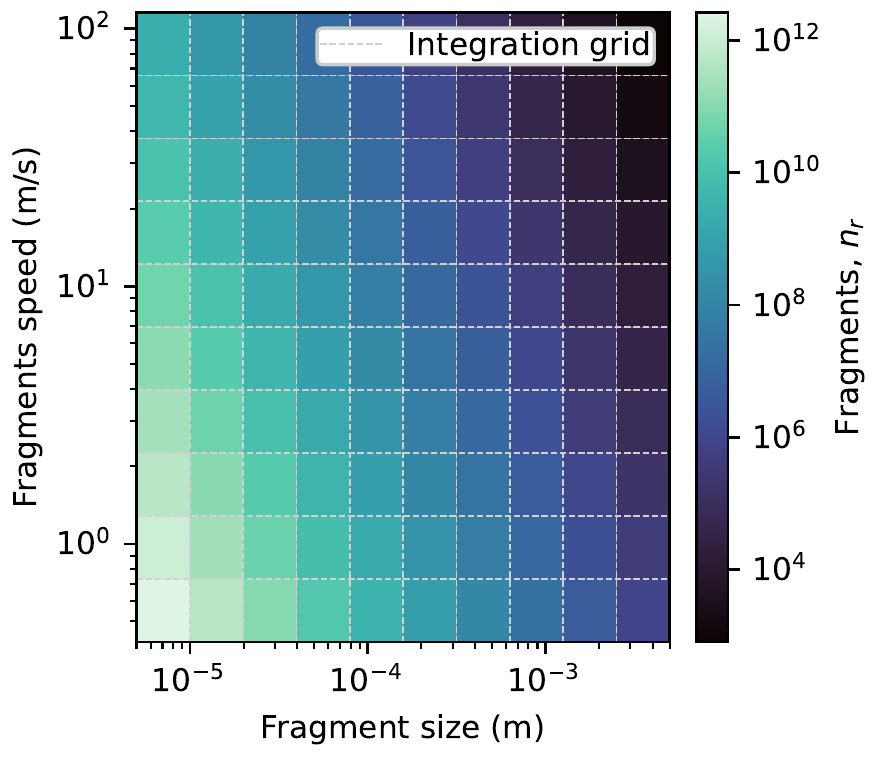}
		\caption{ }
		\label{fig:size-speed-pdf}
	\end{subfigure}
	\hfill
	\begin{subfigure}[b]{0.42\textwidth}
		\centering
		\includegraphics[width=\textwidth]{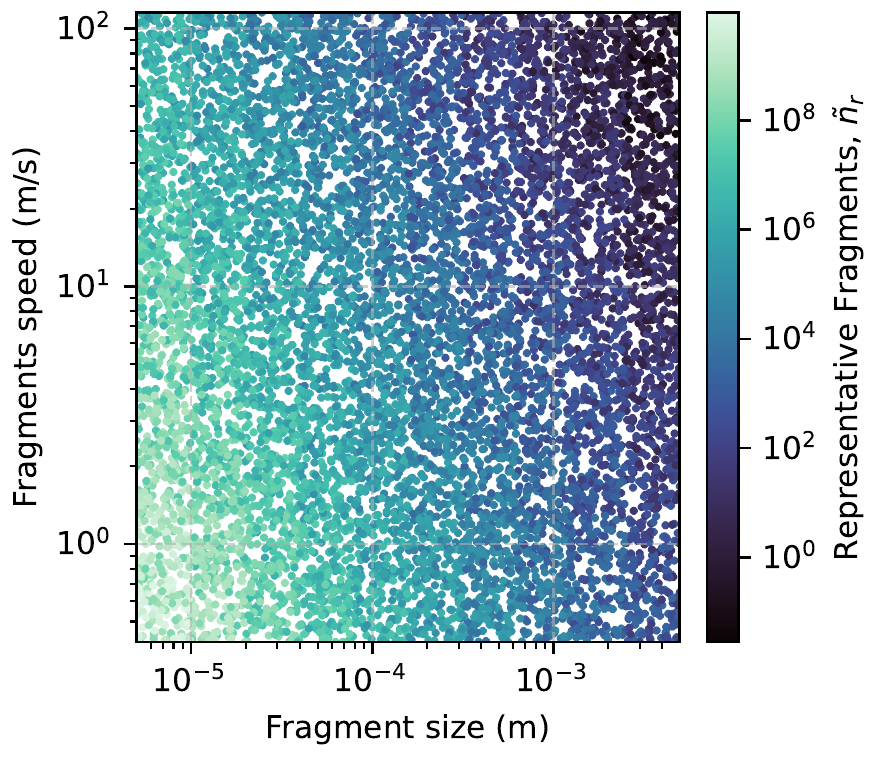}
		\caption{ }
		\label{fig:rep-fragments}
	\end{subfigure}
	\caption{(a) Example of a size-speed distribution with highlighted the integration grid for the representative fragments. (b) Corresponding set of space-filling samples with associated representative fragments.}
	\label{fig:rep-fragments-example}
\end{figure}

\cref{fig:rep-fragments-example} shows an example of the results of the aforementioned procedure for a two-dimensional size-speed distribution. \cref{fig:size-speed-pdf} shows the particle density distribution (it is possible to observe the grid in which the distribution is subdivided), while \cref{fig:rep-fragments} shows ten thousand samples drawn in a space-filling log-log space and the associated number of representative fragments that is proportional to the integral of the distribution of \cref{fig:size-speed-pdf} and the number of samples within each bin.

\section{Dynamics}
\label{sec:dynamics}
The adopted dynamical model is the Photo-gravitational Hill Problem that is the extension of the classical Hill problem to a radiating primary \citep{soldini2017assessing}. The equations of motion are expressed in non-dimensional form in a synodic reference frame centred in the asteroid. The x-axis is along the Sun-asteroid direction, pointing outwards, the z-axis is along the direction of the angular momentum of the asteroid orbit, and the y-axis completes the right-hand system.

\begin{equation}
	\begin{cases}
		\ddot{x} - 2 \dot{y} = - \frac{x}{r^3} + 3 x + \beta \\
		\ddot{y} + 2 \dot{x} = - \frac{y}{r^3} \\
		\ddot{z} = - \frac{z}{r^3} - z
	\end{cases}
\end{equation}

\noindent where $x$, $y$, and $z$ are the non-dimensional particle positions with respect to the centre of the asteroid in the synodic frame, and $r = \sqrt{x^2 + y^2 + z^2}$ is the particle's distance from the centre of the asteroid. The lightness parameter $\beta$ can be expressed as follows \citep{pinto2020}:

\begin{equation}
	\beta = \frac{P_0}{c} \frac{AU^2}{\mu_a^{\frac{1}{3}} \mu_{\rm Sun}^{2/3}} \frac{3 (1 + c_{\rm R})}{2 \rho_{\rm p} d_{\rm p}} \rm .
\end{equation}

Where $P_0$ = 1367 \si{\watt\per\meter\squared} is the solar flux at 1 AU, $c$ is the speed of light, AU is the astronomical unit, $\mu_{\rm Sun}$ and $\mu_a$ are the gravitational parameter of the Sun and the asteroid, respectively, $\rho_{\rm p}$ is the particle density and $d_{\rm p}$ the particle diameter. The reflectivity coefficient, $c_{\rm R}$, is a number between 0 and 1, where 1 is for fully reflective surfaces. Eclipses are taken into account using a cylindrical shadow model via a modified lightness parameter, $\beta^{*}$:

\begin{equation}
	\beta^{*} = \begin{cases}
		\beta \quad & \text{if } x \leq 0 \\
		\beta \cdot f(\sigma) \quad & \text{otherwise}
	\end{cases} \rm ,
\end{equation}

\noindent where $f(\sigma) = \left( 1 + e^{-s \cdot \sigma} \right)^{-1}$ is a sigmoid function with steepness parameter $s$, which, in this work is equal to 8 \citep{pinto2020}. The variable $\sigma = r_x - R_a$, with $r_x = \sqrt{y^2 + z^2}$ distance to the $x$-axis, and $R_a$ mean radius of the asteroid.

\bigbreak
The Photo-gravitational Hill Problem has been selected as the dynamical model for the study because it is a relatively simple model that allows taking into account the main forces acting on the ejected fragments, such as the gravity of the asteroid and the solar radiation pressure. At the same time, it allows maintaining the dynamics sufficiently simple to focus our analysis on the ejecta distribution models and the effects that different modelling choices have on the fate of the ejecta.

\section{Sensitivity analysis}
\label{sec:sensitivity}
The objective of this work is to assess how the modelling decisions concerning the ejecta models affect the overall evolution of the particles around the asteroid. As the initial ejection conditions and the size of the particles are fundamental in shaping the trajectory evolution of the samples, we want to understand what is the impact of specific modelling choices. Specifically, in this work we focus on the characteristics of the ejecta model, while we neglect (and leave to a future study) the effect of the impact location, the size, shape, and orbit of the asteroid. The parameters of the ejecta distributions considered in the sensitivity analysis are:

\begin{itemize}
	\item the minimum particle size, $\smin$;
	\item the slope coefficient of the particle size distribution, $\bar{\alpha}$;
	\item the speed distribution formulation: Housen (\cref{eq:ejection-speed-oblique,eq:speed-housen}) or Richardson (\cref{eq:ejection-speed-oblique,eq:speed-richardson});
	\item the distribution formulation: \emph{position-based} (\cref{subsec:position-dist}) or \emph{speed-based} (\cref{subsec:speed-dist});
	\item the correlation between fragment size and ejection speed (\cref{subsec:correlated-dist,subsec:speed-dist});
	\item the out-of-plane ejection angle distribution: Uniform or Gaussian;
	\item the value of the impact angle, $\phi$.
	\item the effect of the asteroid rotation;
\end{itemize}

The selection of these parameters is based on their influence on the ejecta models and on the level of uncertainty associated to them. In the following sections, each of these points is analysed and their effects on the overall ejecta fate is discussed.
\bigbreak  
\noindent\emph{Target and impactor properties}. For a better comparison among the different parameters, we define common characteristics for the target and the impactor. In addition, we fix some of the parameters of the ejecta models. For the asteroid, we select a spherical body with a diameter of 1 \si{\kilo\meter}, a density of 2.6 \si{\gram\per\centi\meter\cubed}, an albedo of 0.1, and a rotational period of 16 \si{\hour}. The semimajor axis of the asteroid's orbit is 1.755 AU. The rotational axis of the asteroid is perpendicular to its orbital plane. The characteristics of the asteroid are average values from the NASA Small Bodies Database \citep{nasa_sbdb}, except the asteroid radius for which a larger radius has been selected.
For the impactor, we select a mass of 2 \si{\kilo\gram}, a diameter of 0.15 \si{\meter}, and an impact speed of 2 \si{\kilo\meter\per\second} that is a geometry equivalent to the small carry-on impactor used by the Hayabusa2 mission \citep{tsuda2019hayabusa2,tsuda2020hayabusa2}.
The decision of fixing these parameters is here taken to focus the analysis on the modelling decisions relative to the ejecta model and understand the effect they can have in the prediction of the fate of the ejecta and particularly to identify the most influential modelling choices.
\bigbreak  
\noindent\emph{Simulation setup}. The simulations have a time span of 2 months. At the end of the simulation, we refer to the particles as \emph{escaping} if their distance is greater than the Hill sphere, as \emph{impacting} if they re-impact the asteroid's surface, and \emph{orbiting} if they have not impacted nor escaped the asteroid. The impact characteristics are kept constant, with a normal impact ($\phi$ = 90\si{\degree}) on the North pole of the asteroid, for all the simulations except for \cref{subsec:sensitivity-rotation}. A polar impact is specifically selected to remove the dependency on the asteroid rotation in the sensitivity analysis of the other parameters.
\bigbreak  
\noindent\emph{Asteroid material and strength}. Two types of materials are used in the analysis, a sand-like very low-strength material and the moderate-strength Weakly Cemented Basalt (WCB). \cref{tab:materials} shows the reference properties of these materials.

\begin{table}[htb!]
	\centering
	\caption{\label{tab:materials} Material properties \citep{housen2011ejecta}.}
	\begin{tabular}{lcc}
		\hline
		& Sand & WCB \\
		\hline
		\hline
		$\mu$ 		& 0.41 	& 0.46 \\
		$C_1$ 		& 0.55 	& 0.18 \\
		$k$			& 0.3 	& 0.3 \\
		$n_1$		& 1.2 	& 1.2 \\
		$n_2$		& 1.3 	& 1 \\
		$Y_{\rm ref}$ (\si{\mega\pascal}) 	& 0 	& 0.45 \\
		\hline
	\end{tabular}
\end{table}

The coefficients $\mu$, $C_1$, $k$, $n_1$, and $n_2$ are characteristics of the material and determine the shape of the ejecta distributions described in \cref{sec:distributions}. $Y_{\rm ref}$ is the reference strength of the material. These values are typically varied to perform sensitivity analyses with respect to the asteroid's strength. However, as we are focusing on the influence of the ejecta models, the strength values have been fixed to a value of 0 \si{\mega\pascal} for the sand-like material and 5 \si{\kilo\pascal} for the WCB. In this last case, we reduced the value with respect to the reference in \cref{tab:materials} to more closely model weakly cohesive soils similar to regolith \citep{holsapple2012momentum,richardson2007ballistics}. Finally, we also select default parameters for the ejecta distribution. Specifically, the default particle size range is between $\smin = \num{5e-6}$ \si{\meter} and $\smax = \num{5e-3}$ \si{\meter} (i.e., particle diameters between 10 \si{\micro\meter} and 1 \si{\centi\meter}). In addition, we always assume the $\gbar$ coefficient is equal to $3\mu$ (\cref{subsubsec:u-dist-position-based}); therefore, it is only dependent on the material.

\bigbreak  
\noindent\emph{Sampling and statistical analysis}. Given the statistical nature of the sampling procedure described in \cref{sec:sampling} and that we associate to each sample a number of representative fragments, for each analysis, we perform multiple runs. Specifically, we perform 20 runs and in each run we draw 100 000 samples. In this way, we can assess the robustness and variability of the obtained results, particularly concerning the estimate of the number of fragments (via the representative fragments).
Additionally, it is necessary to define the sampling of the distribution. We adopt the procedure described in \cref{sec:sampling}. Because we are interested in the fate of the ejecta in the neighbourhood of the asteroid, we limit the sampling to the fragments with an ejection speed below the escape speed of the asteroid. Therefore, the ejection speed is sampled between $\umin$ and $u_{\rm esc}$. Similarly, as the launch position is connected to the ejection speed, even the sampling in $r$ must be limited. Specifically, we sample between $r_{\rm esc}$ and $\rmax$, where $r_{\rm esc}$ is the launch position corresponding to the escape speed. All the ejection location closer than the "escape radius" have ejection speed greater than the escape speed. For both the escape speed and launch location we use a 16 bins grid with a log-scaling subdivision. In a similar fashion, the size distribution is sampled between the minimum and maximum provided particle sizes ($\smin$, $\smax$) on a 20 bins grid with a log-scaling subdivision. The in-plane component of the ejection angle, $\xi$ is sampled between 0\si{\degree} and 360\si{\degree} on a linear grid with 36 bins. The out-of-plane component, $\psi$, is sampled on a linear grid with 8 subdivisions. The sampling range depends on the model adopted. Between a $\psi_{\rm min}$ and $\psi_{\rm max}$ for a Uniform distribution (see \cref{eq:psi-distribution-uniform} in \cref{subsubsec:outplane-dist-position-based}) and in the range $\mu \pm 3 \sigma$ for the Gaussian model (see \cref{eq:psi-distribution-gaussian} in \cref{subsubsec:outplane-dist-position-based}).

\subsection{Slope coefficient ($\abar$) of the size distribution}
\label{subsec:alpha-sensitivity}
The coefficient $\abar$ of the size distribution (\cref{subsubsec:size-dist}) defines the slope of the distribution that is the number of particles ejected as a function of their size. Larger $\abar$ coefficients produce distributions with a higher portion of small fragments. Because the size distribution is a power law, it can change significantly with the value of the slope coefficient. According to \cite{krivov2003impact}, typical value of the $\abar$ coefficient for basalt and granite targets is between 2.4 and 2.7, and a good choice in many applications is a value between 2.4 and 2.6. Following the results of Krivov, we test the sensitivity of the ejecta fate as function of three values of $\abar \in \{ 2.40,\, 2.55,\, 2.70 \}$. For this test case, the \emph{position-based} formulation has been used (\cref{subsec:position-dist}) with a uniform distribution in the in-plane ejection angle, $\xi$ (normal impact), and a Gaussian distribution for the out-of-plane ejection angle, $\psi$ (\cref{subsubsec:outplane-dist-position-based}). Both target materials of \cref{tab:materials} have been considered.

\bigbreak
First, we compare the fraction of particles orbiting, escaping, and impacting after two months, for the three $\abar$ values considered. \cref{tab:alpha-results-samples,tab:alpha-results-fragments} shows the results for the samples and the estimated fragments, respectively, for the three $\abar$ values and the two materials considered. Because we are have twenty runs for each simulation, we present the results for the average percentage of samples and fragments ($\langle \cdot \rangle$) and percent Relative Standard Deviation (RSD) (i.e., the standard deviation divided by the mean). \cref{tab:alpha-results-samples} shows the statistics for the samples. We can observe that for both materials most of the samples (more than 90\%) re-impact the asteroid. Most of the other samples escape, while only a very small percentage is still orbiting after two months. The results on the Relative Standard Deviation show stable results for both the impacting and escaping samples, while a larger variability is associated with the orbiting samples as they are far fewer. It is interesting to observe that the results are very stable for the different values of $\abar$. Therefore, the influence of the scaling coefficient, $\abar$, on the samples fate appears to be limited. 

\begin{table}[htb!]
	\centering
	\caption{\label{tab:alpha-results-samples} Samples fractions and corresponding percent relative standard deviations for the three $\abar$ coefficients and two materials in exam.}
	\begin{tabular}{l|c|cc|cc|cc}
		\hline
		Material & $\abar$ & $\langle N_{\rm imp} \rangle$ & RSD$_{\rm imp}$ & $\langle N_{\rm esc} \rangle$ & RSD$_{\rm esc}$ & $\langle N_{\rm orb} \rangle$ & RSD$_{\rm orb}$ \\
		\hline
		\hline
		\multirow{3}{*}{Sand} & 2.40 	& 98.48\% & 0.028\% & 	1.51\%	& 1.87\% & 0.006\% & 49.36\%	\\
		& 2.55 	& 98.48\%	& 0.028\% & 1.51\%	& 1.85\% & 0.007\% & 37.25\%	\\
		& 2.70 	& 98.48\%	& 0.039\% & 1.51\%	& 2.58\% & 0.006\% & 44.83\%	\\
		\hline
		\multirow{3}{*}{WCB}  & 2.40 	& 91.07\%	& 0.031\%	& 8.89\%	& 0.31\%	& 0.039\%	& 9.80\%	\\
		& 2.55 	& 91.07\%	& 0.055\%	& 8.89\%	& 0.54\%	& 0.038\%	& 14.08\%	\\
		& 2.70 	& 91.07\%	& 0.060\%	& 8.89\%	& 0.63\%	& 0.038\%	& 16.97\%	\\
		\hline
	\end{tabular}
\end{table}

\cref{tab:alpha-results-fragments} shows equivalent results for the number of fragments estimated via the representative fragments. Comparing \cref{tab:alpha-results-fragments} with \cref{tab:alpha-results-samples} we observe few interesting features. First, the percentage of samples impacting, escaping and orbiting differs from their corresponding samples, in particular for the WCB case. This is a consequence of the representative fragments methodology as a different weight is associated to each sample. The orbiting fragments are a very small percentage in both cases and corresponds to few hundreds or thousands of samples, although their variability is higher as highlighted by the percent RSD. The combination of a distribution-based ejecta model (\cref{sec:distributions}) with the representative fragments sampling methodology (\cref{sec:sampling}) is thus able to characterise the fate of the impacting and escaping ejecta with a small variability.
As for the samples of \cref{tab:alpha-results-samples}, the fate of the ejecta is stable with respect to the change in $\abar$; the main difference resides in the total number of fragments generated by the impact (see \cref{tab:alpha-results-fragments}). In fact, larger $\abar$ generates more fragments. We can thus infer that the selection of the slope coefficient, $\abar$, only minimally influences the ultimate fate of the ejecta; however, it scales the total amount of particles generated.

\begin{table}[htb!]
	\centering
	\caption{\label{tab:alpha-results-fragments} Fragments fractions and corresponding percent relative standard deviations for the three $\abar$ coefficients and two materials in exam.}
	\begin{tabular}{l|c|c|cc|cc|cc}
		\hline
		Material & $\abar$ 	& $N_{\rm tot}$ & $\langle N_{\rm imp} \rangle$ & RSD$_{\rm imp}$ & $\langle N_{\rm esc} \rangle$ & RSD$_{\rm esc}$  & $\langle N_{\rm orb} \rangle$ & RSD$_{\rm orb}$ \\
		\hline
		\hline
		\multirow{3}{*}{Sand} & 2.40 	& \num{3.94e13}	& 99.74\% & 1.23\% & 	0.26\%	& 6.80\% & \num{8.8e-10}\% & 67.17\%	\\
		& 2.55 	& \num{8.11e13}	& 99.75\%	& 1.03\% & 0.25\%	& 8.56\% & \num{1.2e-9}\% & 148.84\%	\\
		& 2.70 	& \num{1.57e14}	& 99.73\%	& 1.59\% & 0.27\%	& 5.67\% & \num{9.5e-10}\% & 224.87\%	\\
		\hline
		\multirow{3}{*}{WCB}  & 2.40 	& \num{6.38e10}	& 76.63\%	& 1.04\% & 23.37\%	& 2.55\%	& \num{4.52e-7}\%	& 46.14\%	\\
		& 2.55 	& \num{1.31e11}	& 76.44\%	& 1.02\% & 23.56\%	& 1.68\%	& \num{3.65e-5}\%	& 297.92\%	\\
		& 2.70 	& \num{2.54e11}	& 75.96\% & 0.57\% & 24.04\%	& 1.87\%	& \num{7.68e-8}\%	& 54.32\%	\\
		\hline
	\end{tabular}
\end{table}

\begin{figure}[htb!]
	\centering
	\begin{subfigure}[b]{0.42\textwidth}
		\centering
		\includegraphics[width=\textwidth]{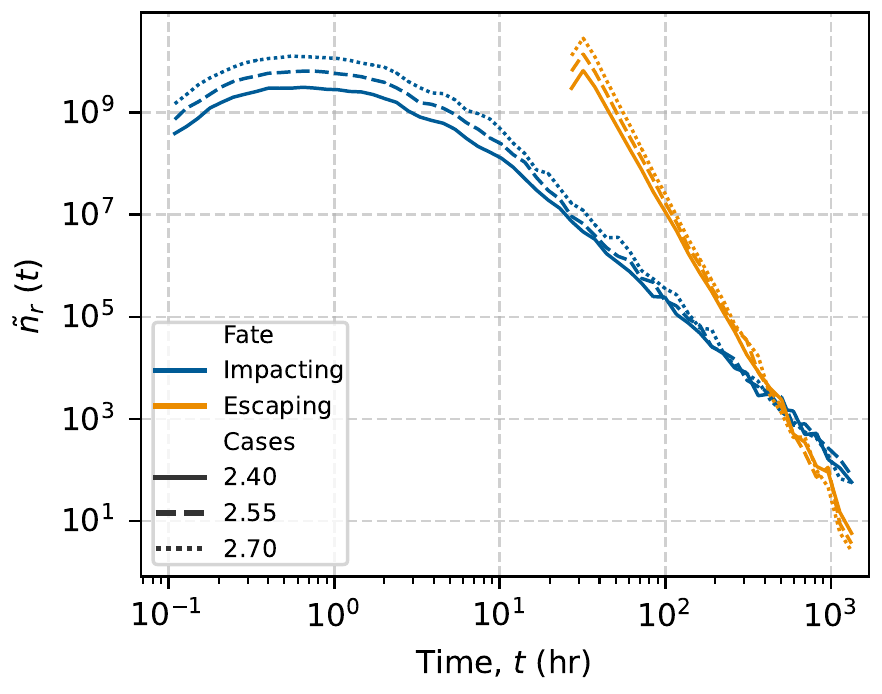}
		\caption{ }
		\label{fig:alpha_diff}
	\end{subfigure}
	\hfill
	\begin{subfigure}[b]{0.42\textwidth}
		\centering
		\includegraphics[width=\textwidth]{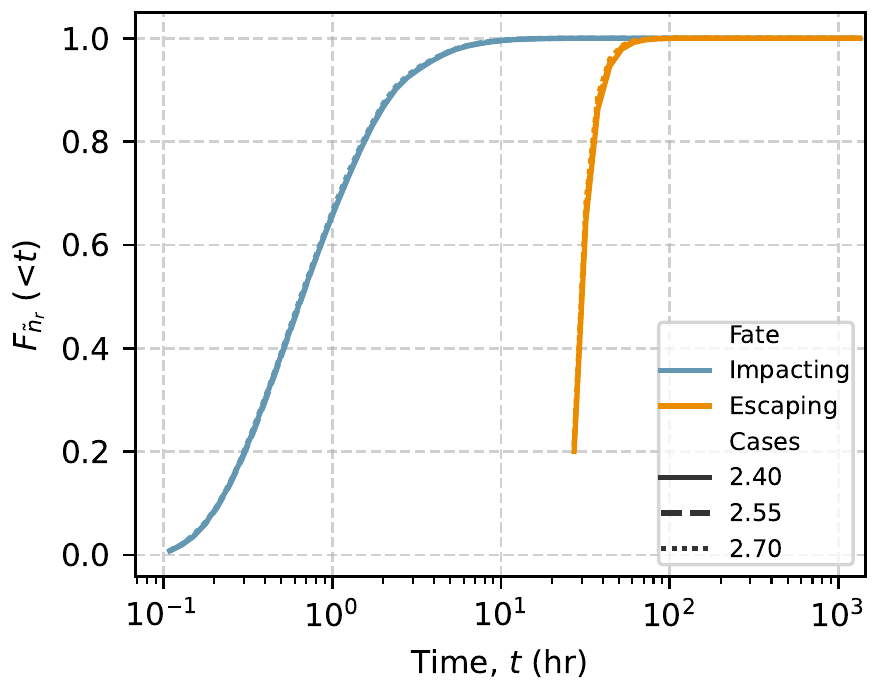}
		\caption{ }
		\label{fig:alpha_cum}
	\end{subfigure}
	\caption{Fragments evolution in time. (a) Number of fragments impacting (blue) and escaping (orange) at time $t$ for the three $\abar$ cases. (b) Corresponding normalised cumulative distribution of the fragments in time.}
	\label{fig:alpha_frag_time}
\end{figure}

Finally, we also look at the fate of the ejecta as a function of time. \cref{fig:alpha_frag_time} shows the results for the WCB material for both the impacting and escaping fragments. Specifically, \cref{fig:alpha_diff} shows the differential distributions of the impacting and escaping particles in time for the three analysed values of $\abar$. At each time instant, we appreciate how many particles escape and impact the asteroid. \cref{fig:alpha_cum} instead shows the corresponding cumulative distribution that is the percentage of particles escaping and impacting within a given time, $t$. From \cref{fig:alpha_cum} we can observe that the behaviour of the normalised distribution is almost identical in all the three cases. The only difference is the absolute number of particles involved, which can be appreciated from \cref{fig:alpha_diff} and \cref{tab:alpha-results-fragments}. Therefore, a variation of $\abar$ within the studied interval does not result in a corresponding variation in the overall behaviour of the fragments.

\subsection{Minimum particle size ($\smin$)}  \label{subsec:sensitivity-smin}
As described in \cref{subsubsec:size-dist}, the size distribution requires definition of the particles size range to be considered. The selection of the upper threshold of the range, $\smax$, is usually selected arbitrarily or as a fraction of the total excavated mass \citep{sachse2015correlation}. The selection of the lower threshold, $\smin$ is somewhat arbitrary; in \cite{yu2017ejecta} a value of 50 \si{\micro\meter} is used, in \cite{yu2018ejecta} a value of 5 \si{\micro\meter} is used instead, while in \cite{sachse2015correlation} the lower threshold is selected based on the sensitivity of the instrument used to detect the particles. Given that the highest particle density correspond to the lower end of the size range, we decided to focus our attention on the selection of $\smin$. Taking as reference the work of Yu and Michel \citep{yu2017ejecta,yu2018ejecta}, we perform the comparison as a function of two values of $\smin \in \{ 5\, \si{\micro\meter},\, 50\, \si{\micro\meter} \}$. The selection of the minimum particle size threshold influences the definition and sampling of the ejecta distribution. Because the integral of the size distribution must satisfy the mass conservation, changing $\smin$ does not only change the minimum fragment size but also the total amount of fragments and the relative percentage of fragments of a given size. For this test case, the \emph{position-based} formulation is used (\cref{subsec:position-dist}) with a uniform distribution for the $\xi$ angle and a Gaussian distribution for the $\psi$ angle (\cref{subsubsec:outplane-dist-position-based}). The results are presented for the WCB material; a similar behaviour can be observed also for sand-like materials.

\begin{figure}[htb!]
	\centering
	\includegraphics[width=0.45\textwidth]{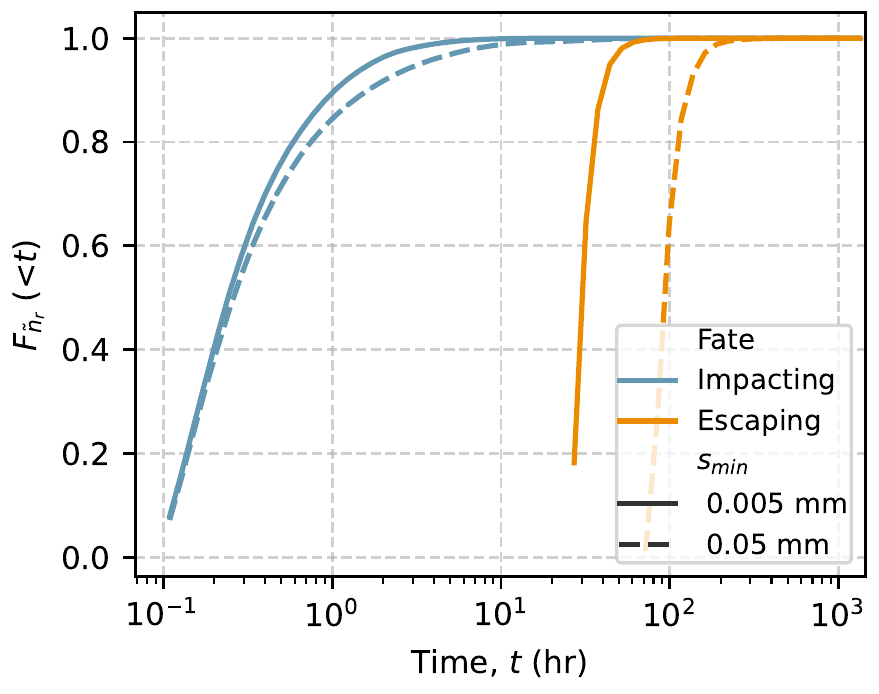}
	\caption{Normalised cumulative distributions of impacting (blue) and escaping (orange) fragments as function of two values of $\smin$ (line style) for the WCB material case.}
	\label{fig:smin_cum_frag_time}
\end{figure}

\cref{fig:smin_cum_frag_time} shows the normalised cumulative distributions of the average impacting and escaping fragments for the two $\smin$ values considered. We can observe that the behaviour of the 0.05 \si{\milli\meter} case shows a less steep curve for the impacting fragments and a delay to higher times for the escaping fragments.  

\begin{figure}[htb!]
	\centering
	\begin{subfigure}[b]{0.42\textwidth}
		\centering
		\includegraphics[width=\textwidth]{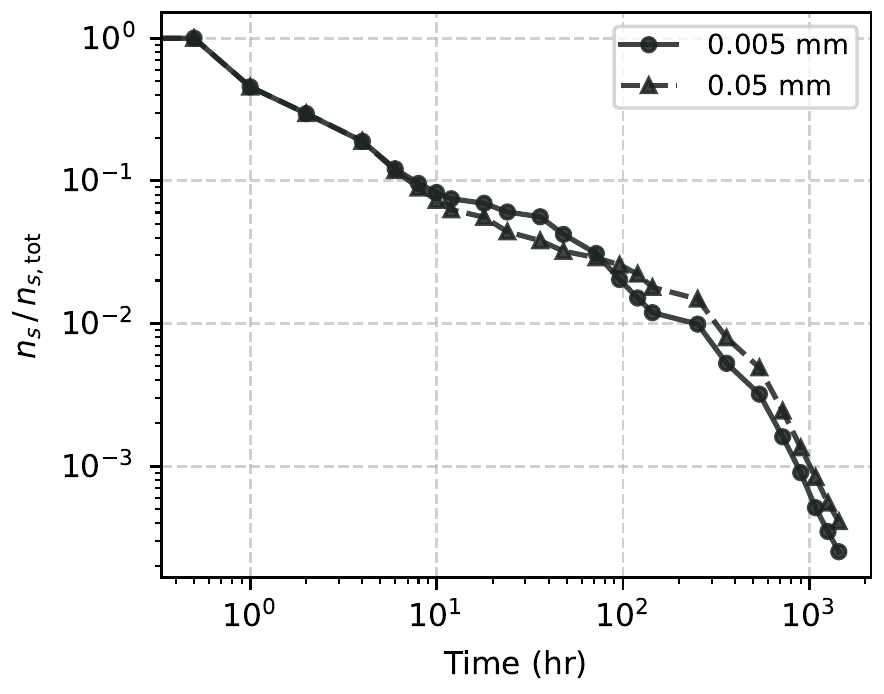}
		\caption{ }
		\label{fig:smin_ns_snap}
	\end{subfigure}
	\hfill
	\begin{subfigure}[b]{0.42\textwidth}
		\centering
		\includegraphics[width=\textwidth]{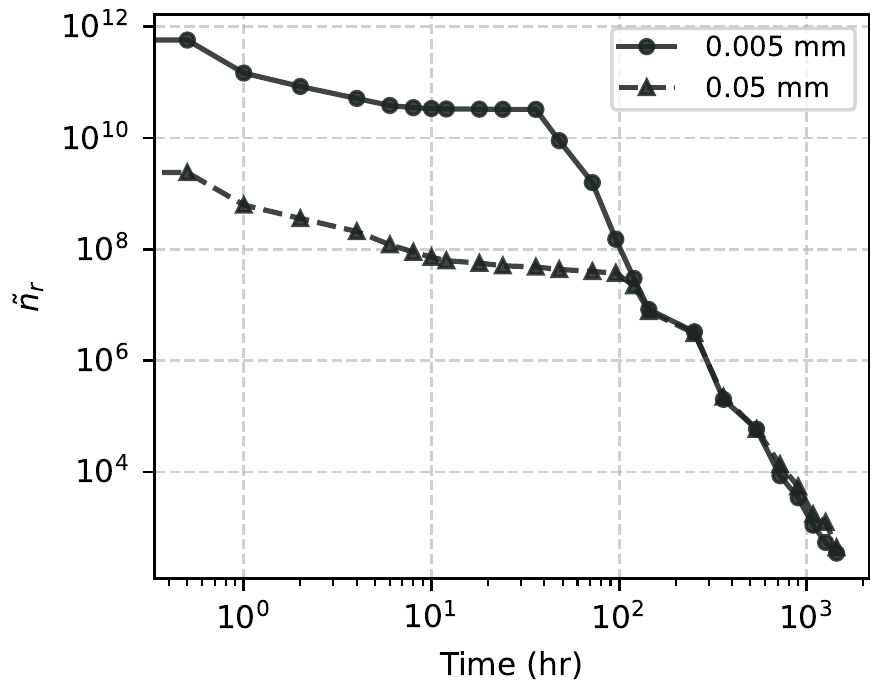}
		\caption{ }
		\label{fig:smin_nr_snap}
	\end{subfigure}
	\caption{Number of samples (a) and fragments (b) at selected snapshot times for $\smin$ = 5 \si{\micro\meter} (solid line with dots) and $\smin$ = 50 \si{\micro\meter} (dashed line with triangles).}
	\label{fig:smin_snap}
\end{figure}

The first behaviour is representative of particles that take longer impact the asteroid and this is because they have, on average, larger diameters. The percentage of impacting particles over the total generated fragments is about 98\% for the 0.05 \si{\milli\meter} case and about 94\% for the 0.005 \si{\milli\meter} case. This latter case, has a higher percentage of escaping particles because their average smaller size causes the fragments to be more easily swept away by the solar radiation pressure. This is also the reason behind the "delayed" behaviour of the cumulative distribution of the escaping particles. In fact, larger particles will take on average longer to escape the neighbourhood of the asteroid. \cref{fig:smin_snap} shows instead the percentage of samples (a) and the number of fragments (b) still orbiting the asteroid at specified snapshots in time. Similarly to what observed in \cref{subsec:alpha-sensitivity}, the behaviour of the samples differs from the evolution of the fragments. While the percentage of samples follows a similar behaviour for both cases, the fragments have a distinct behaviour. Specifically, while the trend is similar, the total number of fragments associated to the samples differs. As expected, a lower number of fragments characterises the 0.05 \si{\milli\meter} case, because a smaller number of particles is required to satisfy the mass conservation equation (\cref{eq:nr}).

From the analysis we can observe that changing the lower threshold of the size range, $\smin$ has two main effects on the generation of the ejecta curtain and its fate around the asteroid. First, the total number of generated fragments varies and decreases as the lower threshold increases. Second, the evolution of the particles results in a lower accretion rate of the impacting fragments and a delayed escape from the sphere of influence of the asteroid of the escaping fragments.

\subsection{Ejecta speed model}  \label{subsec:sensitiviy-speed-model}
As shown in \cref{subsubsec:u-dist-position-based}, the presented ejecta model allows the selection of two different ejecta speed formulations; specifically, \cref{eq:speed-housen} developed by Housen \citep{housen2011ejecta} and \cref{eq:speed-richardson} developed by Richardson \citep{richardson2007ballistics}. To understand the effect of the choice of the ejecta speed formulation, we perform two sets of simulations, changing only the speed model. For this test case, the \emph{position-based} formulation is used (\cref{subsec:position-dist}) with a uniform distribution for the $\xi$ angle and a Gaussian distribution for the $\psi$ angle (\cref{subsubsec:outplane-dist-position-based}). 

Because the fate of an ejected fragment is strongly influenced by its initial velocity, it is interesting to study the effect that a modelling choice can have on the outcome of an impact simulation. In addition, it is interesting to understand if the effects of the ejecta speed model change as a function of the target material. For this reason both materials of \cref{tab:materials} have been considered. \cref{tab:speed-results-fragments} shows the percentage of fragments impacting, escaping and still orbiting after the simulation period of two months, together with the RDS to measure the relative robustness of the prediction. The total number of fragments does not depend on the speed model used; however, it is function of the target material. Specifically, we have a total of approximately \num{2.65e13} fragments for the sand-like material and of \num{1.35e12} fragments for the WCB material. 


\begin{table}[htb!]
	\centering
	\caption{\label{tab:speed-results-fragments} Fragments fractions and corresponding percent relative standard deviations for the Housen and Richardson ejecta speed formulations and the two materials in exam.}
	\begin{tabular}{l|c|cc|cc|cc}
		\hline
		Material & Model & $\langle N_{\rm imp} \rangle$ & RSD$_{\rm imp}$ & $\langle N_{\rm esc} \rangle$ & RSD$_{\rm esc}$ & $\langle N_{\rm orb} \rangle$ & RSD$_{\rm orb}$ \\
		\hline
		\hline
		\multirow{2}{*}{Sand} & Housen 	& 99.80\% & 1.30\% & 	0.19\%	& 4.43\% & \num{2.11e-9}\% & 140.18\%	\\
		& Richardson 	& 98.59\%	& 1.32\% & 1.41\%	& 1.98\% & \num{2.12e-9}\% & 93.72\%	\\
		\hline
		\multirow{2}{*}{WCB}  & Housen 	& 97.63\%	& 0.95\%	& 2.37\%	& 4.86\%	& \num{3.20e-7}\%	& 260.48\%	\\
		& Richardson 	& 55.90\%	& 1.75\%	& 44.10\%	& 0.66\%	& \num{1.11e-7}\%	& 76.03\%	\\
		\hline
	\end{tabular}
\end{table}

The results of \cref{tab:speed-results-fragments} shows some interesting features. If we first focus on the sand-like material, similarly to \cref{subsec:alpha-sensitivity}, we observe that almost the entirety of the fragments eventually re-impact the asteroid. Both the Housen and Richardson models show similar percentages for impacting, escaping, and orbiting fragments with a marginally higher share of escaping fragments for the Richardson case. However, if we look at the WCB row, we observe a substantial difference between the two models. In fact, with the Housen velocity profile we have about 97\% of re-impacting particles, while using the Richardson model, only about 56\% re-impact and almost all the remaining fragments escape. Despite the two speed models differing the most at high speeds (\cref{fig:speed-comparison}), the low-gravity environment of asteroids makes the modelling of low velocity regions of utmost importance. In fact, a minor difference in these small ejection speeds (around tens of centimetres per second) can lead to a substantial difference in the overall fate of the ejecta. \cref{tab:speed-results-fragments} shows that this behaviour is also connected to the material type. For very low strength materials (such as Sand), this effect is limited because both models flatten to very low ejections speeds, considerably lower than the escape velocity of the asteroid.

\begin{figure}[htb!]
	\centering
	\includegraphics[width=0.45\textwidth]{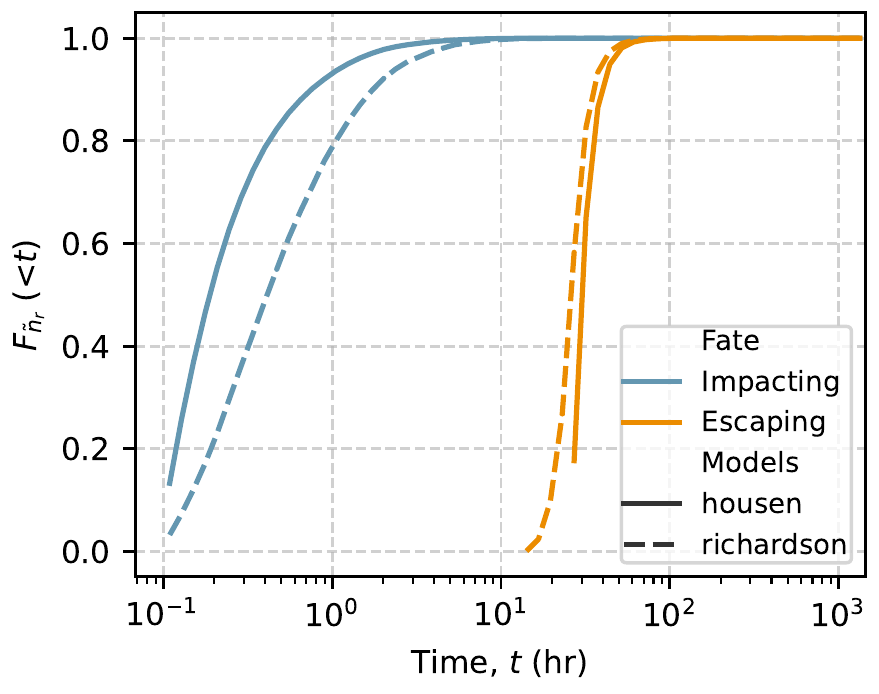}
	\caption{Normalised cumulative distributions of impacting (blue) and escaping (orange) fragments as function of the two ejecta speed models (line style) for the WCB material case.}
	\label{fig:wcb-speed-cum}
\end{figure}

If we now focus on the WCB case, we can compare the normalised cumulative distributions of the impacting and escaping fragments (\cref{fig:wcb-speed-cum}). We observe a more pronounced difference for the impacting particles with respect to the escaping ones. As expected, using the Richardson model, we have particles starting to escape the system before the Housen case. On the other hand, the impacting particles show a less steep cumulative distribution; therefore, they tend to re-impact more gradually and in longer times.

\begin{figure}[htb!]
	\centering
	\includegraphics[width=0.48\textwidth]{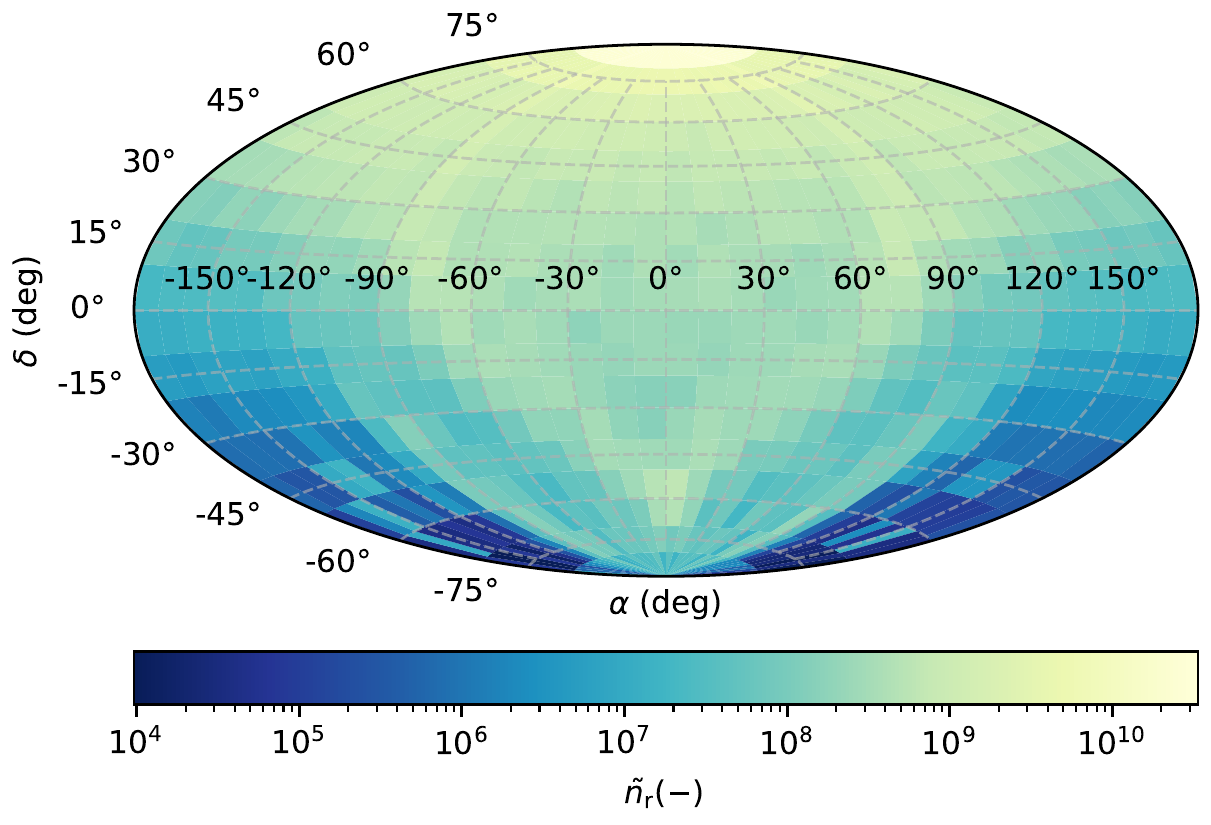}
	\caption{Average number of representative fragments impacting on the asteroid surface represented with a Hammer projection. Each 10\si{\degree}$\times$10\si{\degree} bin in right ascension and declination shows the sum of the representative fragments impacting in it, averaged over the set of 20 simulations performed.}
	\label{fig:speed-wcb-hammer}
\end{figure}

\begin{figure}[htb!]
	\centering
	\begin{subfigure}[b]{0.32\textwidth}
		\centering
		\includegraphics[width=\textwidth]{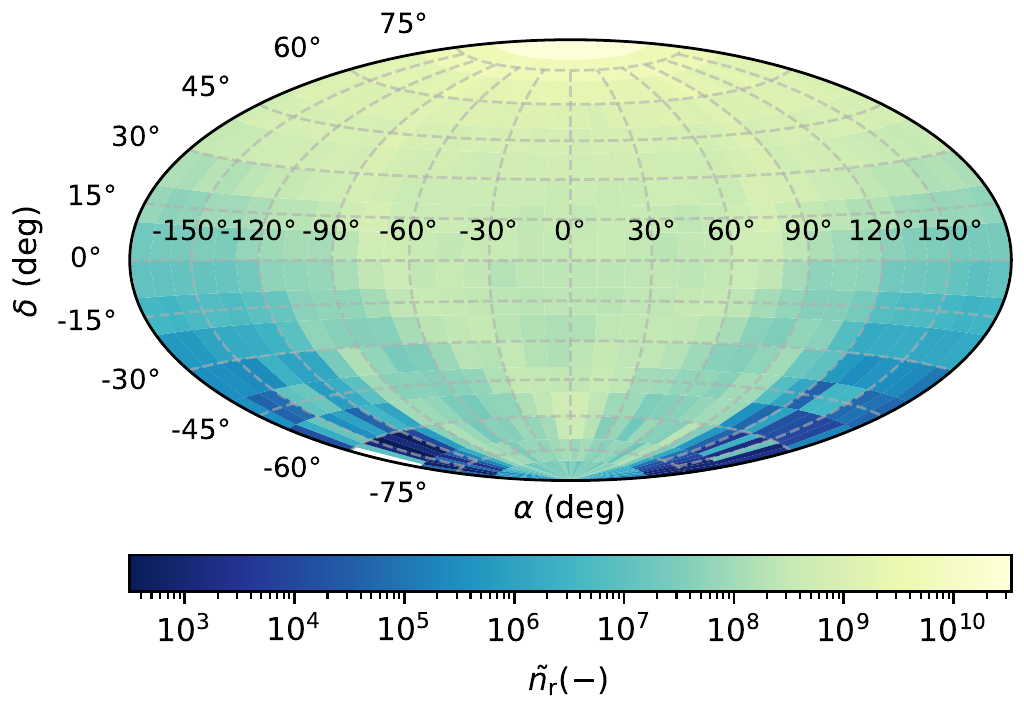}
		\caption{ }
		\label{fig:speed-wcb-hammer-range-0}
	\end{subfigure}
	\hfill
	\begin{subfigure}[b]{0.32\textwidth}
		\centering
		\includegraphics[width=\textwidth]{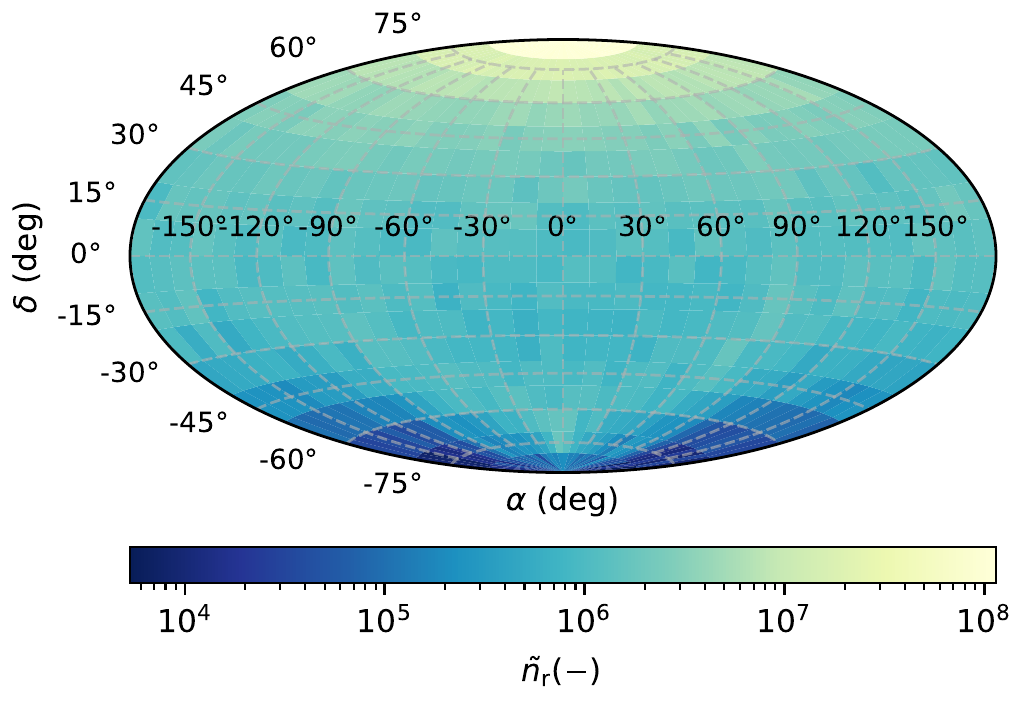}
		\caption{ }
		\label{fig:speed-wcb-hammer-range-1}
	\end{subfigure}
	\hfill
	\begin{subfigure}[b]{0.32\textwidth}
		\centering
		\includegraphics[width=\textwidth]{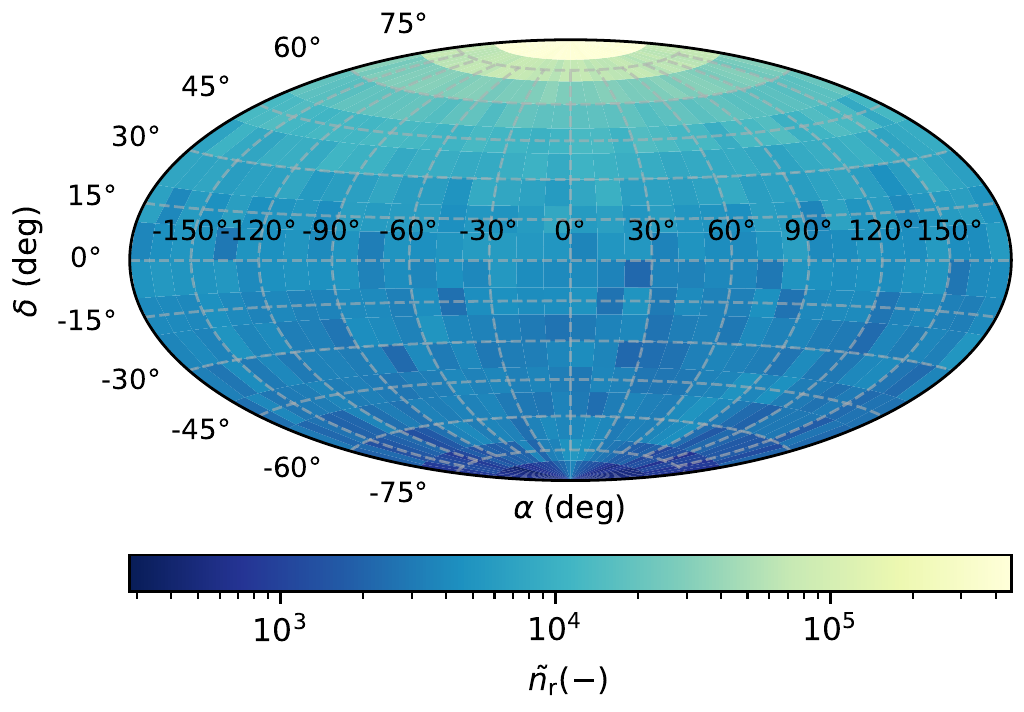}
		\caption{ }
		\label{fig:speed-wcb-hammer-range-2}
	\end{subfigure}
	\caption{Average number of representative fragments impacting on the asteroid surface represented with a Hammer projection. The different plots represent particles belonging to different size ranges. a) $d_p \in \left[ 10\,\si{\micro\metre}, 100\,\si{\micro\metre} \right] $. b) $d_p \in \left[ 100\,\si{\micro\metre} - 1\,\si{\milli\metre} \right]$. c) $d_p \in \left[ 1\,\si{\milli\metre} - 1\,\si{\centi\metre} \right]$.}
	\label{fig:speedpos_nf_ranges}
\end{figure}

\cref{fig:speed-wcb-hammer} shows the distribution of the number of fragments re-impacting the asteroid surface in a 10 deg $\times$ 10 deg grid in right ascension ($\alpha$) and declination ($\delta$) for the Housen speed model. For each grid cell, we compute the total number of representative targets "falling" into the cell and average this value over the 20 impact simulations we perform. This plot is used as a reference to show the distribution of the ejecta blanket for the reference impact scenario to give a better understanding of the following comparisons. In addition, \cref{fig:speedpos_nf_ranges} shows equivalent plots for the, subdividing the fragments' distribution as a function of their size. Specifically, we consider three different size ranges: $d_p \in \left[ 10\,\si{\micro\metre}, 100\,\si{\micro\metre} \right] $, $d_p \in \left[ 100\,\si{\micro\metre} - 1\,\si{\milli\metre} \right]$, and $d_p \in \left[ 1\,\si{\milli\metre} - 1\,\si{\centi\metre} \right]$. As expected, we can observe how the global distribution of \cref{fig:speed-wcb-hammer} tends to follow the distribution of smaller particles as they are in greater quantity. As the particle size increases, the fragments tend to distribute more uniformly on the asteroid's surface as they are less influenced by the effect of solar radiation pressure.

\begin{figure}[htb!]
	\centering
	\includegraphics[width=0.48\textwidth]{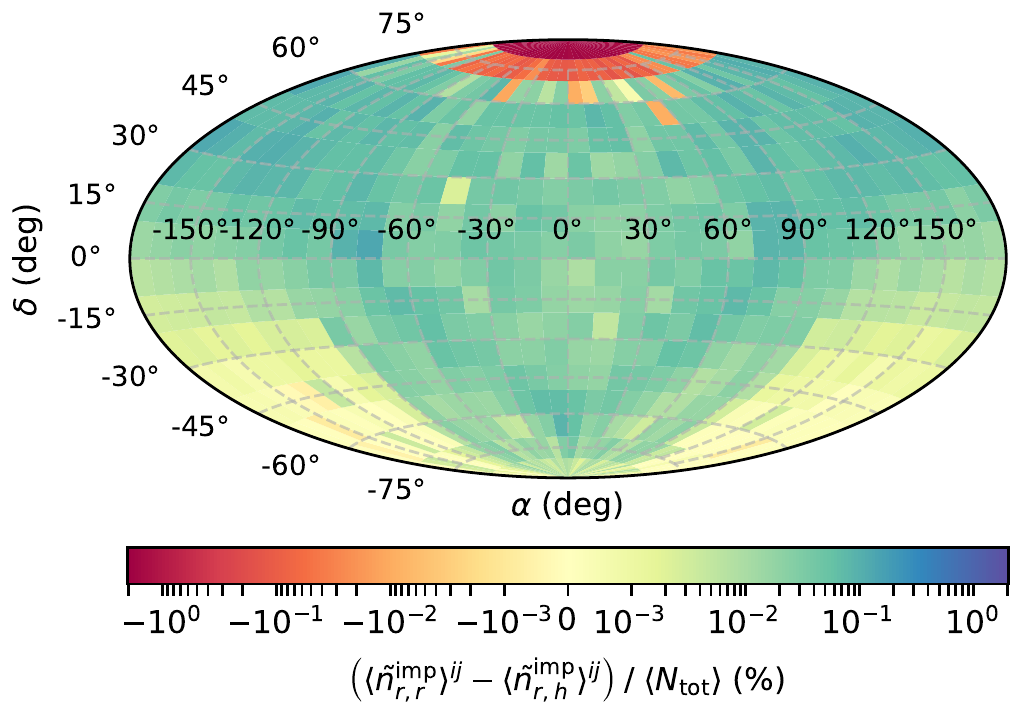}
	\caption{Relative percent difference between the distribution of the average number of impacting representative fragments obtained with the Richardson and the Housen models.}
	\label{fig:speed-wcb-hammer-diff}
\end{figure}

To better understand the difference between the two ejecta speed model, we can perform the difference between these distributions for the two cases.  \cref{fig:speed-wcb-hammer-diff} shows the relative percent difference between a distribution of impacting particles such as \cref{fig:speed-wcb-hammer} for the Richardson and Housen models: the difference between the representative fragments computed for the Richardson case ($\langle \tilde{n}_{r, r}^{\rm imp} \rangle^{ij}$) and the Housen case ($\langle \tilde{n}_{r, h}^{\rm imp} \rangle^{ij}$) is normalised by the total number of fragments, which are identical for both simulations. The light yellow regions correspond to areas where the number of impacting fragments is almost identical for the two models. These are the regions at negative declinations and on the Sun-facing hemisphere of the asteroid. These are also the regions where fewer fragments land as the effect of SRP tends to blow the fragments in direction opposite to the Sun. The red regions identify areas where the Housen model generates more impacts. These are located in the neighbourhood of the impact crater because of the lower ejection speeds predicted by this model. The blue regions instead are characterised by a larger amount of impacts for the Richardson model. The maximum and minimum percent difference are of the order of 1\%. This may seem a small value; however, 1\% of the total number of fragments is in the order of \num{1e9} particles.

The choice of the ejecta speed model can thus influence the overall fate of the ejecta, especially for higher strengths materials such as the WCB. The differences between the ejecta fates is mainly due to the difference in the low-velocity region of the two models. This causes the share of ejecta belonging to the impacting and escaping categories; it also influences the rate of accretion and the distribution of the impacting particles onto the asteroid's surface.

\subsection{Position-based vs. speed-based formulation}  \label{subsec:sensitivity-pos-vel}
\begin{figure}[htb!]
	\centering
	\includegraphics[width=0.45\textwidth]{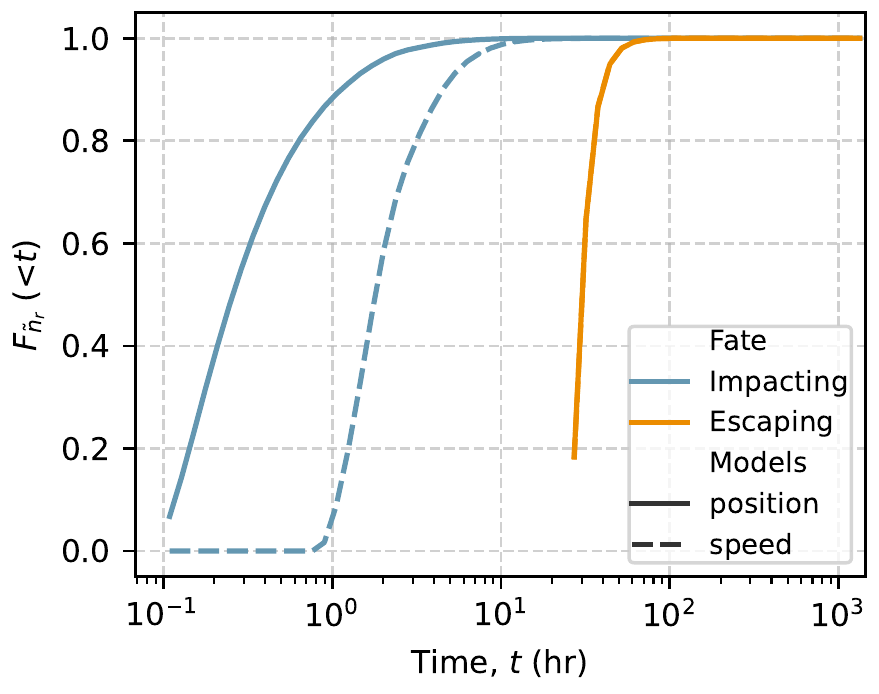}
	\caption{Normalised cumulative distributions of impacting (blue) and escaping (orange) fragments as function of the two ejecta model formulations (line style) for the WCB material case.}
	\label{fig:speedpos-wcb-cum}
\end{figure}

As described in \cref{sec:distributions}, we have introduced two formulation for the developed distribution-based ejecta model; one formulation based on sampling the ejection location, $r$, and identified as \emph{position-based}, and another formulation based on sampling directly the ejection speed, $u$, and identified as \emph{speed-based}. The main difference between the two formulations resides in the presence of a cut-off speed, the \emph{knee-velocity} (\cref{eq:knee-velocity}), in the \emph{speed-based} formulation, below which no mass is ejected. The \emph{knee-velocity} is always larger than the minimum ejection speed of the \emph{position-based} formulation; therefore, a difference between the two formulations is expected. To quantify this difference, we perform two sets of simulations. One set uses the \emph{position-based} formulation (\cref{subsec:position-dist}) with a uniform distribution for the $\xi$ angle and a Gaussian distribution for the $\psi$ angle (\cref{subsubsec:outplane-dist-position-based}). The other set uses the \emph{speed-based} formulation with the corresponding distributions for the $\xi$ and $\psi$ angles (i.e., \cref{subsubsec:inplane-dist-speed-based,subsubsec:outplane-dist-speed-based}, respectively). For this test case only the results for the WCB material are presented.
\bigbreak
Similar to the previous section, we show the normalised cumulative distributions of impacting and escaping fragments in \cref{fig:speedpos-wcb-cum}. We can clearly observe the effect of the \emph{knee velocity} in the cumulative distribution of the impacting particles. The \emph{speed-based} distribution, in fact, shows a significant "delay" in the average impact time of the fragments. After one hour, almost 90\% of the fragments generated with the \emph{position-based} formulation have re-impacted, while only about 5\% of the fragments generated with the \emph{speed-based} formulation. However, it is important to note that the total amount of generated fragments is different for the two formulations, that is, \num{3.816e11} for the \emph{position-based} and \num{2.217e11} for the \emph{speed-based}, which corresponds to about 42\% fewer fragments. In addition, the share of fragments re-impacting the asteroid is different for the two formulations. On one side we have 93.36\% of the total number of fragments for the \emph{position-based} formulation vs. 54.51\% for the \emph{speed-based}.

\begin{figure}[htb!]
	\centering
	\includegraphics[width=0.48\textwidth]{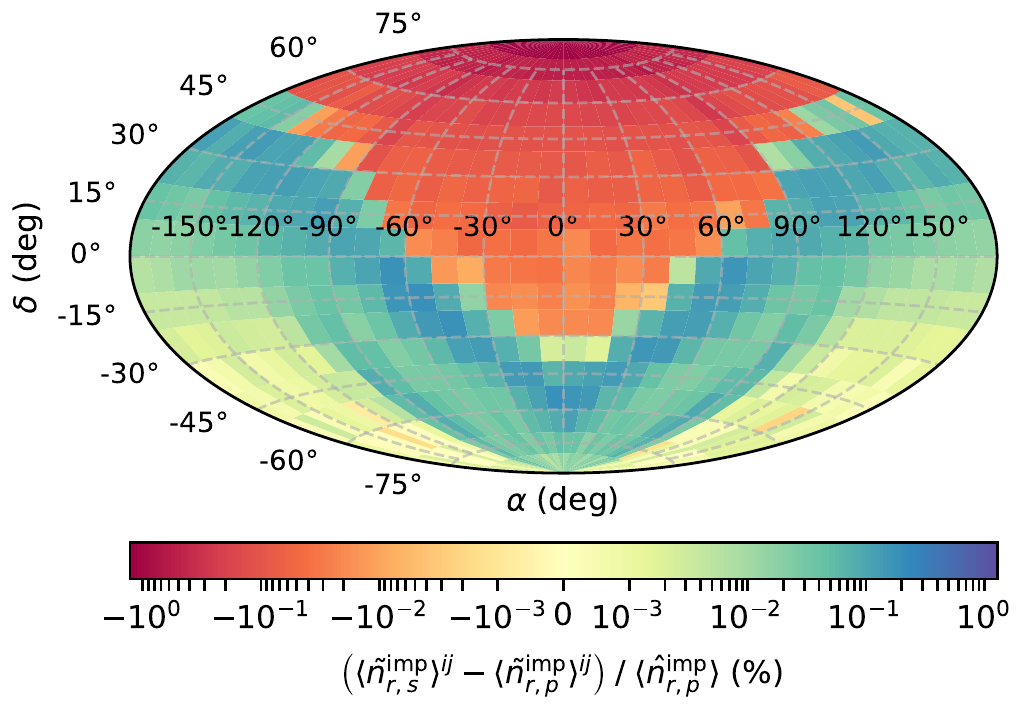}
	\caption{Relative percent difference between the distribution of the average number of impacting representative fragments obtained with the \emph{speed-based} and \emph{position-based} formulations.}
	\label{fig:speedpos-wcb-hammer}
\end{figure}

Analogously to \cref{fig:speed-wcb-hammer}, we show in \cref{fig:speedpos-wcb-hammer} the relative percent difference between the distribution of impacting particles onto a Hammer projection of the asteroid surface obtained with the \emph{speed-based} and \emph{position-based} formulations. In this case, because the total number of fragments is different for the two formulations, we normalise with respect to the total fragments of the \emph{position-based} formulation. The results show how the distribution of impacting fragments is influenced by the choice of the model. We observe a clear pattern in the impact distribution. A first red region closer to the impact location that identifies an area with a larger amount of impacts predicted by the \emph{position-based} model. A second region, highlighted by the blue shades, identifies regions with a higher predicted amount of impacts of \emph{speed-based} model. These area are further away from the impact location because a larger amount of faster particles is generated.

\begin{figure}[htb!]
	\centering
	\begin{subfigure}[b]{0.42\textwidth}
		\centering
		\includegraphics[width=\textwidth]{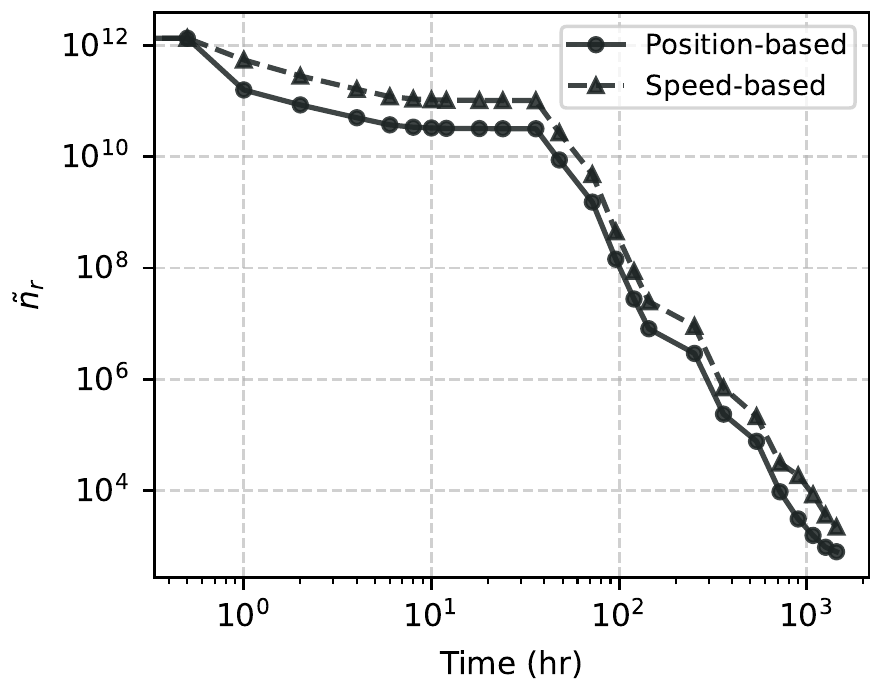}
		\caption{ }
		\label{fig:speedpos_wcb1k_nf}
	\end{subfigure}
	\hfill
	\begin{subfigure}[b]{0.42\textwidth}
		\centering
		\includegraphics[width=\textwidth]{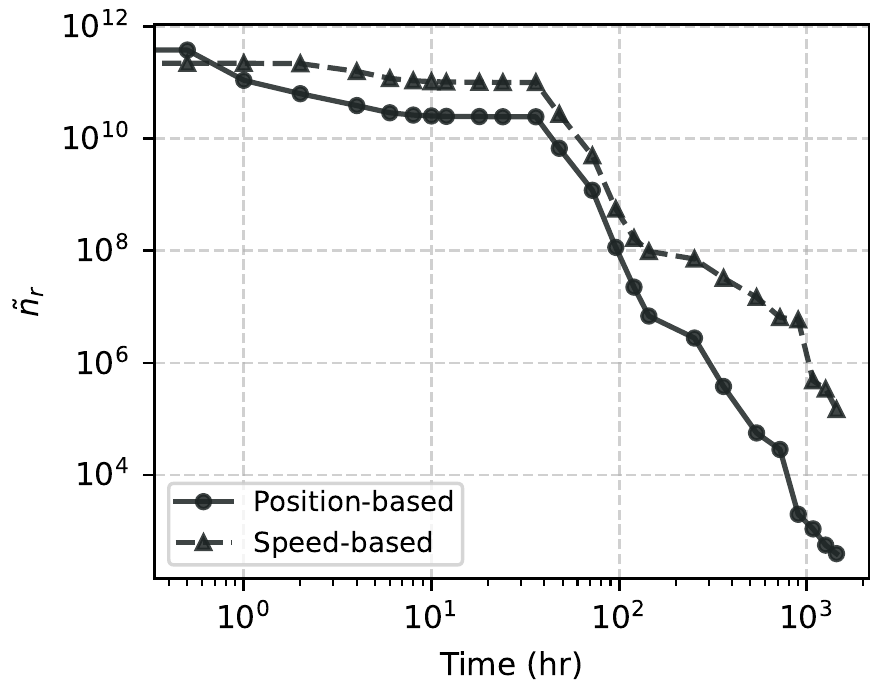}
		\caption{ }
		\label{fig:speedpos_wcb5k_nf}
	\end{subfigure}
	\caption{Number of fragments at selected snapshots for the two distribution formulation. (a) WCB material with strength $Y = 1$ \si{\kilo\pascal} (b) WCB material with strength $Y = 5$ \si{\kilo\pascal}.}
	\label{fig:speedpos_nf}
\end{figure}

Similarly to \cref{subsec:sensitiviy-speed-model}, also the comparison between the \emph{speed-based} and \emph{position-based} formulations is influenced by the magnitude between the \emph{knee velocity} and its difference with respect to the minimum speed of the \emph{position-based} formulation. This in turn is influenced by the type and strength of the target material and its density. Specifically, the higher the strength of the material the higher is the difference in the fate of the ejecta. As mentioned at the beginning of \cref{sec:sensitivity}, the previous results have been obtained for a WCB material with an equivalent strength of 5 \si{\kilo\pascal}. To understand the influence of the strength on this analysis, \cref{fig:speedpos_nf} shows the difference in the evolution of the fragments at selected snapshots for a WCB target with equivalent strength of (a) 1 \si{\kilo\pascal} and  (b) of 5 \si{\kilo\pascal}. We can observe that in the lower strength case the fragment evolution of the two formulations shows a comparable behaviour. Instead, we observe a greater discrepancy once the strength is increased to 5 \si{\kilo\pascal}.

\subsection{Correlated vs. Uncorrelated distribution}  \label{subsec:sensitivity-correlated}
\begin{figure}[htb!]
	\centering
	\includegraphics[width=0.45\textwidth]{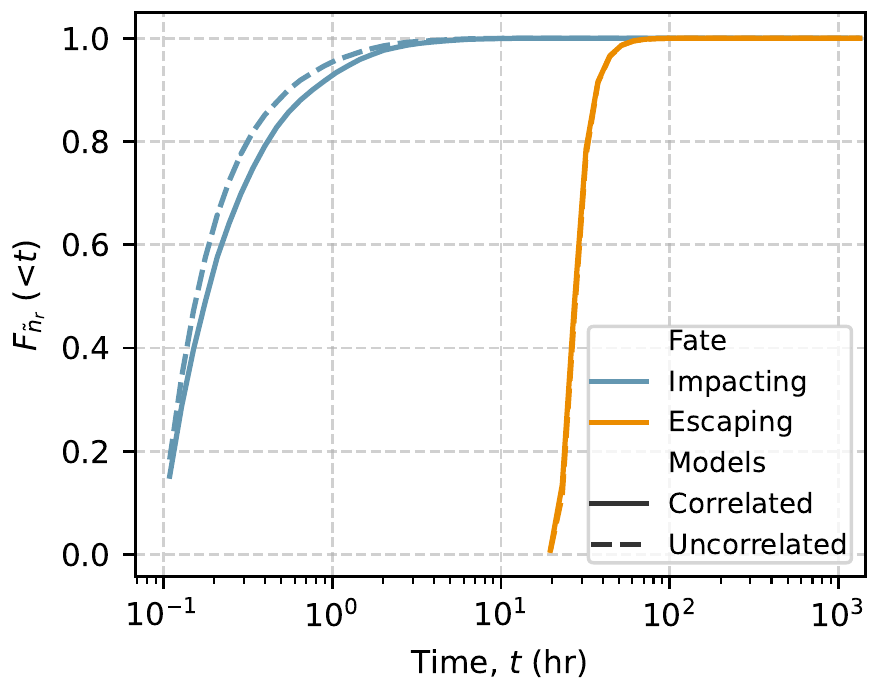}
	\caption{Normalised cumulative distributions of impacting (blue) and escaping (orange) fragments for the correlated and uncorrelated formulations (line style) for the sand-like material case.}
	\label{fig:corruncorr_cum}
\end{figure}

\begin{figure}[htb!]
	\centering
	\includegraphics[width=0.45\textwidth]{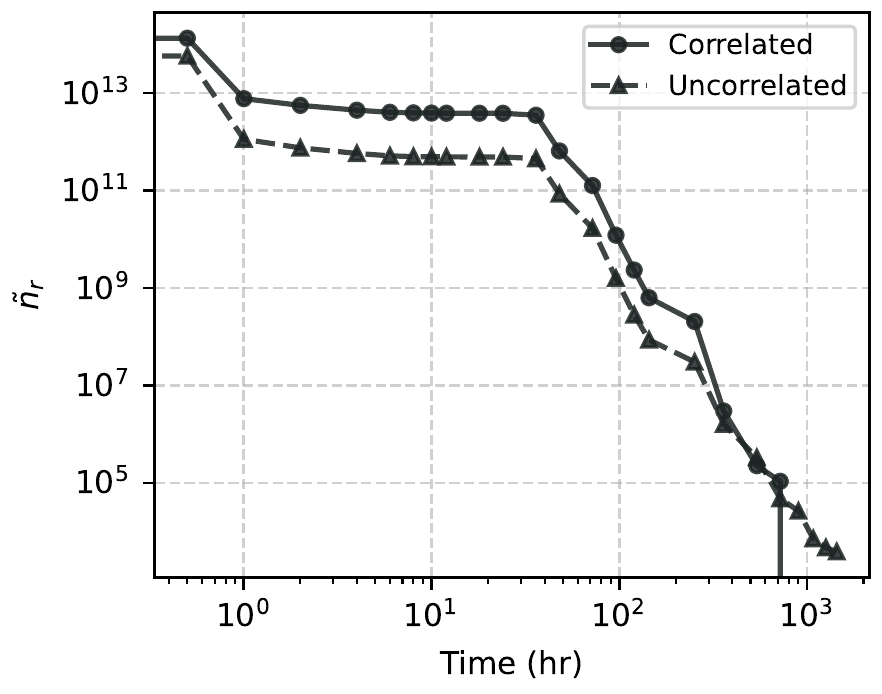}
	\caption{Number of fragments at selected snapshots for the correlated and uncorrelated formulation.}
	\label{fig:corruncorr_nf}
\end{figure}

This section presents the comparison between the speed-averaged formulations with and without the correlation between the particle size and the ejection speed (\cref{subsec:speed-dist,subsec:correlated-dist}). As mentioned in \cref{sec:distributions}, the difference between these two formulations resides in the speed limitation the correlated distribution has for increasing particle size: the larger the particle the lower the maximum admissible ejection speed is. This test case compares the two formulations in terms of the overall ejecta behaviour. Alongside the correlated and uncorrelated size-speed distributions, we use a uniform distribution for the $\xi$ angle and a Gaussian distribution for the $\psi$ angle (\cref{subsubsec:outplane-dist-speed-based}). The material selected for this test case is a strengthless sand-like material.
\bigbreak
Some of the results for the overall ejecta evolution can by observed in \cref{fig:corruncorr_cum,fig:corruncorr_nf}. Here we observe the normalised cumulative distribution for the impacting and escaping fragments and the total number of fragments still orbiting the asteroid as selected snapshots in time. \cref{fig:corruncorr_cum} shows that the escaping particles have almost an identical behaviour over time, while a difference can be observed for the impacting fragments. Specifically, the cumulative distribution associated to the correlated formulation is less steep; therefore, we have a larger percentage of fragments surviving longer around the asteroid, at least in the first stages after the impact.

Other features can be observed in \cref{fig:corruncorr_nf}. First, the total number of fragments is higher for the correlated case: given that larger particles with higher velocities are not admissible, a larger number of smaller fragments is necessary to satisfy the mass conservation constraint. Second, in the latest stages of the propagation (after about 100 hours), the number of fragments for the correlated case decrease more rapidly and no more fragments are present after approximately 700 hours, while fragments are still present for the uncorrelated case. Because in the correlated formulation larger fragments have lower ejection velocities, there is also a smaller probability that they keep orbiting the asteroid for longer timescales.

\begin{figure}[htb!]
	\centering
	\begin{subfigure}[b]{0.48\textwidth}
		\centering
		\includegraphics[width=\textwidth]{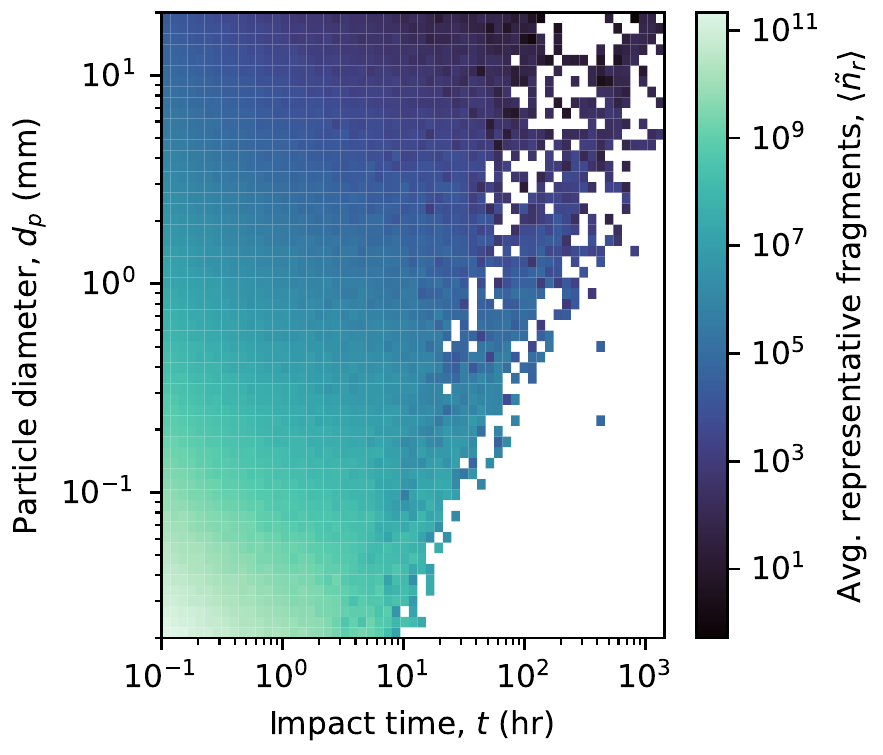}
		\caption{ }
		\label{fig:corruncorr_t_dp_uncorr}
	\end{subfigure}
	\hfill
	\begin{subfigure}[b]{0.48\textwidth}
		\centering
		\includegraphics[width=\textwidth]{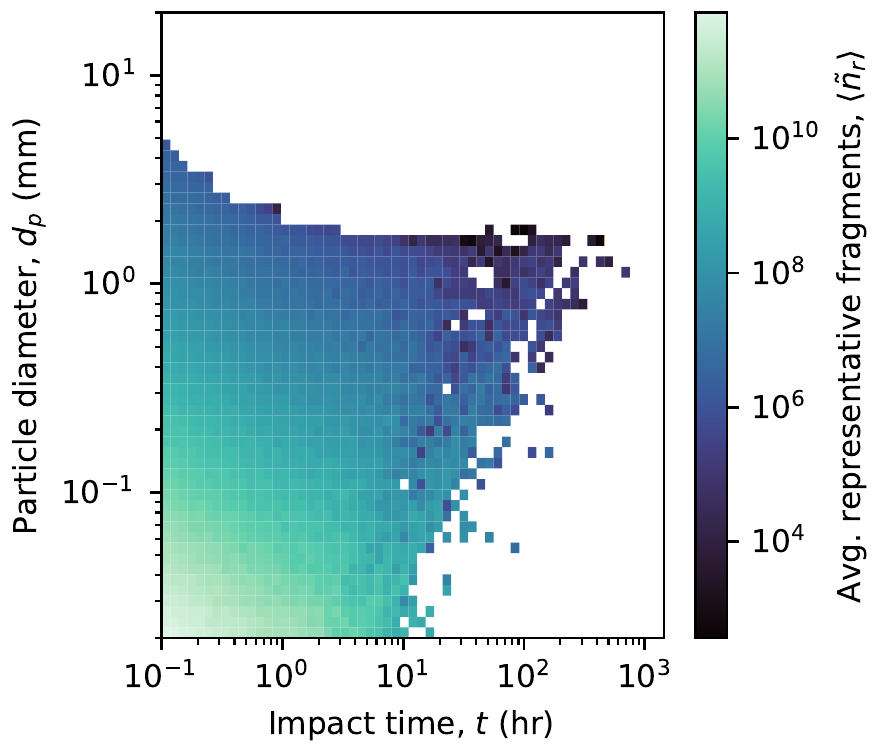}
		\caption{ }
		\label{fig:corruncorr_t_dp_corr}
	\end{subfigure}
	\caption{Average number of representative fragments re-impacting the asteroid as function of the impact time and the particle diameter. (a) Uncorrelated distribution. (b) Correlated distribution.}
	\label{fig:corruncorr_t_dp}
\end{figure}

This feature can be better observed in \cref{fig:corruncorr_t_dp}, which shows the average number of representative fragments re-impacting the asteroid as function of the impact time (on the x-axis) and particle diameter (on the y-axis). \cref{fig:corruncorr_t_dp_corr} clearly shows the effect of the size-speed correlation on larger particles as they present a decreasing upper limit of the impact time as their diameter increases.

Overall, introducing the size-speed correlation seems to have a limited influence on the overall fate of the ejecta. It has to be noted, however, that we are focusing on outcome of a cratering event on small bodies and limited to ejection velocities within the escape speed of the considered target. As this velocity is small, the correlation effect has a more limited influence. In case a wider range of ejection velocities is considered a greater impact may be expected.

\subsection{Out-of-plane ejection angle ($\psi$) distribution}  \label{subsec:sensitivity-psi}
\begin{figure}[htb!]
	\centering
	\includegraphics[width=0.45\textwidth]{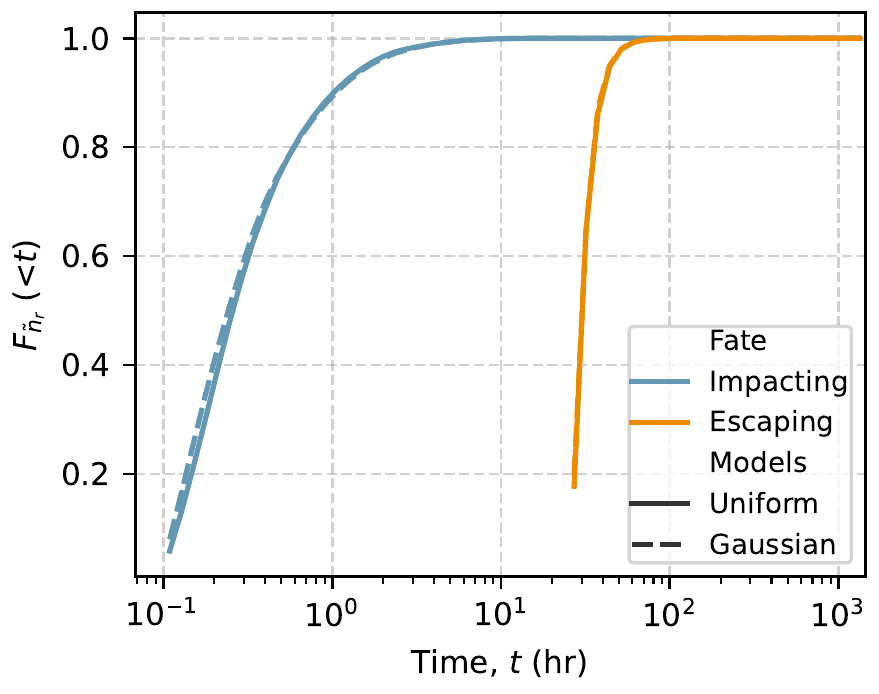}
	\caption{Normalised cumulative distributions of impacting (blue) and escaping (orange) fragments as function of the two ejecta model formulations (line style) for the WCB material case.}
	\label{fig:psi-wcb-cum}
\end{figure}

In this section, we consider the possible difference caused by the selection of different models for the out-of-plane component of the ejection angle, $\psi$. As described in \cref{sec:distributions}, we consider two types of distributions, Uniform and a Gaussian. Again, for the test case, the \emph{position-based} formulation is used (\cref{subsec:position-dist}) with a uniform distribution for the $\xi$ angle and the Housen formulation for the ejection speed (\cref{subsubsec:u-dist-position-based}). The considered material is the WCB (\cref{tab:materials}) Following the previous examples, \cref{fig:psi-wcb-cum} shows the normalised cumulative distribution of the impacting and escaping fragments as function of the out-of-plane ejection angle model. From the figure we can observe only a marginal difference in the overall ejecta fate due to the modelling assumption concerning the ejection angle. The evolution of the escaping fragments is superimposed for the two models, while a small difference can be observed for the impacting fragments.

\begin{figure}[htb!]
	\centering
	\includegraphics[width=0.48\textwidth]{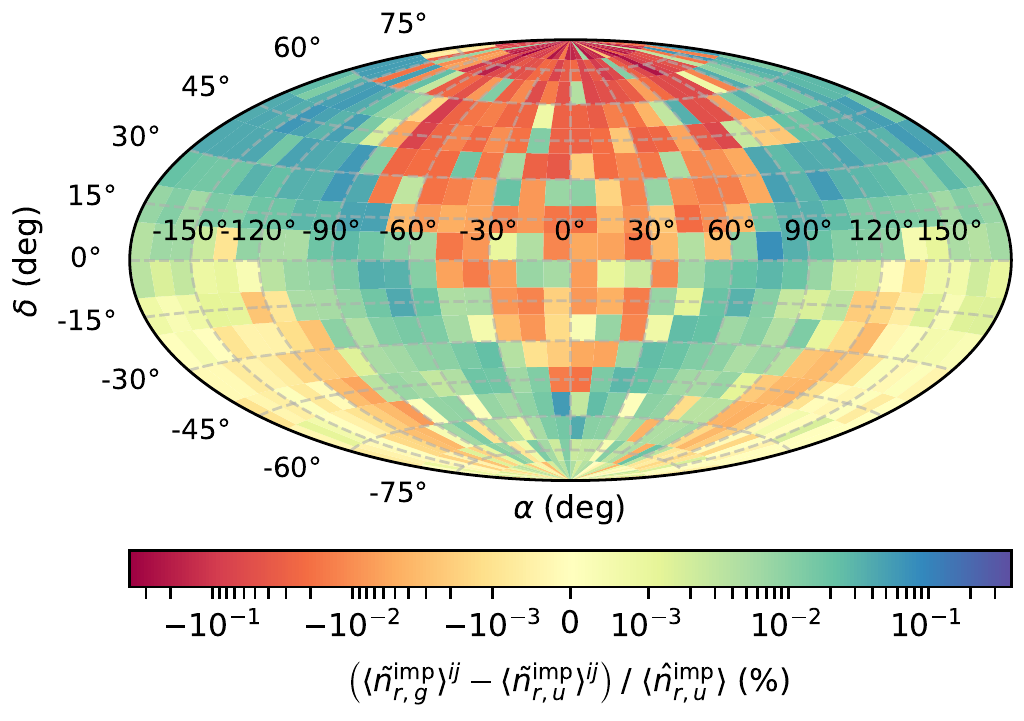}
	\caption{Relative percent difference between the distribution of the average number of impacting representative fragments obtained with the Gaussian and Uniform models of out-of-plane ejection angle ($\psi$) distribution.}
	\label{fig:psi-wcb-hammer}
\end{figure}

\cref{fig:psi-wcb-hammer} shows instead the difference between the distribution on the surface of the asteroid of the average impacting particles in a grid of right ascension and declination between the Gaussian models ($\langle \tilde{n}_{r, g}^{\rm imp} \rangle^{ij}$) and the Uniform model ($\langle \tilde{n}_{r, u}^{\rm imp} \rangle^{ij}$). The difference is again expressed as a relative percentage that is the difference between the number of impact per cell in the Gaussian case minus the one in the Uniform case, divided by the total averaged number of fragments generated in the Uniform case. The obtained distribution shows that there is a central region along the Sun-asteroid direction that tends to have a predominance of fragments generated by the Uniform model. However, we also observe a non-smooth behaviour with few cells that shows a majority of impacts for the Gaussian case. The regions around this central area, instead, shows a majority of fragments generated in the Gaussian case. Nonetheless, we should note that the relative difference is below 1\% of the total amount of fragments (peaks of the order of $\pm$ 0.3\%) that correspond to a number of fragments in the order of \num{1e8}.

Therefore, when looking at the overall fate of the ejecta, the selection of the ejection angle model does not substantially influence the percentage of impacting and escaping particles nor their rate of accretion and escape. On the other hand, it can influence the distribution of the impact location on the asteroid as shown in \cref{fig:psi-wcb-hammer}.

\subsection{Impact angle}  \label{subsec:sensitivity-impact-angle}
In this section, we consider the sensitivity to the impact angle, $\phi$, which defines the inclination of the impactor velocity vector with respect to a plane tangent to the asteroid's surface at the impact point. Therefore, a 90 degrees impact angle corresponds to a normal impact. The effect of the impact angle is twofold. On one hand, the shallower the impact angle is the smaller the normal component of the impact velocity is, which in turn affects the effectiveness of the cratering process. In fact, lower normal components of the impact velocity result in smaller impact craters. On the other hand, the presence of an impact angle different from 90 degrees changes the shape of the ejecta plume, as shown in \cref{sec:distributions}. In this section, we analyse the effects of changing impact angle; specifically, we vary the impact angle from 30$^\circ$ to 90$^\circ$ with increments of 15$^\circ$. The impact location is the Norh pole of the asteroid. The incoming direction of the projectile is along the x-axis of the synodic frame and in the same direction; therefore, moving away from the Sun. The remaining configuration is as follows: the \emph{position-based} formulation is used (\cref{subsec:position-dist}) with a \emph{Lobed} distribution for the $\xi$ angle (\cref{eq:xi-distribution}), a Gaussian distribution for the $\psi$ angle (\cref{eq:psi-distribution-gaussian}), and the Housen formulation for the ejection speed (\cref{subsubsec:u-dist-position-based}). The considered material is the WCB (\cref{tab:materials}) with an equivalent strength of 5 \si{\kilo\pascal}.

\begin{figure}[htb!]
	\centering
	\begin{subfigure}[b]{0.42\textwidth}
		\centering
		\includegraphics[width=\textwidth]{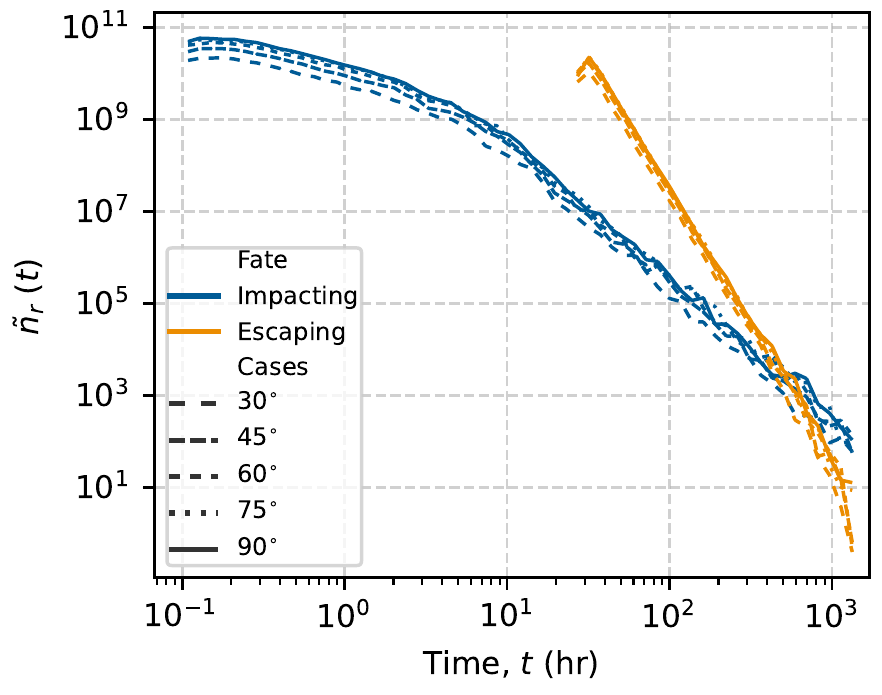}
		\caption{ }
		\label{fig:phi_diff}
	\end{subfigure}
	\hfill
	\begin{subfigure}[b]{0.42\textwidth}
		\centering
		\includegraphics[width=\textwidth]{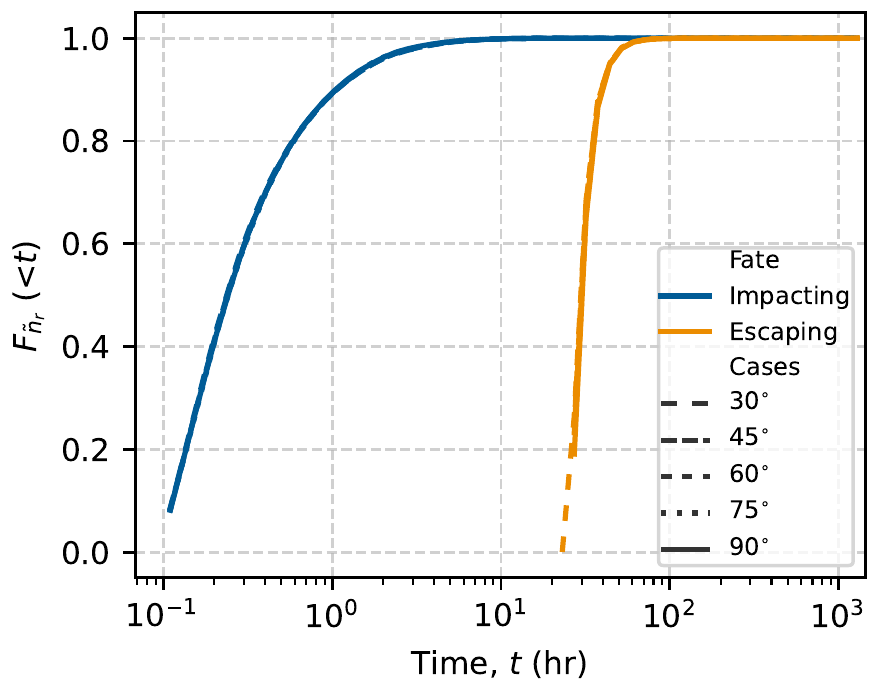}
		\caption{ }
		\label{fig:phi_cum}
	\end{subfigure}
	\caption{Fragments evolution in time. (a) Number of fragments impacting (blue) and escaping (orange) at time $t$ for the analysed impact angles. (b) Corresponding normalised cumulative distribution of the fragments in time.}
	\label{fig:phi_frag_time}
\end{figure}

\cref{fig:phi_frag_time} shows the fate of the ejecta as a function of time for both the impacting and escaping fragments. Specifically, \cref{fig:phi_diff} shows the differential distributions of the impacting and escaping particles in time for the five analysed values of impact angle, $\phi$. \cref{fig:phi_cum} instead shows the corresponding cumulative distribution that is the percentage of particles escaping and impacting within a given time, $t$. From \cref{fig:phi_cum} we can observe that the behaviour of the normalised distribution is almost identical in all the cases. A behaviour that is also corroborated by the similar differential trends that can be observed in \cref{fig:phi_diff}. The only difference we can observe is the absolute number of particles involved. Therefore, we observe that a variation of the impact angle does not substantially change the overall behaviour of the fragments, but it is most probably a cause of local changes in the spatial distribution of the fragments.

\begin{figure}[htb!]
	\centering
	\begin{subfigure}[b]{0.42\textwidth}
		\centering
		\includegraphics[width=\textwidth]{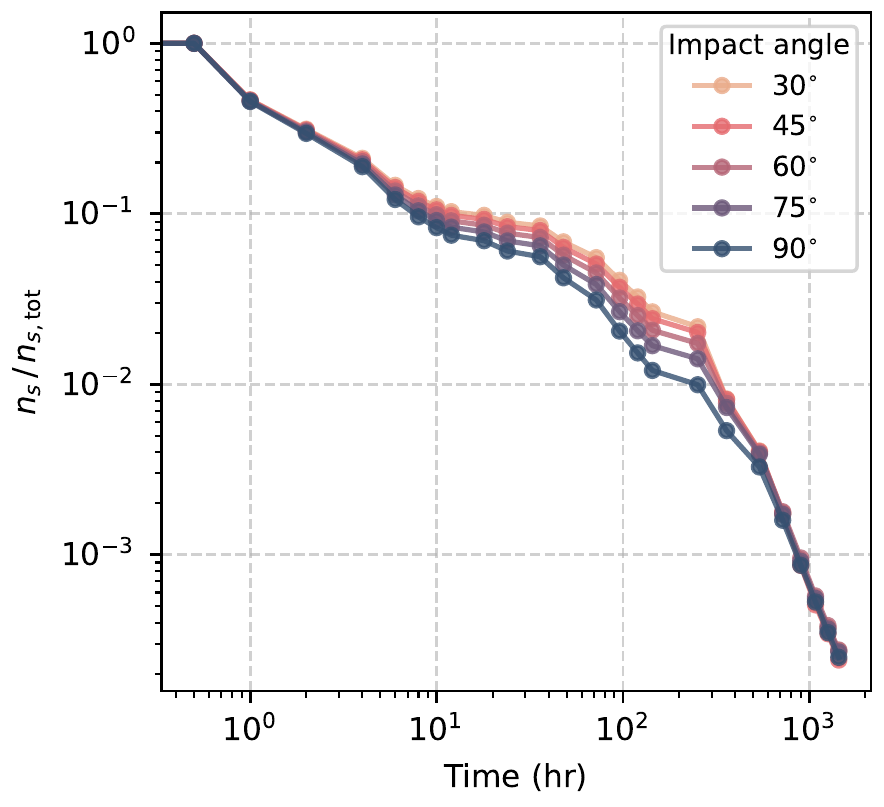}
		\caption{ }
		\label{fig:phi_ns_snap}
	\end{subfigure}
	\hfill
	\begin{subfigure}[b]{0.42\textwidth}
		\centering
		\includegraphics[width=\textwidth]{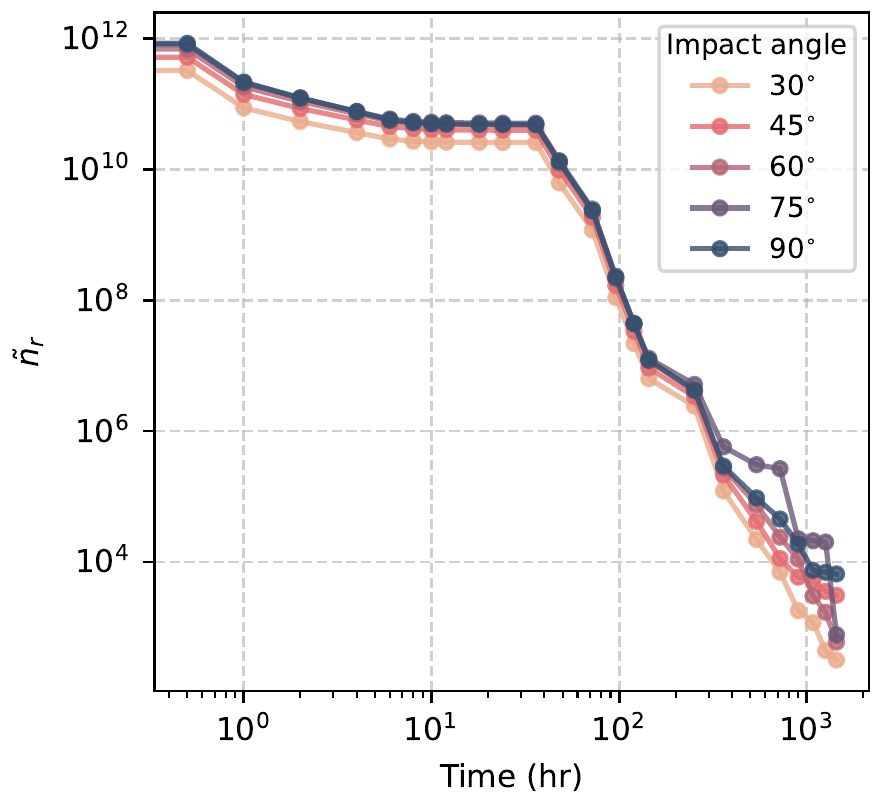}
		\caption{ }
		\label{fig:phi_nf_snap}
	\end{subfigure}
	\caption{Number of samples (a) and fragments (b) at selected snapshot times for the different impact angles analysed.}
	\label{fig:phi_snap}
\end{figure}

\cref{fig:phi_snap} shows instead the percentage of samples (a) and the number of fragments (b) still orbiting the asteroid at specified snapshots in time. In this case, we observe a difference in the samples evolution in the central portion of the analysed time frame (between about 10 and 200 hours), with a steeper decrease of samples as the impact angle increases. This behaviour, however, does not directly translate into a significant change in the evolution of the number of fragments. As shown in \cref{fig:phi_nf_snap}, the overall trend is similar among the different impact angles. However, we again observe the difference in the total number of fragments produced by the varying impact angle. In fact, the total number of ejected fragments is a function of the normal component of the impact velocity, which changes as the sine of the impact angle. Consequently, we observe a reduction of generated fragments as the impact angle decreases.

\subsection{Asteroid rotation}  \label{subsec:sensitivity-rotation}
\begin{figure}[htb!]
	\centering
	\includegraphics[width=0.45\textwidth]{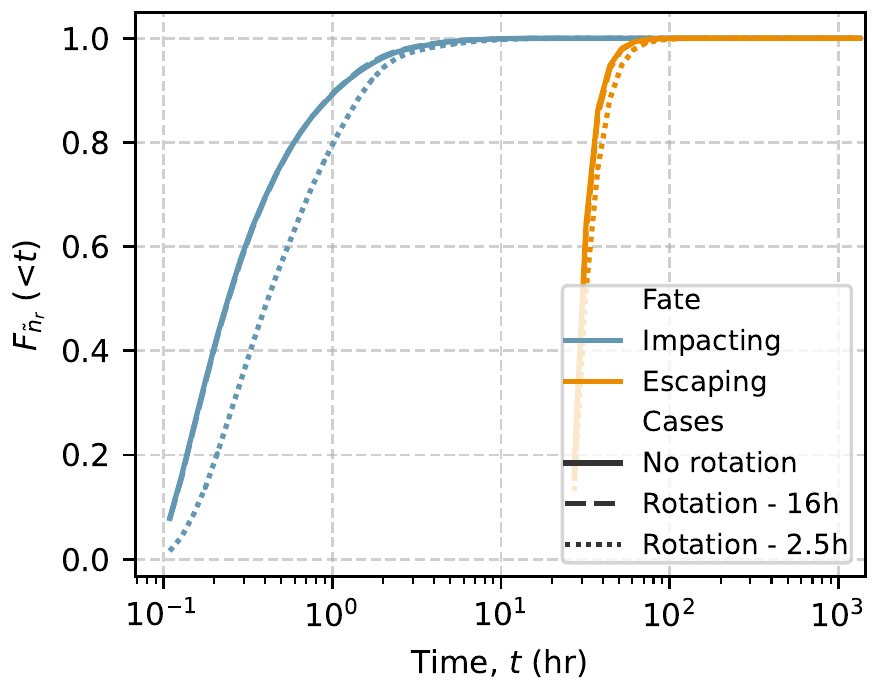}
	\caption{Normalised cumulative distributions of impacting (blue) and escaping (orange) fragments as function of the rotational state (line style).}
	\label{fig:rot_cum}
\end{figure}

In this section, we consider the sensitivity to the asteroid rotation. While it is not strictly related to the ejecta modelling, including the contribution of the asteroid rotation can be regarded as a modelling assumption. The effect of the asteroid rotation is an additional velocity change that must be added to the ejection speed vector defined by the quantities sampled from the ejecta model distribution. To evaluate the effect of the rotational speed, we change the impact conditions with respect to the previous test cases. In this case, we adopt an equatorial impact (instead of a polar impact), to maximise the effect of the rotation (still assuming the rotational axis of the asteroid is perpendicular to the orbital plane). The impact location has therefore a latitude of 0\si{\degree} and a longitude of 90\si{\degree} (measured from the x-axis of the synodic system). As in the previous cases, the ejecta model formulation is the \emph{position-based} one with a uniform distribution for the $\xi$ angle and a Gaussian distribution for the $\psi$ angle (\cref{subsubsec:outplane-dist-position-based}). The ejection speed model is the Housen formulation (\cref{eq:speed-housen}). The considered material is again the WCB with an equivalent strength of 5 \si{\kilo\pascal}. Three different rotational states have been considered: \emph{no-rotation}, a rotational period of 16\si{\hour} (average over the asteroid database), and a rotational period of 2.5\si{\hour} (lower end of the admissible rotational period for rubble pile asteroids, before the family of very fast rotating asteroids \citep{pravec2002asteroid}).
\bigbreak
The results of the simulations shows that the share of impacting and escaping fragments is almost identical when we compare the slow rotational period of 16\si{\hour} with the case without rotation, corresponding to approximately 94.5\% of impacting and 5.5\% of escaping fragments, respectively. When we consider the fast-rotating case with a period of 2.5\si{\hour}, the partition changes to an 87.6\% of impacting fragments and a 12.4\% of escaping ones. If we look at the velocity contribution due to the asteroid rotation for a point at the equator, we have an addition of approximately 0.17 \si{\meter\per\second} for a period of 2.5\si{\hour} and of approximately 0.027 \si{\meter\per\second} for the 16\si{\hour} period. These contributions correspond to the 29\% and 4.5\% of the asteroid escape velocity, respectively. We can thus expect an increasing effect of the rotational period on the behaviour of the ejecta as the period decreases.
If we look at the different behaviours in time, \cref{fig:rot_cum} shows the normalised cumulative distributions for the impacting and escaping fragments. Also from this plot, we observe that differences are visible only for the fast rotating case, whose contribution, in this case, is to change the frequency at which the particles re-impact the asteroid and escape the system. 

\begin{figure}[htb!]
	\centering
	\includegraphics[width=0.48\textwidth]{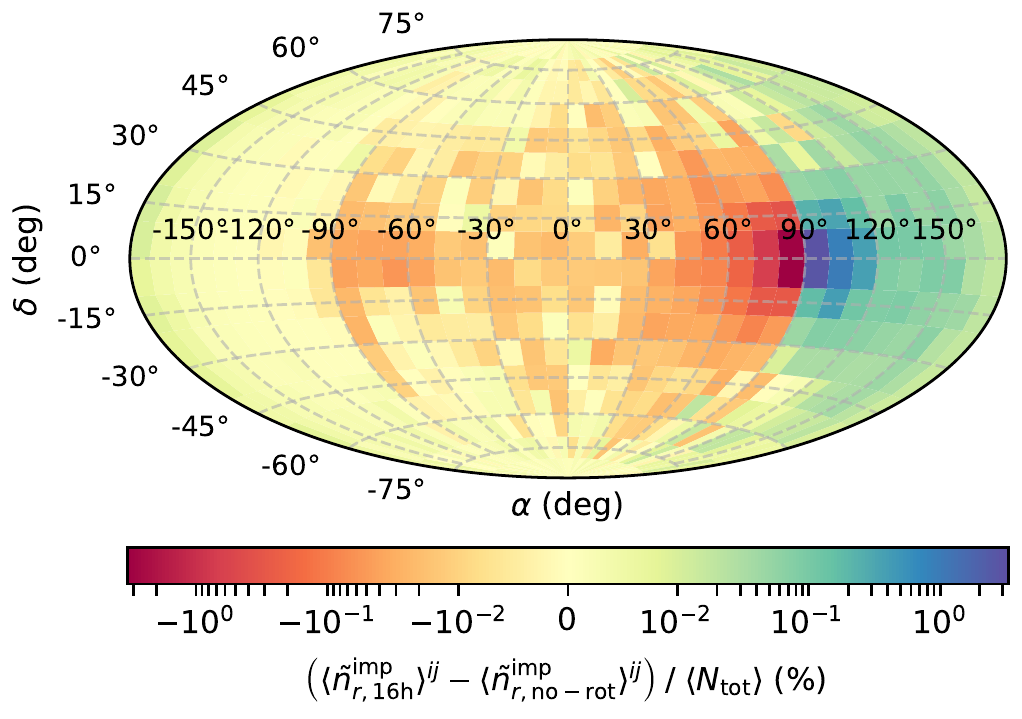}
	\caption{Relative percent difference between the distribution of the average number of impacting representative fragments obtained with the 16h period case and the no-rotation case.}
	\label{fig:rot16-wcb-hammer-diff}
\end{figure}

\begin{figure}[htb!]
	\centering
	\includegraphics[width=0.48\textwidth]{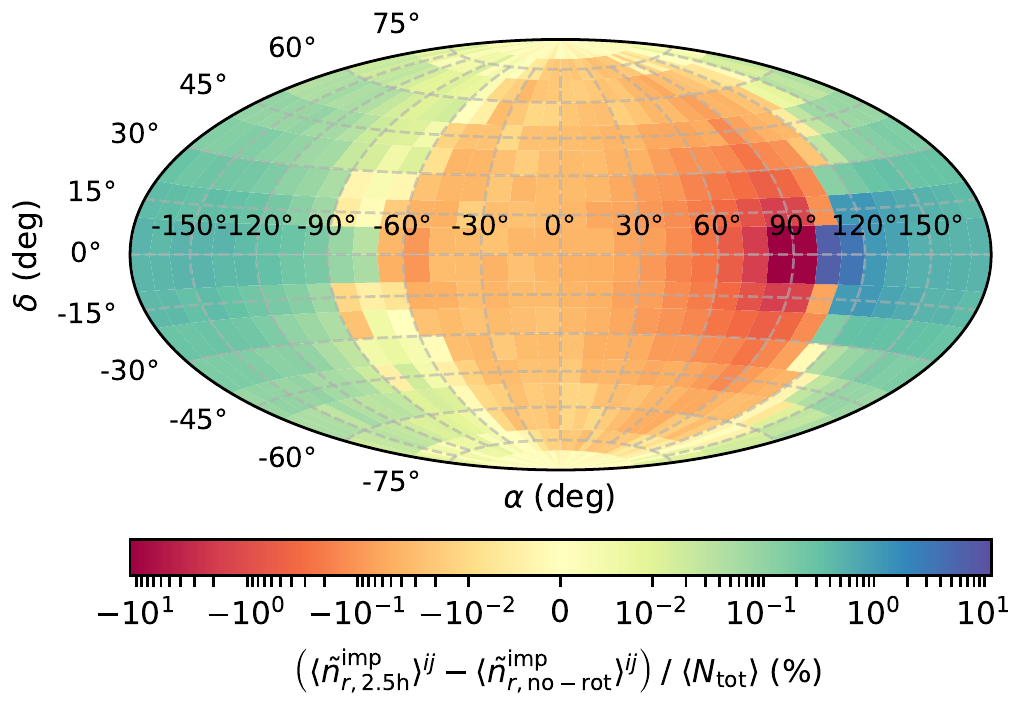}
	\caption{Relative percent difference between the distribution of the average number of impacting representative fragments obtained with the 2.5h period case and the no-rotation case.}
	\label{fig:rot2.5-wcb-hammer-diff}
\end{figure}

Similarly to \cref{fig:psi-wcb-hammer}, \cref{fig:rot16-wcb-hammer-diff,fig:rot2.5-wcb-hammer-diff} show the relative difference between the distribution of impacting fragments onto the asteroid's surface, normalised with respect to the total number of ejected fragments. \cref{fig:rot16-wcb-hammer-diff} shows the difference between the 16h period and the no-rotation case, while \cref{fig:rot2.5-wcb-hammer-diff} shows the difference between the 2.5h period and the no-rotation case. We first notice that even the 16h produces differences in the distribution of fragments despite the effect on the overall fate of the ejecta being limited. We observe a neat separation between to regions in correspondence of the impact point ($\alpha = 90^\circ$, $\delta = 0^\circ$). Along the Eastward direction we have a predominance of fragments for the case that includes the rotational period, while the opposite is observed Westwards. 

\cref{fig:rot2.5-wcb-hammer-diff} shows that as the rotational speed increases, the shift is more pronounced and the difference between the distributions becomes higher. In fact, we pass from a maximum difference of 2-3\% of the total fragments in \cref{fig:rot16-wcb-hammer-diff} to about 10\% in \cref{fig:rot2.5-wcb-hammer-diff}. This in turn corresponds to an order of magnitude difference between the number of fragments, from \num{1e9} to \num{1e10}.

\section{Conclusions and Discussion}
\label{sec:conclusions}

In this work, the development of a modular distribution-based ejecta model is presented. The model is a combination of probability distribution functions that characterise the main parameters governing the ejection phenomenon, i.e., the particle size, the ejection position and speed, and the ejection direction (in-plane and out-of-plane angles). Combining the single distributions, a full four-dimensional characterisation of the ejecta cloud can be obtained as a function of the target and the impactor properties. From the probability density functions, we obtain analytical expressions of the corresponding cumulative distribution functions, which are the base for sampling the distribution and determining the associated \emph{representative fragments}. Three different formulations of the ejecta model have been presented in this work; however, the modularity of the presented model can be leveraged to introduce new and improved correlations, as long as they can be expressed as PDFs or conditional distributions. In addition, the presented analytical formulation can be exploited to estimate the number of particles satisfying specific conditions: by finding the ranges of conditions that satisfy a given scenario (e.g., particles trapped in quasi-stable orbits), the ejecta distribution can be integrated to find the number of particles associated to it.

The developed ejecta distributions are used to compare different modelling techniques and assumptions and assess the sensitivity of the overall ejecta fate. In this work, we focus on few aspects of the ejecta modelling and analyse the contribution of some relevant parameters: the minimum particle size that defines the range of the distribution, the slope coefficient of the size distribution, the type of model used for the ejection speed and the ejection angle, the type of distribution formulation used and the possible correlation between the size and speed of the fragments. We perform the sensitivity analysis focusing on the overall behaviour of the ejecta cloud. This translated to the analysis of the share of particles escaping, re-impacting, and still orbiting the asteroid after a two-months period and in understanding the particles' behaviour in time as a whole. The analysis thus took into account the cumulative distribution in time of the impacting and escaping particles. In addition, an interesting aspect was the distribution of the re-impacting particles onto the asteroid surface.

From the analyses, we observed interesting results that can better inform our future choices about modelling impact crater events and the resulting ejecta clouds. First, we observed that the selection of the slope coefficient, $\abar$, the type of distribution of the out-of-plane ejection angle, $\psi$, and the size-speed correlation do not significantly influence the overall fate of the ejecta in terms of fragment evolution in time; although, a minor influence of the size-speed correlation can be observed for re-impacting fragments. The correlated formulation also generates a higher number of generated fragments and a lower presence of surviving fragments in the latest stages of the simulation. It must be noted that the difference introduced by the size-speed correlation may be higher when the considered speed range is larger that is when the considered asteroid has a bigger gravitational parameter. in fact, the greater the gravitational parameter, the higher is the escape speed, which is the upper limit chosen for the ejection speed in this study. The asteroid rotation has also a marginal influence on the ejecta evolution, as long as the rotational speed is limited. \cref{subsec:sensitivity-rotation}, in fact, showed that for faster rotations the contribution can be relevant in influencing both the rate of impact and escape of the fragments and their distribution on the asteroid surface when they re-impact. Similarly, the type of distribution of the out-of-plane ejection can have an influence on the impacting particles distribution (\cref{fig:psi-wcb-hammer}). Our analyses also showed a limited effect of the impact angle on the overall ejecta fate as the main contribution was the variation of the total number off ejected fragments as a consequence of a smaller normal component of the impact speed as the impact angle decreases. However, further analyses on the effect of the impact angle should be performed to assess the effect it may have on the local distribution of the ejecta plume, especially in the instants right after the impact event. Changing the minimum size of the particles, $\smin$, affected the overall number of generated fragments and their behaviour in time. The effect was more pronounced for escaping particles that reached escape condition in double the time and decreased their rate of escape. As the particles were on average bigger, also the rate of re-impact resulted reduced. Finally, both the selection of the distribution formulation (from the \emph{position-based} and the \emph{speed-based} models) and the type of model of the ejecta speed had the greatest influence on the behaviour of the ejecta, particularly on the impacting particles. A strong difference in the rate of accretion has been observed (\cref{fig:speedpos-wcb-cum,fig:wcb-speed-cum}) and in the distribution of the impacting particles on the asteroid surface (\cref{fig:speedpos-wcb-hammer,fig:speed-wcb-hammer-diff}). Additionally, the difference between the \emph{position-based} and the \emph{speed-based} formulations showed to be dependent on the target material \cref{fig:speedpos_nf}. This is mainly due to the difference in the cut-off speed of the \emph{speed-based} formulation; in fact, the \emph{position-based} formulation showed lower influence form the target strength. In the end, regardless of the type of analysis performed, an effective and robust analysis of the fate of the ejecta cannot ignore the intrinsic uncertainties introduced by the modelling assumptions.

From a perspective of initial conditions generation, the \emph{position-based} ejecta formulation guarantees the highest flexibility as it allows generating samples from the entirety of the ejection speed spectrum. This characteristics allows the analysis also of low velocity particles, which is of particular importance when considering low-gravity environments such as the ones of asteroids. Given the small escape velocities, a large number of small and slow particles may be generated by an hypervelocity impact and is thus important to be able to generate initial conditions also for these particles. In the case of the \emph{position-based} formulation, we can also select the ejection speed model we want to adopt. As we have seen in the sensitivity analysis, this choice indeed influences the fate of the particles. However, it is not trivial to decide which model should be preferred among the ones proposed by Housen and Richardson. Additional tests and real mission data is probably required to make a more informed decision and, possibly, better tune the parameters defining these models. When looking at the \emph{speed-based} formulations, instead, we incur in the risk of neglecting parts of the low-velocity portion of the ejecta cloud. Therefore, this formulation is less suitable to analyse the fate of the ejecta in the neighbourhood of asteroids, especially when considering smaller asteroids and higher strength materials. However, the model can still capture the bulk of the ejecta cloud, by modelling most of the ejected mass. In addition, it can be useful when only the information on the ejection speed is required or of interest, without passing through the ejection location inside the crater. In fact, most of the time, the ejection location can be approximated with a point on the asteroid's surface.
Finally, the idea of using a combination of distributions to model the overall ejecta cloud in a modular fashion can be leveraged to plug-in new and more accurate models, but also to tailor the parameters of the models given new inputs from actual impacts; for example, by comparing the simulated results to images of impact cratering events such as the ones of Hayabusa2 and DART.

\section*{Acknowledgements}
This project has received funding from the European Union’s Horizon 2020 research and innovation programme under the Marie Sklodowska-Curie grant agreement No 896404 - CRADLE.


\appendix

\section{Coefficients derivation for the correlated distribution}
\label{app:corr-params}
The correlated distribution of \cref{eq:su-corr-sachse} has two additional parameters with respect to the uncorrelated one, specifically, $b$ and $\bbar$. The computation of the parameters is thus more complex, also given the different codependencies between them. For the correlated distribution we follow the approach of Sachse \citep{sachse2015correlation}, which introduces three additional coefficients ($\alpha$, $\beta$, $\gamma$) with the following meaning:

\begin{align}
	p_s(s) = \int_{\umin}^{\umax} p_{su} (s, u) \, d\!u \: &\sim \: s^{-1-\alpha} \label{eq:alpha} \\
	p_{\bar{u}}(s) = \frac{1}{p_s(s)} \int_{\umin}^{\umax} u \, p_{su} (s, u) \, d\!u \: &\sim \: s^{-\beta} \label{eq:beta} \\
	m(u) = \frac{4}{3} \pi \rho \int_{\smin}^{\smax} s^3 p_{su} (s, u) \, d\!s \: &\sim \: u^{-1-\gamma} \label{eq:gamma}
\end{align}

Therefore, $\alpha$ regulates the amount of ejected particles, $\beta$ the slope of the size-averaged ejection speed, and $\gamma$ the amount of ejected mass as a function of the speed. The parameter $\alpha$ depends on the target material and must be greater than 0; typically $\alpha$ ranges from 1.5 for loose to 3 for solid targets. The parameter $\beta$ is always greater than zero and, usually smaller than 1, while $\gamma$ depends on the target material and ranges between 1 for porous to 2 for non-porous materials \citep{sachse2015correlation}. Sachse also shows that $\alpha$, $\beta$ and $\gamma$ are related to $\abar$, $\bbar$ and $\gbar$ as follows:

\begin{align}
	\abar &= \alpha \\
	\bbar &= \frac{3 - \alpha - \beta}{\gamma - 1} \label{eq:bbar_vs_beta} \\
	\gbar &= \gamma + \frac{\alpha - 3}{\bbar}
\end{align}

Given the limits in $\alpha$, $\beta$ and $\gamma$, we also have range limitations for $\abar$, $\bbar$ and $\gbar$ that need to be taken into account. Specifically, we impose that $\abar$ and $\gbar$ must satisfy the following mathematical limitations: $0 < \abar < 3$ and $0 <\gbar < 1$ \citep{sachse2015correlation}.

In general, to fully specify \cref{eq:su-corr-sachse}, we can start by defining a value for the maximum ejection speed and the minimum particle size. Both these quantities can be reasonably assumed or computed. For the maximum ejection speed, we can refer to \cref{eq:speed-housen} or \cref{eq:speed-richardson}, while for the minimum size, this can be reasonably assumed (10-100 \si{\micro\meter}) or assumed equal to the minimum size detectable by an instrument \citep{sachse2015correlation}. We can then select the values for $\abar$ and $\gbar$ as they depend on the target material. To fully define the distribution we are left with  $A$, $\bbar$, and $b$, which must be selected to satisfy the mass conservation equation.

In addition, $b$ is related to $\bbar$ by the Heaviside function because the Heaviside function returns one only if its argument is greater than zero. Practically, this expression limits the maximum ejection speed as a function of the particle size. We thus have: $b = u \cdot s^{\bbar}$. As we have already selected the minimum particle size, $\smin$, and the maximum speed, $\umax$, we can substitute them into this expression to find a relation between $b$ and $\bbar$. As discussed in \cref{subsubsec:u-dist-speed-based}, we can also specify the minimum ejection speed. By doing so and using the aforementioned expression for $b$ with the values of $\smax$ and $\umin$, we can obtain the desired value for $\bbar$. In fact, we have:

\begin{equation}  \label{eq:b_equating}
	\left( \frac{\smax}{\smin} \right)^{\bbar} = \left( \frac{\umax}{\umin} \right) \rm .
\end{equation}

As we are fixing the size range ($\smin$, $\smax$) and the ejection speed range ($\umin$ and $\umax$), we can invert \cref{eq:b_equating} to obtain $\bbar$ as follows:

\begin{equation}  \label{eq:bbar_solution}
	\bbar = \log_\zeta \eta \rm ,
\end{equation}

\noindent where $\zeta = \frac{\smax}{\smin}$ and $\eta = \frac{\umax}{\umin}$. At this point it is important to verify the validity of this solution. In fact, following \cref{eq:bbar_vs_beta}, $\bbar$ is a function of $\alpha$, $\beta$, and $\gamma$. The limitations in ranges of $\alpha$ and $\gamma$ impose limits also on $\beta$ and, as a consequence, on $\bbar$. Therefore, the value of $\bbar$ must be such that $\beta$ satisfies the following expression: $0 < \beta < \beta_{\rm max}$, where:


\begin{equation}  \label{eq:beta_max}
	\beta_{\rm max} = (3 - \alpha) \cdot \left( 1 - \frac{\gamma -1}{\gamma} \right)
\end{equation}

At this point, we can obtain the value of the last parameter, $A$, by using the mass conservation equation as follows:

\begin{equation} \label{eq:mass_conservation_corr}
	M_{\rm tot} = \frac{4}{3} \pi \rho \int_{\smin}^{\smax} s^3 p_s(s) \, d\!s = \int_{\umin}^{\umax} m(u) \, d\!u
\end{equation}

\begin{figure}[htb!]
	\centering
	\includegraphics[width=2.6in]{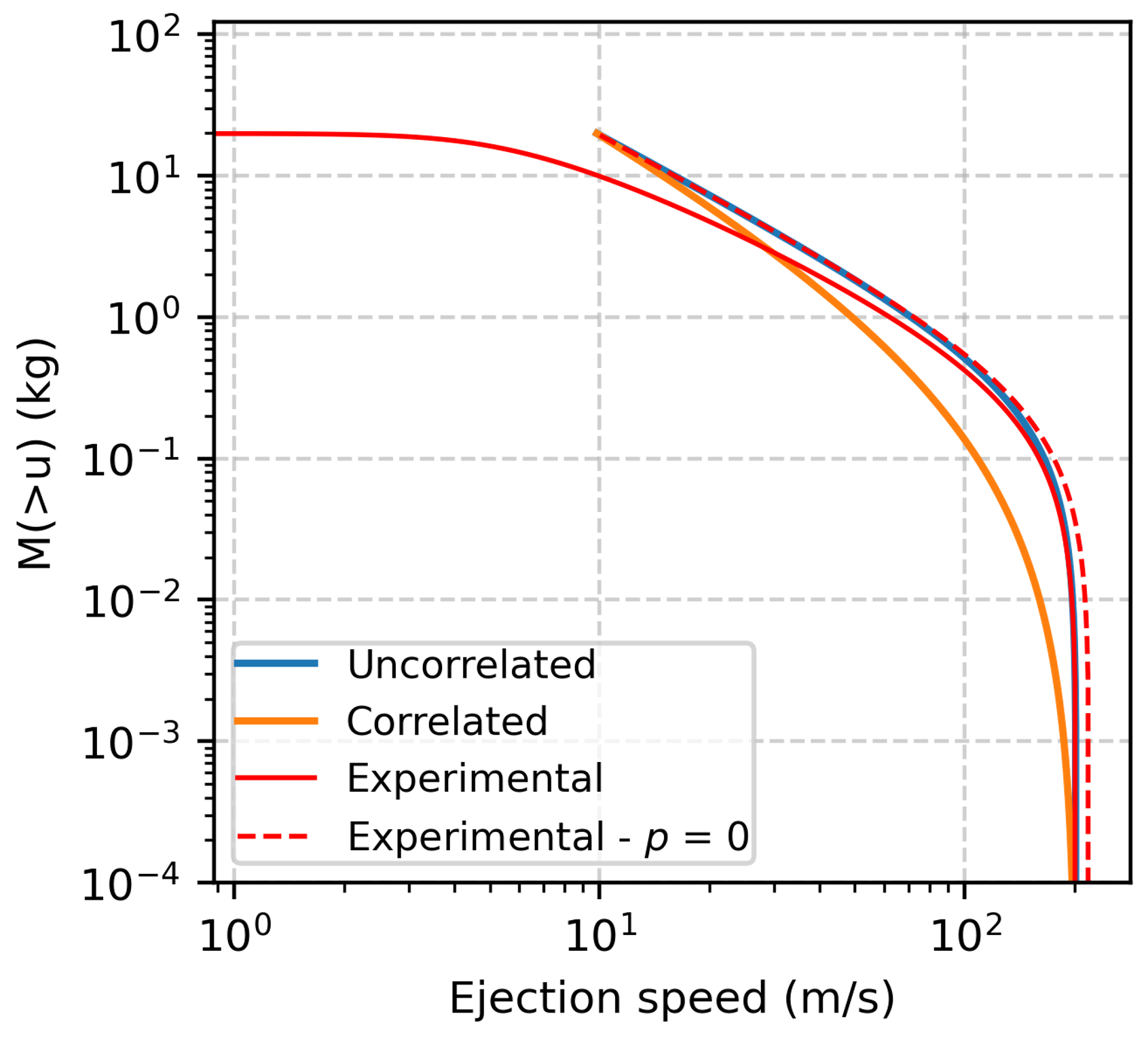}
	\caption{Comparison of cumulative ejection mass vs particle speed.}
	\label{fig:dist_comparison}
\end{figure}

An example of the characteristics of the distribution is given in \cref{fig:dist_comparison}, where we show the cumulative mass distribution as function of the ejection speed. In red, the two experimentally derived expressions \citep{housen2011ejecta}, one considering the material porosity (solid line) and the other without (dashed line) (\cref{eq:speed-housen} with $p \neq 0$ and $p=0$, respectively). It is possible to observe that the cumulative distribution derived from the uncorrelated case (blue line) closely matches this experimental correlation. The correlated case, instead, shows a steeper behaviour that is consistent with the limitations on the maximum velocity vs particle size.

\begin{figure}[htb!]
	\centering
	\includegraphics[width=2.6in]{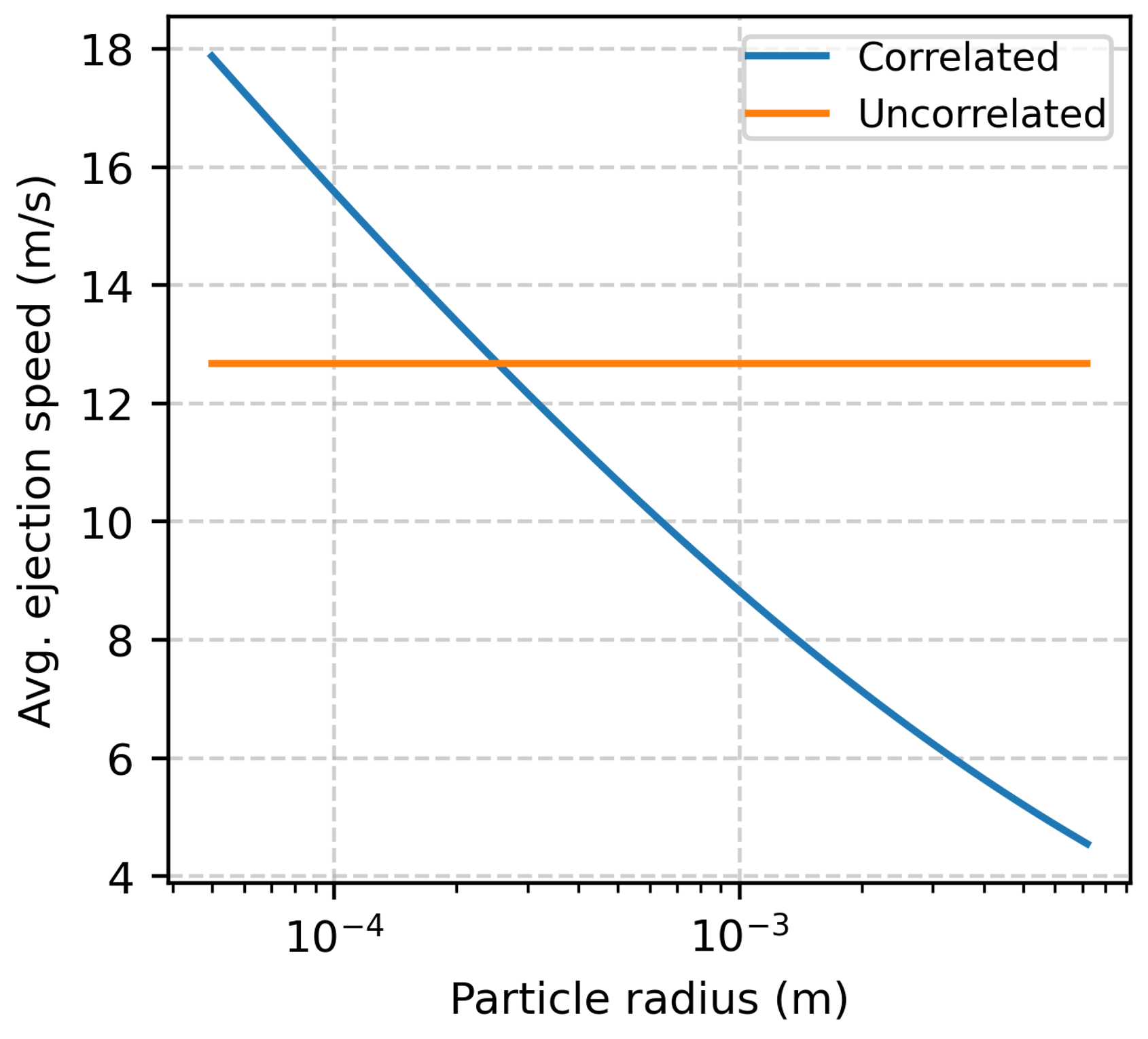}
	\caption{Comparison between the mean ejection speed for correlated and uncorrelated distributions.}
	\label{fig:umean_comparison}
\end{figure}

In addition, \cref{fig:umean_comparison} shows the average speed vs. particle size for the ejecta. We clearly see the different behaviour of the two distributions for which larger particles have, on average, smaller ejection speeds.

\section{Cumulative distributions}
\label{app:cumulative-dist}

The sampling of the ejecta distribution is based on the continuous description of the presented ejecta formulation. As we are expressing the ejecta distribution as a combination of probability density functions and conditional probability density functions (e.g., \cref{eq:conditional-position-based,eq:conditional-speed-based,eq:conditional-speed-based-corr}), we can use the corresponding Cumulative Distribution Functions (CDFs) to sample the distributions. In fact, by inverting the CDF we can directly sample the distribution. For example, let us have the PDF of the generic variable $\theta$, $p_\theta (\theta)$. Obtaining the relevant CDF and inverting it we have:

\begin{equation}
	{\rm CDF}_\theta^{-1} (\theta) = \Theta \quad\quad {\rm with } \; 0 \leq \Theta \leq 1
\end{equation}

By randomly sampling $\Theta$ between 0 and 1 we obtain the samples corresponding to $p_\theta (\theta)$. We use this procedure for all the distribution presented in this work, paying attention to the sampling order when conditional distributions are combined. For example, to sample \cref{eq:conditional-position-based}, we can first sample the size, $s$, and the launch position, $r$, distributions as they are independent from all the others. We then use the sampled values of $r$ to draw samples from the conditional distribution of the in-plane ejection angle, $\xi$. Finally, we perform an analogous procedure for the conditional distribution of the out-of-plane ejection angle, $\psi$, which depends on the samples of $r$ and $\xi$.
\bigbreak
The cumulative distributions are also used for the sampling procedure of \cref{sec:sampling} and the computation of the representative fragments. As an example, let us consider \cref{eq:conditional-position-based}, in the variables $s$, $r$, $\xi$, $\psi$. We want to compute the number of fragments in a generic four-dimensional cell of the state space, whose limits can be expressed as: $\mathcal{D} = \left[ s_0, s_1 \right] \cap \left[ r_0, r_1 \right] \cap \left[ \xi_0, \xi_1 \right] \cap \left[ \psi_0, \psi_1 \right]$. To do so, we can exploit the cumulative distribution functions as follows:

\begin{align}
	n_f (\mathcal{D}) = &\Big( P_s (s_1) - P_s (s_0) \Big) \cdot \Big( P_r (r_1) - P_r (r_0) \Big) \cdot \Big( P_{\xi|r} (\xi_1|\bar{r}) - P_{\xi|r} (\xi_0|\bar{r}) \Big) \cdot \nonumber \\
	& \cdot \Big( P_{\psi|\xi, r} (\psi_1| \bar{\xi}, \bar{r}) - P_{\psi|\xi, r} (\psi_0| \bar{\xi}, \bar{r}) \Big)
\end{align}

where $P_s$, $P_r$, $P_{\xi|r}$, and $P_{\psi|\xi, r}$ are the conditional distributions in size, launch position, in-plane and out-of-plane ejection angles, respectively. Because the conditional distribution in $\xi$ requires the values of $r$ and the one in $\psi$ the values of both $r$ and $\xi$, we need to specify their values. In this work, the values are assumed as the mean of the corresponding interval (e.g., $\bar{r} = \left( r_0 + r_1 \right) / 2$).

\bigbreak
The following sections contain a review of all the CDF derived for the models presented in \cref{sec:distributions}, except for the equations based on Normal distributions (i.e., \cref{eq:xi-distribution-gaussian,eq:psi-distribution-gaussian,eq:psi-dist-gaussian-speed-based}), whose sampling can be performed relying on algorithms available in most scientific programming languages (in this work the Python package \emph{scipy} is used \citep{2020SciPy-NMeth}.)

\subsection{Position-based formulation}
In this section, the CDF relative to the \emph{position-based} formulation are reported.
\bigbreak
\noindent\emph{Size distribution} \\
\cref{eq:size-cdf} is the CDF of the particle size distribution. Inverting the distribution to obtain $s$, we can sample the particle size according to the power law.

\begin{equation}  \label{eq:size-cdf}
	P_s (s) = \frac{N_r}{N_{\rm tot}} \cdot \left( \smin^{-\abar} - s^{-\abar} \right)
\end{equation}

\bigbreak
\noindent\emph{Launch position distribution} \\
\cref{eq:position-cdf} is the CDF of the launch position distribution. Inverting the distribution for $r$ and drawing random samples between 0 and 1 for $P_r$ allows sampling the corresponding distribution.

\begin{equation}  \label{eq:position-cdf}
	P_r (r) = k \cdot \frac{\rho}{M_{\rm tot}} \cdot (r^3 - \rmin^3)
\end{equation}

\bigbreak
\noindent\emph{In-plane ejection angle distribution} \\
\cref{eq:xi-cdf-pos} is the CDF of the in-plane ejection angle distribution of \cref{eq:xi-distribution}. Unfortunately, inverting this equation is not possible; however, the variable $\xi$ can still be sampled by first drawing a random sample for $P_{\xi|r}$ between 0 and 1, and then solving the equation numerically for $\xi$. As this is a conditional distribution, we first need to sample the launch position $r$.

\begin{align}  \label{eq:xi-cdf-pos}
	P_{\xi|r} (\xi | r; \phi) = \frac{1}{2\pi} \cdot & \bigg[ \xi + \frac{\cos \phi}{C_0} \Big( C_1 \sin \xi - C_2 \sin 3\xi + \nonumber \\
	& - C_3 \sin 5\xi + C_4 \sin 7\xi + C_5 \sin 9\xi \Big) \cdot \left( 1 - \frac{r^8}{\rmax^8} \right) \bigg],
\end{align}

where $C_0 = 25200$, $C_1 = -10710$, $C_2 = -2730$, $C_3 = -378$, $C_4 = 45$, and $C_5 = 35$.

\bigbreak
\noindent\emph{Out-of-plane ejection angle distribution} \\
\cref{eq:psi-cdf-uniform} is the CDF of the uniform model of the out-of-plane ejection angle distribution (\cref{eq:psi-distribution-uniform}). This expression can be easily inverted and solved for $\psi$ to sample the distribution, provided we first sample of $r$ and $\xi$ as this is a conditional distribution.

\begin{equation} \label{eq:psi-cdf-uniform}
	P_{\psi|\xi, r} (\psi| \xi, r; \phi) = \frac{\sin \left(\psi + K_\psi(\xi, r; \phi) \right) - \sin \left(\psi_{\rm min} + K_\psi (\xi, r; \phi) \right)}{\sin \psi_{n, \rm max} - \sin \psi_{n, \rm min}}
\end{equation}

\subsection{Speed-based formulation}  \label{subsec:app-cdf-speed}
In this section, we present the analogous CDF for the \emph{speed-based} formulation (\cref{subsec:speed-dist}).

\bigbreak
\noindent\emph{In-plane ejection angle distribution} \\
\cref{eq:xi-cdf-speed} is the CDF for the $\xi$ distribution described by \cref{eq:xi-dist-lobed-speed-based}. Analogously to \cref{eq:xi-cdf-pos}, the sampling of $\xi$ requires solving numerically the equation. In this case, however, the distribution is not conditioned.

\begin{equation}   \label{eq:xi-cdf-speed}
	P_{\xi} (\xi) = \frac{1}{C_6 \pi} \cdot \left[ C_7 \xi + \phi \cdot \left( C_1 \sin \xi - C_2 \sin 3\xi - C_3 \sin 5\xi + C_4 \sin 7\xi + C_5 \sin 9\xi \right) \right],
\end{equation}

where $C_1 = -10710$, $C_2 = -2730$, $C_3 = -378$, $C_4 = 45$, $C_5 = 35$, $C_6 = 56700$, and $C_7 = 28350$. 

\bigbreak
\noindent\emph{Out-of-plane ejection angle distribution} \\
\cref{eq:psi-cdf-speed} is the CDF of the $\psi$ distribution for the uniform model. Solving the expression for $\psi$ allows sampling the distribution in a spherically uniform fashion inside the specified range.

\begin{equation}  \label{eq:psi-cdf-speed}
	P_{\psi|\xi} (\psi |\xi ; \phi) = \frac{ \sin{\left( \psi - \bar{K}_\psi (\xi; \phi) \right)} - \sin{\left( \psi_{n, \rm min} \right)} }{ \sin{\psi_{n, \rm max}} - \sin{\psi_{n, \rm min}} }
\end{equation}

\bigbreak
\noindent\emph{Speed distribution} \\
\cref{eq:u-cdf-speed} is the CDF for the speed distribution. This is a conditional distribution; provided $\xi$ and $\psi$ are sampled before, inverting the following expression for $u$ allows the sampling of the speed distribution.

\begin{equation}  \label{eq:u-cdf-speed}
	P_{u | \psi, \xi} (u | \psi, \xi; \phi) = \frac{K_u}{\gbar} \cdot \frac{\sin{\left( \psi - \bar{K}_\psi (\xi; \phi) \right)}}{\sin \psi} \cdot \left( u_{\rm min}^{-\gbar} - u^{-\gbar} \right) \rm ,
\end{equation}

\noindent where $K_u = \frac{\gbar}{u_{n, \rm min}^{-\gbar} - u_{n, \rm max}^{-\gbar}}$ as for \cref{eq:un-dist-speed-based}.

\subsection{Correlated speed-based formulation}

For the case of the correlated distribution, the sampling of the ejection angles, $\xi$ and $\psi$ are analogous to \cref{subsec:app-cdf-speed}. The main difference resides in the sampling of the size-speed distribution (\cref{eq:su-corr-oblique}). To sample this distribution, we rearrange it as:

\begin{equation}
	p_{su|\xi, \psi} (s, u | \xi, \psi) = p_{u |s, \xi, \psi} (u | s, \xi, \psi) \cdot p_{s | \xi, \psi} (s | \xi, \psi)\rm,
\end{equation}

and find the CDF of the two conditional distributions as follows:

\begin{equation} \label{eq:cdf-corr-s}
	P_{s | \xi, \psi} (s | \xi, \psi) = \frac{A}{\gbar} \cdot \left[ \left( \frac{\umin}{| J |} \right)^{-\gbar} \cdot \frac{1}{\abar} \left( \smin^{-\abar} - s^{-\abar} \right) + b^{-\gbar} \cdot \frac{ s^{-\abar + \bbar \gbar} - \smin^{- \abar + \bbar \gbar} }{ \abar - \bbar \gbar } \right] \rm,
\end{equation}

\begin{align} \label{eq:cdf-corr-u}
	P_{u |s, \xi, \psi} (u | s, \xi, \psi) = &\frac{ \left( \unmin b s^{-\bbar} u \right)^{-\gbar} }{ \left( \unmin^{-\gbar} - b^{-\gbar} s^{\bbar \gbar} \right) | J |^{-\gbar}} \cdot \Bigg\{ \left( \frac{u}{| J |} b s^{-\bbar} \right)^{\gbar} - \left( \frac{\umin}{| J |} u \right)^{\gbar} + \nonumber \\
	& \left[ \left( \frac{\umin}{| J |} u \right)^{\gbar} - \left( \umin b s^{-\bbar} \right)^{\gbar} \right] \cdot \Theta \left[ b s^{-\bbar} - \frac{u}{| J |} \right]  \Bigg\} \rm,
\end{align}

where $ |J| = \frac{\sin{\left( \psi - \bar{K}_\psi (\xi; \phi) \right)}}{\sin \psi}$ as in \cref{eq:speed-based-tranformation}. Both \cref{eq:cdf-corr-s,eq:cdf-corr-u} cannot be inverted and must be solved numerically to sample them. Again, as we have a combination of conditional probability distributions, we first sample $\xi$, then $\psi$, then $s$, and, finally, $u$.

\printcredits

\bibliographystyle{cas-model2-names}

\bibliography{references}

%

\end{document}